\documentclass[]{spie}  

\usepackage{amsmath,amsfonts,amssymb}
\usepackage{graphicx}
\usepackage{subcaption}
\usepackage{bm}
\usepackage{float}
\usepackage{booktabs}
\usepackage{amsmath}
\usepackage[colorlinks=true, allcolors=blue]{hyperref}

\title{Updating the SCExAO/CHARIS polarimetric calibration following the Nasmyth beam-switcher upgrade}

\author[1]{Thomas McIntosh} 
\author[1]{Manxuan Zhang}
\author[1]{Briley L. Lewis}    
\author[2,3]{Miles Lucas}
\author[1]{Maxwell A. Millar-Blanchaer}
\author[1]{Jaren Ashcraft}
\author[4,3]{Kyohoon Ahn}
\author[5]{Jeffrey Chilcote}
\author[6,3]{Thayne Currie}
\author[7,3]{Vincent Deo}
\author[8]{Yoshiyuki Doi}
\author[9]{Tyler Groff}
\author[3,12,2,11]{Olivier Guyon}
\author[3]{Takashi Hattori}
\author[3,12]{Tomoyuki Kudo}
\author[15,16,17]{Kellen Lawson}
\author[3,12]{Julien Lozi}
\author[8]{Yosuke Minowa}
\author[3]{Yuhei Takagi}
\author[10]{Rob G. Van Holstein}
\author[14,13,3]{Sébastien Vievard}

\affil[1]{University of California, Santa Barbara, Santa Barbara, CA 93106, United States}
\affil[2]{Steward Observatory, University of Arizona, Tucson, AZ 85719, United States}
\affil[3]{National Astronomical Observatory of Japan, Subaru Telescope, 650 North Aohoku Place, Hilo, HI 96720, USA}
\affil[4]{Korea Astronomy and Space Science Institute (KASI), Daejeon 34055, Republic of Korea}
\affil[5]{Department of Physics and Astronomy, University of Notre Dame, 225 Nieuwland Science Hall, Notre Dame, IN 46556 USA}
\affil[6]{Department of Physics and Astronomy, University of Texas at San Antonio, San Antonio, TX 78006, USA}
\affil[7]{Optical Sharpeners, Manosque, France}
\affil[8]{National Astronomical Observatory of Japan, Mitaka, Tokyo 181-8588, Japan}
\affil[9]{NASA Goddard Space Flight Center, Greenbelt, MD 20771}
\affil[10]{European Southern Observatory, Vitacura, Santiago, Chile}
\affil[11]{College of Optical Sciences, University of Arizona, Tucson, AZ 85721, USA}
\affil[12]{Astrobiology Center, 2-21-1, Osawa, Mitaka, Tokyo 181-8588, Japan}
\affil[13]{Space Science and Engineering Initiative, College of Engineering, University of Hawaii, Hilo, HI 96720, USA}
\affil[14]{Institute for Astronomy, University of Hawaii, Hilo, HI 96720, USA}
\affil[15]{Center for Space Sciences and Technology, University of Maryland, Baltimore County, 1000 Hilltop Circle, Baltimore, MD 21250, USA}
\affil[16]{Astrophysics Science Division, NASA-GSFC, 8800 Greenbelt Rd, Greenbelt, MD 20771, USA}
\affil[17]{Center for Research and Exploration in Space Science and Technology, NASA-GSFC, 8800 Greenbelt Rd, Greenbelt, MD 20771, USA}

\authorinfo{Further author information: (Send correspondence to Thomas McIntosh)\\E-mail: thomasmcintosh@ucsb.edu}

\begin{document} 
\maketitle

\begin{abstract}
Subaru/SCExAO/CHARIS enables near-infrared integral field spectropolarimetry. Quantitative polarimetry is useful for a variety of science cases, particularly measurements related to dust grain properties in circumstellar disks. This capability requires correcting for polarization effects from the optical path via a Mueller matrix model. We present an updated model accounting for the recently installed SCExAO near-infrared wavefront sensor YJH50 dichroic beamsplitter and the major Subaru Nasmyth beam-switcher upgrade. Using internal light source measurements from before and after the beam switcher installation, we find an elliptical retarder model for the image derotator improves polarimetric accuracy over the previous linear retarder model. We additionally find the YJH50 dichroic produces faint polarization effects that we cannot characterize with our Mueller matrix modeling capabilities, and that the Nasmyth beam-switcher has minimal polarization effects other than inducing a sign flip in $Q$ and $V$ polarized light. Using unpolarized standard star calibration measurements, we fit the diattenuation of Subaru's tertiary mirror as a function of wavelength and find that the diattenuation has increased since the previous CHARIS calibration. We calculate that the polarimetric accuracy of the model in the degree of linear polarization ranges from 0.02\% to 0.12\% for a 1\% polarized target. This model update will soon be incorporated into CHARIS's data processing pipeline, and should be used for any polarimetric data taken after the Nasmyth beam-switcher update in October 2025. Additionally, we provide the code for this calibration as part of an open-source Python package for polarimetric calibration called \texttt{pyPolCal}, enabling straightforward re-calibration of the system after any future changes, e.g. the recent recoating of M3.
\end{abstract}


\section{INTRODUCTION}
\label{sec:intro}
Polarimetric differential imaging (PDI) is a high-contrast imaging technique that enables light from circumstellar disks to be separated from their host stars \cite{Kuhn_2001, Currie2023a, follette2023intro}. Unpolarized starlight scattering off small particles induces a linear polarization, creating a natural separation from the unpolarized host starlight. This separation allows the starlight to be subtracted from the image. PDI in the near-infrared (NIR) has been successful in detecting faint circumstellar disks and characterizing their morphologies \cite{perrin2014polarimetry, esposito2020debris}; quantitative NIR polarimetry facilitates an even wider range of science cases for circumstellar disks and beyond. For example, young, hot exoplanets emit thermal radiation with a polarization that depends on the atmosphere's temperature gradient and the distribution of cloud particles \cite{dekok2011characterizing}. With circumstellar disks, polarization fraction measurements and/or the scattering phase function can reveal dust grain properties such as porosity, which can inform planet formation models \cite{2024MNRAS.533.2473C,dykes2024scexao}. For example, porosity is a proposed mechanism allowing grains to overcome the so-called radial drift barrier, where centimeter- to decimeter-sized grains spiral into their central star due to drag forces \cite{Okuzumi_2012,Garcia_2020}. 

NIR polarimeters have been or will be deployed on many high-contrast imaging platforms. Example systems include the Gemini Planet Imager/GPI 2.0 \cite{Chilcote2020-cf,Millar-Blanchaer2016-nn,gpisloan,brucegpi}, VLT/NaCo (decommissioned) \cite{deregt2024polarimetric}, VLT/SPHERE/IRDIS \cite{de_Boer_2020,van_Holstein_2020}, Subaru/IRCS \cite{honda2022subaru}, Subaru/HiCIAO (decommissioned) \cite{Suzuki2009-hm}, and the recently commissioned Keck/NIRC2-Pol \cite{lewis2026nirc}. 

Polarimetry with the Coronagraphic High Angular Resolution Imaging Spectrograph (CHARIS) at the Subaru Telescope provides \textit{JHK} integral field spectropolarimetry capabilities not replicated by other instruments \cite{Lawson2021}. CHARIS is located behind Subaru's extreme adaptive optics (AO) platform SCExAO, which feeds light into multiple high-contrast imaging instruments \cite{Lozi_2018,Jovanovic_2015}. The instrument's spectropolarimetric mode facilitates coronagraphic polarimetry across 22 wavelength bins in its broadband \textit{JHK} mode or 20 wavelength bins in its so-called ``high-resolution'' modes in \textit{J}, \textit{H}, or \textit{K}. CHARIS's broadband \textit{JHK} polarimetric capabilities are supported by a comprehensive Mueller matrix model for polarimetric calibration \cite{hart2021characterizationinstrumentalpolarizationeffects}. CHARIS polarimetry has provided crucial insights into the structure and dust properties of protoplanetary disks \cite{dykes2024scexao, mullin2026time} and supporting evidence in favor of protoplanets embedded in these disks (i.e. AB Aurigae b \cite{Currie2022,dykes2024scexao}).

A primary challenge of quantitative polarimetry is calibrating the instrument. As shown in the Fresnel equations, an electromagnetic wave interacting with a surface will change the amplitude and phase of its orthogonal electric field components. In an astronomical polarimeter, this means that interactions with components in the optical path alter the target's incident polarization state. These effects are often significant, rendering many quantitative science cases impossible without a detailed Mueller matrix model. A Mueller matrix modifies a Stokes vector, which represents light's polarization state. We assign a Mueller matrix to each optical component, allowing us to fully describe the instrument's polarization effects and reconstruct a target's incident polarization state.

Ref. \citenum{vanholstein2020calibrationinstrumentalpolarizationeffects} first characterized CHARIS's polarization effects using a polarized internal calibration light source, fitting Mueller matrix model parameters for each wavelength bin. Subsequently, Ref. \citenum{hart2021characterizationinstrumentalpolarizationeffects} expanded on their work by repeating internal calibrations and using unpolarized standard stars to calibrate the effects of the telescope mirrors. They derived physically motivated models as a function of wavelength, greatly reducing the number of parameters to fit. 

In the five years since the most recent calibration, there have been significant changes to the Subaru infrastructure and SCExAO/CHARIS, particularly the installation of the Nasmyth beam-switcher (NBS) and new optics to enable near-infrared wavefront sensing. These changes to the optical path could have significant polarization effects, and therefore motivate recalibration. The NBS was installed in 2025, directing light from the AO3k adaptive optics system (formerly AO188) to science instruments like SCExAO \cite{zheng2022optical}. Dichroic beamsplitters that send light to the near-infrared wavefront sensor (NIRWFS) were also recently installed \cite{lozi2024ao3ksubaruonskyresults}, and are known to have significant polarimetric effects \cite{heath2020}. Recalibration is also justified by incidental changes over time in CHARIS's existing optics; for example, small amounts of dust or debris that collect on an optic over time could affect how it polarizes light. Characterizing these new optics and revisiting the model for existing optics are necessary to preserve CHARIS's quantitative polarimetric capabilities. Additionally, previous works did not provide a streamlined process for calibrating CHARIS, and the source code for both previous calibrations was not publicly available, making recalibration difficult. Since even minor changes over time in optics can affect model accuracy, open-source code and a detailed description of the fitting process are necessary to allow for regular calibrations and thus preserve polarimetric accuracy. 

In this work, we present an update to the Mueller matrix model for CHARIS and develop open-source software to allow for regular calibration. In Section \ref{sec:conventions-defns}, we describe the mathematical conventions we adopt for modeling the polarization of light and its interactions with optics. In Section \ref{sec:charis-optical-path}, we describe CHARIS's optical path and how its dual-channel polarimeter functions. In Section \ref{sec:model}, we outline the Mueller matrix model of the optical path and the open source software we created to carry out polarimetric calibration. In Section \ref{sec:internal-cal}, we present the procedure and results of internal source calibrations. In Section \ref{sec:onsky-cal}, we update the Mueller matrix model for Subaru's tertiary mirror using unpolarized standard stars. Finally, in Section \ref{sec:pol-eff}, we calculate the polarimetric accuracy and efficiency of the updated model, and we discuss our results and conclusions in Section \ref{sec:conclusion}.

\section{CONVENTIONS AND DEFINITIONS}
\label{sec:conventions-defns}
\subsection{Stokes Vectors}
\label{subsec:stokesvec}
The polarization of an electromagnetic wave is defined by the orientation of its electric field vector in space. We quantify a light beam's observable polarization using a time-averaged vector sum of the orientations of its constituent waves. A Stokes vector is convenient for this task:
\begin{equation}
    \mathbf{S}=\begin{pmatrix}
        I \\
        Q \\
        U \\
        V
    \end{pmatrix}
\end{equation}

$I$ is the beam's total intensity. We define $Q$ as the difference in intensities between vertical and horizontal linear polarizations, $U$ as the difference in intensities between $\pm45^\circ$ linear polarizations (where the angles are defined as counterclockwise from the vertical when looking into the beam), and $V$ as the difference in intensities between right-handed circular and left-handed circular polarizations. Linear polarization is when the electric field vector oscillates in a plane at some specified angle, and circular polarization is when the electric field vector traces out a spiral. Elliptical polarization is a combination of the two.
\begin{equation}
\label{eq:stokes-defn}
    \begin{pmatrix} I \\Q \\ U \\ V \end{pmatrix} = \begin{pmatrix} \langle I_{\mathrm{pol}}\rangle+\langle I_{\mathrm{unpol}}\rangle \\\langle I_{\mathrm{V}} \rangle - \langle I_{\mathrm{H}} \rangle \\ \langle I_{+45^\circ} \rangle - \langle I_{-45^\circ} \rangle \\ \langle I_{\text{RCP}} \rangle - \langle I_{\text{LCP}} \rangle \end{pmatrix}
\end{equation}
The brackets represent a time average. The V and H subscripts stand for vertical and horizontal linear polarization. RCP denotes right-handed circular polarization, and LCP denotes left-handed circular polarization. The sign conventions of $Q$, $U$, and $V$ can vary across disciplines; here, we adopt the convention outlined in Section 2 of Ref. \citenum{de_Boer_2020}, as written above in Equation \ref{eq:stokes-defn}. 

We define the degree of linear polarization (DoLP) as the fraction of a beam's intensity that is linearly polarized: 
\begin{equation}
    \label{eq:dolp}
    \mathrm{DoLP}= \frac{\sqrt{Q^2+U^2}}{I}
\end{equation}
We also define the angle of linear polarization (AoLP) as:
\begin{equation}
\label{eq:aolp}
\mathrm{AoLP} = \frac{1}{2}\arctan{\bigg(\frac{U}{Q}\bigg)}
\end{equation}

\subsection{Mueller Matrices}\label{subsec: MM}
Mueller matrices allow us to model how an incoming Stokes vector is modified by CHARIS's optical path. We use Mueller matrices to describe two main effects: retardance and diattenuation. Retardance is when a phase shift is introduced between orthogonal electric field components; this process converts one type of polarization to another, leaving incident unpolarized light unaffected. A linear retarder introduces a phase shift between orthogonal linear polarizations, transforming the polarization state. A circular retarder introduces a phase shift between left-hand circular and right-hand circular polarizations. Elliptical retardance is a combination of linear and circular retardance. Diattenuation is when one electric field component is attenuated differently than another, causing a preferential absorption or transmission of specific polarization states. Diattenuation causes so-called instrumental polarization, where telescope and/or instrument optics introduce additional linear polarization, converting unpolarized light into linearly polarized light. The equation below shows how a general Mueller matrix transforms a Stokes vector: \cite{hart2021characterizationinstrumentalpolarizationeffects} \begin{equation}
\begin{pmatrix}
I_{\mathrm{out}} \\
Q_{\mathrm{out}} \\
U_{\mathrm{out}} \\
V_{\mathrm{out}}
\end{pmatrix}
=
\begin{pmatrix}
I \to I & Q \to I & U \to I & V \to I \\
I \to Q & Q \to Q & U \to Q & V \to Q \\
I \to U & Q \to U & U \to U & V \to U \\
I \to V & Q \to V & U \to V & V \to V
\end{pmatrix}
\begin{pmatrix}
I_{\mathrm{in}} \\
Q_{\mathrm{in}} \\
U_{\mathrm{in}} \\
V_{\mathrm{in}}
\end{pmatrix},
\end{equation}
\subsection{Polarimetric Accuracy}
\label{subsec:polacc}
We assess how well the model fits the data using an adaptation of the procedure from Ref. \citenum{van_Holstein_2020}. We first compute the corrected sample standard deviation of residuals ($s_{\mathrm{res}}$) of the model's fit to a calibration dataset:
\begin{equation}
\label{eq:sres}
    s_{\mathrm{res}} = \sqrt{\frac{\sum_{i=1}^{n} r_i^2}{n - k}}
\end{equation}
 where $r_i$ are the residuals, $n$ is the number of data points, and $k$ is the number of model parameters fit to the dataset. To calculate the polarimetric accuracy of the model, we compute the standard error of the mean:
 \begin{equation}
 \label{eq:sem}
     s_{\bar{x}}=\frac{s_{\mathrm{res}}}{\sqrt{n}}
 \end{equation}
 where $n$ is the number of data points. Previous CHARIS calibrations use $s_{\mathrm{res}}$ as the polarimetric accuracy \cite{vanholstein2020calibrationinstrumentalpolarizationeffects,hart2021characterizationinstrumentalpolarizationeffects}; however, we determine that $s_{\bar{x}}$ is more appropriate, as it scales with the number of calibration data points. We note that spatial variability of the incident light's polarization across the detector could affect the accuracy in practice, although that effect is not accounted for here. We denote $s_{\mathrm{\bar{x}}}$ of the polarized calibration dataset as $s_{\mathrm{rel}}$. This is the error that scales with polarized intensity. We denote $s_{\mathrm{\bar{x}}}$ of the internal and on-sky unpolarized calibration datasets as $s_{\mathrm{unpol}}$ and $s_{\mathrm{sky}}$, respectively. The absolute error that scales with overall intensity is 
 \begin{equation}
 \label{eq:sabs}
     s_{\mathrm{abs}}=\sqrt{s^2_{\mathrm{unpol}}+s^2_{\mathrm{sky}}}
 \end{equation} 
 We then compute the polarimetric accuracy of the model in $Q$ and $U$ for a hypothetical target as:
 \begin{equation}
     s_Q=s_{\mathrm{abs}}+s_{\mathrm{rel}}|q|
 \end{equation}
 \begin{equation}
     s_U=s_{\mathrm{abs}}+s_{\mathrm{rel}}|u|
 \end{equation}
$q$ and $u$ are Stokes $Q$ and $U$ divided by the total intensity $I$. Since we are interested in the polarimetric accuracy as a function of DoLP ($p$) and AoLP ($\chi$) we compute:
\begin{equation}
    |q|=|p\cos{2\chi}|
\end{equation}  
\begin{equation}
    |u|=|p\sin{2\chi}|
\end{equation}
Then, we calculate the polarimetric accuracy of the model for a hypothetical target using Gaussian error propagation:
\begin{equation}
\label{eq:s_p}
    s_p= \sqrt{\frac{q^2s_Q^2+u^2s_U^2}{q^2+u^2}}
\end{equation}
 \begin{equation}
 \label{eq:s_chi}
     s_{\chi}=\frac{\sqrt{q^2s_U^2+u^2s_Q^2}}{2(q^2+u^2)}
 \end{equation}
Targets with a faint polarization signal have an $s_p$ and $s_{\chi}$ dominated by $s_{\mathrm{abs}}$. Targets that are polarized on the order of tens of percent, like some circumstellar disks, have a greater dependence on $s_{\mathrm{rel}}$ in $s_p$ and $s_{\chi}$. To calculate the total error on a DoLP or AoLP measurement, the observer must propagate this error in the model with measurement noise. 

\section{CHARIS OPTICAL PATH}
\label{sec:charis-optical-path}

\begin{figure}
    \centering
    \includegraphics[width=0.8\linewidth]{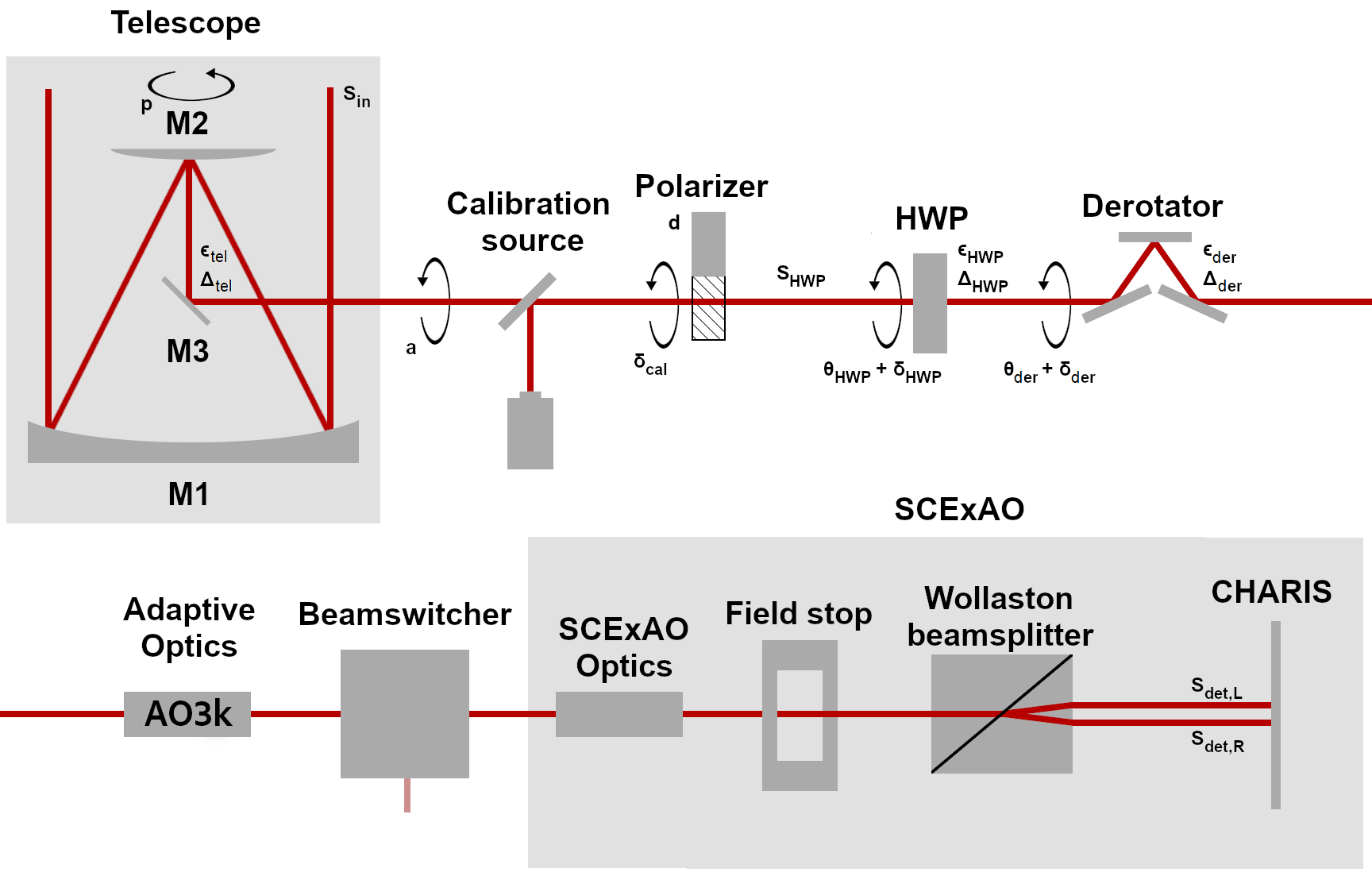}
    \caption{Overview of components in the CHARIS optical path that are relevant to the spectropolarimetric mode, adapted from Ref. \citenum{vanholstein2020calibrationinstrumentalpolarizationeffects}. The red beam is the path of the incident Stokes vector. The black circular arrows and their accompanying $\theta$ and $\delta$ represent rotating components. $p$ and $a$ represent the parallactic and altitude rotations. We clarify the distinction between $\theta$ and $\delta$ rotation angles in Subsection \ref{subsec:math-description}. $\epsilon$ represents diattenuation and $\Delta$ represents retardance. For on-sky calibrations, the calibration source and polarizer are removed from the optical path.}
    \label{fig:path}
\end{figure}

Figure \ref{fig:path} shows components of the optical path that are relevant for CHARIS's spectropolarimetric mode. Light is reflected from Subaru's tertiary mirror (M3) directly into the optical path. We can neglect polarization effects from M1 and M2 due to their geometric symmetries. For internal calibrations, a calibration source injects 100\% $-Q$ polarized light into the optical path via an insertable linear polarizer, bypassing the telescope. CHARIS's spectropolarimetric capability uses dual-channel polarimetry, using a removable half-wave plate (HWP) to rotate the polarization state for calibration and a Wollaston prism as a polarizing beamsplitter. This design of polarimeter is only sensitive to linear polarization; any light converted to circular polarization is lost information. The HWP is directly downstream of the telescope; following the HWP is the image derotator (also referred to as a k-mirror), which is part of the AO3k facility adaptive optics system \cite{lozi2024ao3ksubaruonskyresults}. AO3k contains multiple removable dichroic beamsplitters that allow light to be simultaneously sent to a science instrument and the NIRWFS \cite{Lozi2023AO3k}. In this calibration update, we characterize the YJH50 dichroic: This dichroic sends 50\% of \textit{y}- to \textit{H}-band light to the NIRWFS, and the other 50\% to the science path. It additionally sends 100\% of \textit{K}-band light to the science path. Downstream of AO3k is the recently installed NBS, which feeds light into one of three science instruments: NINJA, IRCS, or SCExAO \cite{zheng2022optical}. As of July 2026, SCExAO feeds light into five instruments: CHARIS, VAMPIRES, REACH, FAST PDI, and the recently-commissioned FIRST-PL \cite{vampires,reach,fastpdi, Jovanovic_2015}. In CHARIS, a Wollaston prism splits light into two orthogonal polarizations, resulting in two images of the target on the detector per wavelength bin. The Wollaston prism and HWP allow us to determine Stokes $Q$ and $U$ through the double differencing method. 

By subtracting the two images from the Wollaston prism on either side of the detector, we find the \textit{single difference}:
\begin{equation}
\label{singledif}
    X^{\pm}=I_{\mathrm{det,R}}-I_{\mathrm{det,L}}
\end{equation}
The minus superscript denotes a $45^\circ$ rotation of the HWP relative to the plus, which flips the sign of the polarization.  We denote the plus superscript as the positive single difference and the minus superscript as the negative single difference. Note that the order of subtraction is a convention that can vary by instrument. Ours is consistent with the calibration source emitting $-Q$ polarized light in our coordinate system defined in Subsection \ref{subsec:stokesvec}. Note that the two previous CHARIS polarimetric calibration proceedings each use different conventions for single differencing \cite{hart2021characterizationinstrumentalpolarizationeffects, vanholstein2020calibrationinstrumentalpolarizationeffects}. The convention we adopt is from Ref. \citenum{hart2021characterizationinstrumentalpolarizationeffects} and is used in the CHARIS data processing pipeline (DPP), which we discuss in Subsection \ref{subsec:internal-procedure} \cite{hart2021characterizationinstrumentalpolarizationeffects}. $X_{\theta_{\mathrm{HWP}}^+}$ denotes a double difference, where ${\theta_{\mathrm{HWP}}^+}$ is the HWP angle of the positive single difference $X^+$ (Equation \ref{singledif}), defined as follows:
\begin{equation}
\label{doublediff}
    X_{\theta_{\mathrm{HWP}}^+}=\frac{1}{2}(X^+-X^-)
\end{equation}
Total intensity is derived from sums. We define the \textit{single sum} as:
\begin{equation}
\label{singlesum}
    I_{X^{\pm}}=I_{\mathrm{det,R}}+I_{\mathrm{det,L}}
\end{equation}
and the \textit{double sum} as: 
\begin{equation}
\label{doublesum}
    I_{X_{\theta_{\mathrm{HWP}}^+}}=\frac{1}{2}(I_{X^+}+I_{X^-})
\end{equation}
Using this double sum, we define the \textit{normalized double difference},
\begin{equation}
\label{normdoublediff}
    x_{\mathrm{\theta_{\mathrm{HWP}}^+}}=\frac{X_{{\theta_{\mathrm{HWP}}^+}}}{I_{X_{\theta_{\mathrm{HWP}}^+}}}
\end{equation}
which is the quantity used for calibration.
For the critical angles 0$^\circ$/45$^\circ$ and 22.5$^\circ$/67.5$^\circ$, the double difference is equivalent to the Stokes parameter $Q$ or $U$. $Q$ and $U$ correspond to $0^\circ$ and $22.5^\circ$ positive single difference HWP orientations in an ideal system, respectively, as $Q = \frac{1}{2}(Q^+-Q^-) = \frac{1}{2}[(I_{\rm det, R}(0^\circ)-I_{\rm det, L}(0^\circ))-(I_{\rm det, R}(45^\circ)-I_{\rm det, L}(45^\circ))]$ and $U = \frac{1}{2}(U^+-U^-) = \frac{1}{2}[(I_{\rm det, R}(22.5^\circ)-I_{\rm det, L}(22.5^\circ))-(I_{\rm det, R}(67.5^\circ)-I_{\rm det, L}(67.5^\circ))]$.
We define double differences more generally here, as done in previous works \cite{van_Holstein_2020,hart2021characterizationinstrumentalpolarizationeffects}, because  calibration requires data at HWP angles beyond the standard critical angles needed to extract Stokes $Q$ and $U$. 

Double differencing removes diattenuation from non-rotating components downstream of the HWP and differential effects from post-processing \cite{de_Boer_2020, van_Holstein_2020}. This can be understood by noting that while rotating the HWP by $45^\circ$ flips the sign of polarization upstream of the HWP, any instrumental polarization (diattenuation) introduced downstream is unaffected, as shown below:
\begin{equation}
\label{eq:doubledif_pe}
   X_{\theta_{\mathrm{HWP}}^+}=\frac{1}{2}(X^+-X^-)=\frac{1}{2}((X_{\theta_{\mathrm{HWP}}^+}+IP)-(-X_{\theta_{\mathrm{HWP}}^+}+IP))=X_{\theta_{\mathrm{HWP}}^+}
\end{equation}
where $\mathrm{IP}$ is instrumental polarization downstream of the HWP from nonrotating components.

\section{MODEL OF THE OPTICAL SYSTEM}
\label{sec:model}
\subsection{Mathematical Description}
\label{subsec:math-description}
We model the CHARIS optical path using a chain of Mueller matrices,
\begin{equation}
    \mathbf{S_{\mathrm{det,L/R}}}=M_{\mathrm{n}}\cdot M_{\mathrm{n-1}}\dots M_2 \cdot M_1\cdot \mathbf{S_{\mathrm{in}}}
\end{equation}
where each Mueller matrix describes an optical component. We can further simplify the model by only assigning Mueller matrices to components that share a reference frame. For example, we do not need a Mueller matrix for every mirror on the derotator because they all rotate together \cite{van_Holstein_2020}. We assign Mueller matrices to the following components: M3 (for on-sky observations only), a fold mirror used to direct the calibration lamp into the optical path (for unpolarized internal source calibrations only), the calibration polarizer (for polarized internal source calibrations only), the HWP, the derotator, the YJH50 dichroic beamsplitter, the NBS, and the Wollaston prism.  We construct the chain with Mueller matrices for each of these components and rotation matrices $T(\Theta)$. These rotation matrices describe rotations between the reference frames of subsequent optics \cite{vanholstein2020calibrationinstrumentalpolarizationeffects}. We define $\Theta=\theta+\delta$, where $\theta$ is the intended rotation of the reference frame and $\delta$ is a small, fitted misalignment angle, as done in previous CHARIS calibrations \cite{vanholstein2020calibrationinstrumentalpolarizationeffects,hart2021characterizationinstrumentalpolarizationeffects}:
\begin{equation}
   T(\Theta)= \begin{pmatrix}
1 & 0 & 0 & 0 \\
0 & \cos(2\Theta) & \sin(2\Theta) & 0 \\
0 & -\sin(2\Theta) & \cos(2\Theta) & 0 \\
0 & 0 & 0 & 1
\end{pmatrix}
\end{equation}
For on-sky data, the CHARIS Mueller matrix model is:
\begin{equation}
\label{onskymodel}
    \begin{aligned}
    \mathbf{S_{\mathrm{det,L/R}}}=M_{\mathrm{Wol}} \cdot M_{\mathrm{NBS}}\cdot M_{\mathrm{dichroic}}\cdot T(-\Theta_{\mathrm{der}})\cdot M_{\mathrm{der}} \cdot
    \\
  T(\Theta_{\mathrm{der}})\cdot T(-\Theta_{\mathrm{HWP}})\cdot M_{\mathrm{HWP}}\cdot T(\Theta_{\mathrm{HWP}})\cdot T(-a)\cdot M_{\mathrm{tel}}\cdot T(p)\cdot \mathbf{S_{\mathrm{in}}}
    \end{aligned}
\end{equation}
The rotation matrices surrounding $M_{\mathrm{tel}}$ are for altitude ($a$) and parallactic ($p$) angles. Note that in the most recent CHARIS polarimetric calibration, they report using the positive altitude angle ($a$) \cite{hart2021characterizationinstrumentalpolarizationeffects}. Using $+a$ produces a sign flip in our model that is not consistent with either CHARIS calibration, so we use $-a$ as done in Ref. \citenum{vanholstein2020calibrationinstrumentalpolarizationeffects}. Equation \ref{onskymodel} completely describes the optical path. Thus, inverting this matrix allows us to recover the input Stokes vector $\mathbf{S_{\mathrm{in}}}$. 

For polarized internal calibrations, the CHARIS Mueller matrix model requires the addition of the calibration polarizer, $M_{\rm cal}$, as follows: 
\begin{equation}
\label{internalmodel}
    \begin{aligned}
    \mathbf{S_{\mathrm{det,L/R}}}=M_{\mathrm{Wol}} \cdot M_{\mathrm{NBS}}\cdot  M_{\mathrm{dichroic}}\cdot T(-\Theta_{\mathrm{der}})\cdot M_{\mathrm{der}} \cdot
    \\
  T(\Theta_{\mathrm{der}})\cdot T(-\Theta_{\mathrm{HWP}})\cdot M_{\mathrm{HWP}}\cdot T(\Theta_{\mathrm{HWP}})\cdot T(-\delta_{\mathrm{cal}})\cdot M_{\mathrm{cal}}\cdot \mathbf{S_{\mathrm{in}}}
    \end{aligned}
\end{equation}
$M_{\mathrm{cal}}$ converts the $100\%$ unpolarized input Stokes vector ($\mathbf{S_{\mathrm{in}}}$) into $100\%$ $-Q$ polarized light. For unpolarized calibrations, the calibration polarizer is removed, but we must account for diattenuation from the fold mirror, which is at a $45^\circ$ angle of incidence with respect to the calibration source. We do not include the fold mirror in the polarized internal calibrations since the calibration polarizer is downstream of the fold mirror. For unpolarized internal calibrations, the CHARIS Mueller matrix model is then:
\begin{equation}
\label{internalmodelunpol}
    \begin{aligned}
    \mathbf{S_{\mathrm{det,L/R}}}=M_{\mathrm{Wol}} \cdot M_{\mathrm{NBS}}\cdot  M_{\mathrm{dichroic}}\cdot T(-\Theta_{\mathrm{der}})\cdot M_{\mathrm{der}} \cdot
    \\
  T(\Theta_{\mathrm{der}})\cdot T(-\Theta_{\mathrm{HWP}})\cdot M_{\mathrm{HWP}}\cdot T(\Theta_{\mathrm{HWP}})\cdot  M_{\mathrm{mirror}}\cdot  \mathbf{S_{\mathrm{in}}}
    \end{aligned}
\end{equation}
\subsection{Component Mueller Matrices}
\label{subsec:comp_mueller_matrices}
We now discuss candidate Mueller matrix models. We present the best fit Mueller matrices for each CHARIS optic in Section \ref{sec:internal-cal}. In previous polarimetric calibration proceedings, such as Refs. \citenum{hart2021characterizationinstrumentalpolarizationeffects, vanholstein2020calibrationinstrumentalpolarizationeffects, van_Holstein_2020}, the authors only used linear retarders and diattenuators in their models. We show in Subsection \ref{subsubsec:pol-internal} that an elliptical retarder model for the derotator performs better than a linear retarder model. We define an elliptical retarder as:
\begin{equation}
\label{ellipticalretarder}
\renewcommand{\arraystretch}{1.5}
 M_{\mathrm{ER}}=  \begin{pmatrix}
1 & 0 & 0 & 0 \\
0 & \frac{\Delta_{\mathrm{H}}^2 + (\Delta_{45}^2 + \Delta_{\mathrm{R}}^2)C}{\Delta^2} & \frac{\Delta_{45}\Delta_{\mathrm{H}}T + \Delta_{\mathrm{R}}S\Delta}{\Delta^2} & \frac{\Delta_{\mathrm{H}}\Delta_{\mathrm{R}}T - \Delta_{45}S\Delta}{\Delta^2} \\
0 & \frac{\Delta_{45}\Delta_{\mathrm{H}}T - \Delta_{\mathrm{R}}S\Delta}{\Delta^2} & \frac{\Delta_{45}^2 + (\Delta_{\mathrm{R}}^2 + \Delta_{\mathrm{H}}^2)C}{\Delta^2} & \frac{\Delta_{\mathrm{R}}\Delta_{45}T + \Delta_{\mathrm{H}}S\Delta}{\Delta^2} \\
0 & \frac{\Delta_{\mathrm{H}}\Delta_{\mathrm{R}}T + \Delta_{45}S\Delta}{\Delta^2} & \frac{\Delta_{\mathrm{R}}\Delta_{45}T - \Delta_{\mathrm{H}}S\Delta}{\Delta^2} & \frac{\Delta_{\mathrm{R}}^2 + (\Delta_{45}^2 + \Delta_{\mathrm{H}}^2)C}{\Delta^2}
\end{pmatrix}
\end{equation}
where $\Delta_{H}$ is horizontal linear retardance, $\Delta_{45}$ is $45 ^\circ$ retardance, and $\Delta_R$ is right-handed circular retardance \cite{Chipman2018-be}. The following shorthand is also used: $C=\cos{\Delta}$, $S=\sin{\Delta}$, $T=1-\cos{\Delta}$ and $\Delta=\sqrt{\Delta_H^2+\Delta_{45}^2+\Delta_{R}^2}$. The disadvantage of fitting an elliptical retarder instead of a linear retarder is that it increases the number of model parameters, raising the risk of overfitting. A linear retarder is a special case of an elliptical retarder, where $\Delta_R=0$. We still use this linear retarder model for some components, which we derive here by defining a Mueller matrix that combines diattenuation $\epsilon$ and linear retardance $\Delta$. A linear retarder Mueller matrix is given by: 
\begin{equation}
\label{linret}
  M_{\mathrm{R}}(\Delta)=  \begin{pmatrix}
1 & 0 & 0 & 0 \\
0 & 1 & 0 & 0 \\
0 & 0 & \cos\Delta & \sin\Delta \\
0 & 0 & -\sin\Delta & \cos\Delta
\end{pmatrix}
\end{equation}
Diattenuators are defined using transmission coefficients $p_x$ and $p_y$, corresponding to how much the $x$ and $y$ electric field components are attenuated \cite{goldstein2017polarized}: 
\begin{equation}
    M_{\mathrm{D}}(p_x,p_y) = \frac{1}{2} 
    \renewcommand{\arraystretch}{1.3}
    \begin{pmatrix}
p_x^2 + p_y^2 & p_y^2 - p_x^2 & 0 & 0 \\
p_y^2 - p_x^2 & p_x^2 + p_y^2 & 0 & 0 \\
0 & 0 & 2p_x p_y & 0 \\
0 & 0 & 0 & 2p_x p_y
\end{pmatrix} \quad 0 \leq p_{x,y} \leq 1.
\end{equation}
We express this matrix in the basis defined in Subsection \ref{subsec:stokesvec}, then re-arrange the matrix as a function of one parameter:
\begin{equation}
\label{eq:diattenuation}
    \epsilon=\frac{p_y^2-p_x^2}{p_x^2+p_y^2}
\end{equation}
ranging from (-1,1). We neglect reductions in the beam's intensity since our calibration data are normalized. Multiplying this rearranged Mueller matrix with Equation \ref{linret} and assuming our matrix doesn't reduce the incident beam's intensity yields the linear diattenuator-retarder model used in the previous CHARIS polarimetric calibrations \cite{hart2021characterizationinstrumentalpolarizationeffects,vanholstein2020calibrationinstrumentalpolarizationeffects}:
\begin{equation}
\label{diattenuatorretarder}
    M_{\mathrm{DR}}(\epsilon,\Delta)=\begin{pmatrix}
1 & \epsilon & 0 & 0 \\
\epsilon & 1 & 0 & 0 \\
0 & 0 & \sqrt{1 - \epsilon^2} \cos \Delta & \sqrt{1 - \epsilon^2} \sin \Delta \\
0 & 0 & -\sqrt{1 - \epsilon^2} \sin \Delta & \sqrt{1 - \epsilon^2} \cos \Delta
\end{pmatrix}
\end{equation}
 We define the Mueller matrix for a Wollaston beamsplitter as:
\begin{equation}
\label{Wollaston}
    M_{\mathrm{Wol}}(\eta)=\frac{1}{2} 
\begin{pmatrix} 
1 & \eta & 0 & 0 \\ 
\eta & 1 & 0 & 0 \\ 
0 & 0 & 0 & 0 \\ 
0 & 0 & 0 & 0 
\end{pmatrix}
\end{equation}
$\eta$ defines the beam; it changes sign depending on whether we are describing the ordinary or extraordinary beam. In our case, the right side of the detector is the ordinary beam ($+\eta$) and the left side of the detector is the extraordinary beam ($-\eta$). We can fit $|\eta|$ as a modulation efficiency term, accounting for imperfections in the Wollaston. A perfect Wollaston prism has $|\eta|=1$, and $|\eta|<1$ amounts to a reduction in polarized intensity. 

\subsection{Physical Models}
\label{subsec:physmodel}
We implement physical models from Ref. \citenum{hart2021characterizationinstrumentalpolarizationeffects} for the retardance of the HWP and the diattenuation of M3, which we summarize here. We calculate the HWP retardance ($\Delta_{\mathrm{HWP}}$) as a function of the wavelength of the incident light and the widths of its layers: 
\begin{equation}
\label{eq:hwp_physical_model}
\Delta_{\mathrm{HWP}}(\lambda,w_{\mathrm{SiO}_2},w_{\mathrm{MgF_2}}) = \frac{2\pi}{\lambda} \left[ w_{\mathrm{SiO}_2} \left[ n_{\mathrm{e,SiO}_2}(\lambda) - n_{\mathrm{o,SiO}_2}(\lambda) \right] - w_{\mathrm{MgF}_2} \left[ n_{\mathrm{e,MgF}_2}(\lambda) - n_{\mathrm{o,MgF}_2}(\lambda) \right] \right]
\end{equation}
$w_{\mathrm{SiO}_2}$ and $w_{\mathrm{MgF_2}}$ are the widths of the $\mathrm{SiO}_2$ and $\mathrm{MgF_2}$ layers. The subscripts on the refractive indices $n$ denote which layer and which axis; $e$ is the fast axis and $o$ is the slow axis. We obtain the refractive indices as a function of wavelength for $\mathrm{SiO}_2$ and $\mathrm{MgF_2}$ from Refs. \citenum{Ghosh1999DispersionequationCF} and \citenum{Dodge:84}, respectively.

The diattenuation and retardance of M3 can be calculated with its refractive index via the Fresnel equations, but we do not have a measurement of M3's refractive index. M3 is a silver-coated fold mirror, so we calculate its complex refractive index $n=\hat{n}+i\kappa$ using the dielectric functions of silver:
\begin{equation}
\begin{aligned}
\hat{n} &= \frac{1}{\sqrt{2}}\sqrt{\sqrt{\epsilon_{r1}(\lambda)^2 + \epsilon_{r2}(\lambda)^2} + \epsilon_{r1}(\lambda)}, \\
\kappa &= \frac{1}{\sqrt{2}}\sqrt{\sqrt{\epsilon_{r1}(\lambda)^2 + \epsilon_{r2}(\lambda)^2} - \epsilon_{r1}(\lambda)},
\end{aligned}
\end{equation}
$\epsilon_{r1}$ and $\epsilon_{r2}$ are the dielectric functions, which we approximate with a linear function in log-log space:
\begin{equation}
\label{eq:m3_physical_model}
\begin{aligned}
\epsilon_{r1}(\lambda,b_1,m_1) &= -10^{b_1} \lambda^{m_1}, \\
\epsilon_{r2}(\lambda,b_2,m_2) &= 10^{b_2} \lambda^{m_2}
\end{aligned}
\end{equation}
We can thus calculate the diattenuation and retardance of M3 as a function of wavelength by finding $b_1,m_1,b_2$ and $m_2$.

\subsection{Python Mueller Matrix Modeling with \texttt{pyPolCal}}
\label{subsec:pypolcal} The CHARIS Mueller Matrix model described herein was implemented in a newly-developed open-source polarimetric calibration package made for SCExAO/CHARIS and SCExAO/VAMPIRES, with potential to be adapted to other systems. The workflow is as follows. First, a user defines a \texttt{system\_mueller\_matrix} dictionary containing the Mueller matrix chain. These dictionaries are converted into matrices using the \texttt{pyMuellerMat} package \cite{pyMuellerMat}. The user then defines starting guesses and bounds. The user can fit by wavelength bin or globally with either \texttt{scipy.optimize} or \texttt{emcee.EnsembleSampler} \cite{foreman2013emcee,virtanen2020scipy}. The fitting functions account for parameters that vary per frame (such as the HWP and derotator angles) using \texttt{configuration\_lists} that contain these parameters. This work has added functions that automatically extract these parameters from CHARIS FITS files and perform aperture photometry using the \texttt{Photutils} package \cite{Bradley2025-oh}. \texttt{pyPolCal} additionally includes a number of ways to assess goodness of fit. The plotting script includes functions to simultaneously plot modeled and observed double differences as a function of HWP or derotator angle. There are also functions to calculate polarimetric accuracy and efficiency, which we discuss in Section \ref{sec:pol-eff}. Finally, the package contains all the Jupyter notebooks used to carry out the work of this proceeding and tutorial notebooks to enable future calibrations. \texttt{pyPolCal} is under development to create its full functionality, where it will be easily adaptable to other instruments beyond CHARIS and VAMPIRES. However, in its current state, the package is ready to use for any further updates to the CHARIS Mueller matrix model if more calibration data become available or other changes are made to the optical path necessitating recalibration. A link to \texttt{pyPolCal} can be found in Ref. \citenum{pypolcal}.

\section{POLARIZED AND UNPOLARIZED INTERNAL SOURCE CALIBRATIONS} 
\label{sec:internal-cal}
\subsection{Calibration Procedure and Data Reduction}   
\label{subsec:internal-procedure}
In July 2025, before the NBS installation, we obtained two datasets using an internal calibration source: one with the YJH50 dichroic, and one without. We inserted the calibration polarizer for both datasets, generating $100\%$ $-Q$ polarized light. For each dataset, we rotated the derotator from  $45^\circ$ to $132.5^\circ$ in steps of $12.5^\circ$. For each derotator angle, we rotated the HWP from $0^\circ$ to $78.75^\circ$ in steps of $11.25^\circ$ (here referred to as a HWP cycle). Four normalized double differences (Equation \ref{normdoublediff}) are extracted from each HWP cycle. 

In October 2025, after the NBS installation, we obtained two additional internal calibration datasets using the same HWP angles and derotator angles described above, both with the YJH50 dichroic inserted; one dataset was without the calibration polarizer (referred to as unpolarized internal calibration data) and one was with the calibration polarizer (referred to as polarized internal calibration data, where the input is $100\%$ -Q light). We use the polarized dataset to characterize retardance and the unpolarized dataset to characterize diattenuation \cite{hart2021characterizationinstrumentalpolarizationeffects}. Table \ref{tab:internal_calibration_datasets} summarizes all internal calibration data used in this work. All datasets for this work were obtained in CHARIS's ``broadband'' \textit{JHK} mode.
\begin{table}[htbp]

    \centering
    \caption{Summary of the datasets used for internal source calibrations.}
    \label{tab:internal_calibration_datasets}
    \resizebox{\textwidth}{!}{%
    \begin{tabular}{@{}cllllcc@{}}
        \toprule
        \textbf{Dataset} & \textbf{Date} & \textbf{Polarization} & \textbf{YJH50 Pickoff} & \textbf{NBS Status} & \textbf{\# of HWP Cycles} & \textbf{Derotator Angle Range} \\
        \midrule
        1 & 2025-07-15 & $100\%$ $-Q$   & Out & Not Installed & 8 & $45.0^\circ - 132.5^\circ$ \\
        2 & 2025-07-15 & $100\%$ $-Q$   & In  & Not Installed & 8 & $45.0^\circ - 132.5^\circ$ \\
        3 & 2025-10-05 & $100\%$ $-Q$   & In  & Installed     & 8 & $45.0^\circ - 132.5^\circ$ \\
        4 & 2025-10-05 & Unpolarized    & In  & Installed     & 8 & $45.0^\circ - 132.5^\circ$ \\
        \bottomrule
    \end{tabular}%
    }
\end{table}

\begin{figure}
    \centering
    \includegraphics[width=0.5\linewidth]{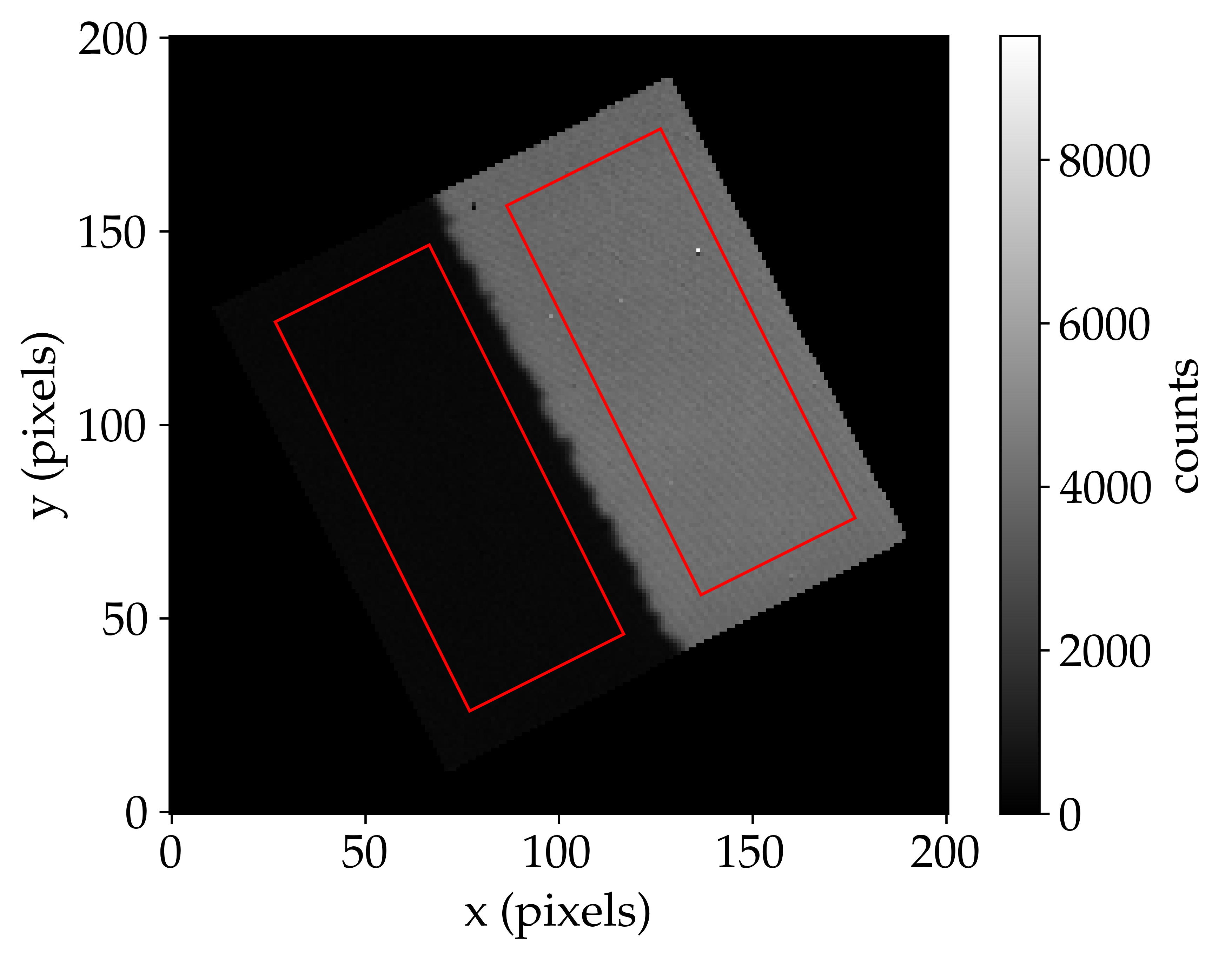}
    \caption{One slice of a polarized internal calibration data cube with apertures overlaid. The right and left apertures correspond to $I_{\mathrm{det,R}}$ and $I_{\mathrm{det,L}}$ in Equations \ref{singledif} and \ref{singlesum}.
    \label{fig:internalapertures}}
\end{figure}

We first use the CHARIS Data Extraction Pipeline (DEP) \cite{brandt2017datareductionpipelinecharis} to convert raw frames into data cubes of 22 images per exposure, one for each wavelength bin. We then perform aperture photometry to obtain single differences and sums (Equations \ref{singledif} and \ref{singlesum}). We place the apertures such that they exclude edge effects from the detector, as done in Ref. \citenum{hart2021characterizationinstrumentalpolarizationeffects} and shown in Figure \ref{fig:internalapertures}. 
The aperture sums for each beam give fluxes $I_{\mathrm{det,R}}$ and $I_{\mathrm{det,L}}$. We calculate errors on these fluxes using $\sigma=\sqrt{I_{\mathrm{det,L}}}$. We then obtain the single differences and sums using Equations \ref{singledif} and \ref{singlesum}. For each frame, we extract the HWP angle and derotator angle from the FITS headers. For CHARIS, there are two FITS headers related to the HWP angle: \texttt{RET-ANG1} and \texttt{RET-POS1}. \texttt{RET-POS1} accounts for HWP tracking offsets and should be used for on-sky data. We use \texttt{RET-ANG1} for internal calibrations, which does not account for these offsets. We compute the normalized double differences (Figure \ref{fig:ddbyimr}) $x_{0^\circ}$, $x_{11.25^\circ}$, $x_{22.5^\circ}$, and $x_{33.75^\circ}$ from the single differences and sums by following the procedure in Section \ref{sec:charis-optical-path}. We calculate errors on the normalized double differences using Gaussian error propagation of the flux errors. Due to the large aperture size shown in Figure \ref{fig:internalapertures}, these errors are small. 

\subsection{Fitting Procedure}
\label{subsec:internal-fitting}
We first detail the model parameters and fitting procedure for the polarized internal calibrations. We assign different Mueller matrices from Subsection \ref{subsec:comp_mueller_matrices} to $M_{\mathrm{Wol}}$, $M_{\mathrm{NBS}}$, $M_{\mathrm{dichroic}}$, $M_{\mathrm{der}}$, $M_{\mathrm{HWP}}$, and $M_{\mathrm{cal}}$ in the polarized internal calibration model (Equation \ref{internalmodel}). We tested several candidate Mueller matrices for each component, the failures of which are detailed in Subsection \ref{subsec:internal-results} to justify our choices. Our final Mueller matrix assignments are as follows: 
\begin{itemize}
\item \textbf{Wollaston Prism} ($M_{\mathrm{Wol}}$): Modeled using Equation \ref{Wollaston}, with modulation efficiency ($|\eta|$) as a free parameter.
\item \textbf{Nasmyth beam-switcher} ($M_{\mathrm{NBS}}$): Modeled as a diagonal matrix $\mathrm{diag}(1,-1,1,-1)$, derived from ray tracing. This is excluded from the fits to Datasets 1 and 2 (Table \ref{tab:internal_calibration_datasets}) as they are from before the NBS install.
\item \textbf{YJH50 Dichroic} ($M_{\mathrm{dichroic}}$): Set to the identity matrix, as we could not reliably fit a Mueller matrix to its crosstalk. 
\item \textbf{Derotator} ($M_{\mathrm{der}}$): Modeled as an elliptical retarder (Equation \ref{ellipticalretarder}) with free parameters $\Delta_H, \Delta_{45},$ and $\Delta_R$.
\item \textbf{Half-Wave Plate} ($M_{\mathrm{HWP}}$): Modeled as a linear retarder (Equation \ref{diattenuatorretarder}, $\epsilon=0$). Using Equation \ref{eq:hwp_physical_model}, we model the retardance ($\Delta$) using the layer thicknesses ($w_{\mathrm{MgF2}}$ and $w_{\mathrm{SiO2}}$), which are free parameters.
\item \textbf{Calibration Polarizer} ($M_{\mathrm{cal}}$): Modeled as a perfect linear diattenuator (Equation \ref{diattenuatorretarder}, $\Delta=0$ and $\epsilon=-1$).
\end{itemize}
Additionally, we fit a misalignment angle ($\delta$) to the HWP, derotator, and calibration polarizer. We fit the derotator elliptical retardance components and Wollaston modulation efficiency \textit{per wavelength} (once per wavelength bin). We fit offset angles and the HWP layer thicknesses \textit{globally} (across all wavelength bins simultaneously). Thus, we have 5 globally fit parameters and 4 parameters fit per wavelength, giving our model a total of 93 parameters. CHARIS's internal calibration Mueller matrix model is then a function of these fitted parameters, the derotator angle, and the HWP angle. See Table \ref{tab:internal_parameters} for a summary of all model free parameters.
\begin{table}[htbp]
    \centering
    \caption{Summary of free parameters for the internal calibration Mueller matrix model. Parameters designated as ``per wavelength'' are fit once per wavelength bin, while ``global'' parameters are fit across all wavelength bins simultaneously.}
    \label{tab:internal_parameters}
    \begin{tabular}{@{}lllc@{}}
        \toprule
        \textbf{Component} & \textbf{Parameter Description} & \textbf{Symbol} & \textbf{Fit Scope} \\
        \midrule
        Derotator & Elliptical retardance components & $\Delta_H, \Delta_{45}, \Delta_R$ & Per Wavelength \\
        Wollaston Prism & Modulation efficiency & $|\eta|$ & Per Wavelength \\
        \midrule
        HWP & $\mathrm{MgF_2}$ layer thickness & $w_{\mathrm{MgF2}}$ & Global \\
        HWP & $\mathrm{SiO}_2$ layer thickness & $w_{\mathrm{SiO2}}$ & Global \\
        Calibration Polarizer & Misalignment angle & $\delta_{\mathrm{cal}}$ & Global \\
        Derotator & Misalignment angle & $\delta_{\mathrm{der}}$ & Global \\
        HWP & Misalignment angle & $\delta_{\mathrm{HWP}}$ & Global \\
        \bottomrule
    \end{tabular}
\end{table}

We begin by fitting the HWP retardance (a nuisance parameter that is discarded in the global fit), derotator elliptical retardance components, and Wollaston modulation efficiency per wavelength with \newline \texttt{scipy.optimize.least\_squares} \cite{virtanen2020scipy}, setting all misalignment angles to zero. We fit all datasets separately. The misalignment angles are degenerate with the fitted retardances. Accordingly, we initially set these angles to zero, as the optics should be precisely aligned on the SCExAO bench. We use the fits from Ref. \citenum{hart2021characterizationinstrumentalpolarizationeffects} as initial guesses for the HWP retardance and derotator horizontal retardance, and we use $|\eta|=1$  for the initial guess of the Wollaston modulation efficiency. The bounds are highly permissive, spanning all physically allowed parameter values: $(0,2\pi)$ for retardances and $(0,1)$ for the Wollaston modulation efficiency. 

We then fit all global parameters using \texttt{scipy.optimize.minimize}. For this step, we implement the per-wavelength fit model parameters described in the above paragraph into the models for the derotator and Wollaston prism. We use the fit from Ref. \citenum{hart2021characterizationinstrumentalpolarizationeffects} as the initial guess for the HWP layer thicknesses and $0^\circ$ as the initial guess for the misalignment angles. For HWP layer thicknesses, we set the bounds to be from 0.6 to 1.4 times the fits from Ref. \citenum{hart2021characterizationinstrumentalpolarizationeffects}. We set the bounds for misalignment angles to be from $-5^\circ$ to $5^\circ$, since we assume these are small offsets. We additionally performed a Markov Chain Monte Carlo simulation to explore degeneracies in the model, which we describe in Appendix \ref{appendix:mcmc}.

For the unpolarized internal calibrations, we assign Mueller matrices to $M_{\mathrm{Wol}}$, $M_{\mathrm{NBS}}$, $M_{\mathrm{dichroic}}$, $M_{\mathrm{der}}$, $M_{\mathrm{HWP}}$, and $M_{\mathrm{mirror}}$ in the unpolarized internal calibration model (Equation \ref{internalmodelunpol}). Our final Mueller matrix assignments are as follows: We use the model parameter fits from the polarized internal calibrations for $M_{\mathrm{Wol}}$, $M_{\mathrm{NBS}}$, $M_{\mathrm{dichroic}}$, $M_{\mathrm{der}}$, and $M_{\mathrm{HWP}}$ to model the optical path's retardance. We do not fit any retardances with the unpolarized calibrations. We model the fold mirror ($M_{\mathrm{mirror}}$) as a linear diattenuator (Equation \ref{diattenuatorretarder}, $\Delta=0$) with the diattenuation ($\epsilon$) as a free parameter. We found we can neglect diattenuation from all other components, which we justify in the following subsection. We fit the fold mirror's diattenuation per wavelength using \texttt{scipy.optimize.least\_squares}. We use an initial guess of $0$ and bounds of $(-1,1)$. When the derotator is at $45^\circ$, the SNR of the polarized signal is too low to produce a reliable fit, so we remove exposures with that derotator angle from our calibration data.

We use the method outlined in Appendix E of Ref. \citenum{van_Holstein_2020} to estimate uncertainties on fitted parameters. For this calculation to be valid, model parameters must be uncorrelated \cite{hart2021characterizationinstrumentalpolarizationeffects}. The errors due to photon noise in the calibration data described in Subsection \ref{subsec:internal-procedure} are very small. For this reason, we follow the procedure from Ref. \citenum{hart2021characterizationinstrumentalpolarizationeffects} and neglect them during fitting. While there are likely other unknown measurement errors, we have no way of estimating these. 

\subsection{Results and Discussion}
\label{subsec:internal-results}

\subsubsection{Polarized Internal Calibrations}
\label{subsubsec:pol-internal}

We published the model parameters that we fit per wavelength with uncertainties to Zenodo, which can be found at the link in Ref. \citenum{McIntosh2026-pg}. We now discuss the fit to Dataset 3 (Table \ref{tab:internal_calibration_datasets}), which we implement into the CHARIS Mueller matrix model. Table \ref{tab:fitted_values} shows all of the globally fit model parameters. We found that the globally fit parameters are correlated, which we discuss in Appendix \ref{appendix:mcmc}; therefore, we do not include uncertainty estimates.

\begin{table}[H]
    \centering
    \caption{Model parameters fit globally (across all wavelength bins simultaneously) for the internal calibration model, fit using Dataset 3 (Table \ref{tab:internal_calibration_datasets}).}
    \label{tab:fitted_values}
    \begin{tabular}{|l|c|}
        \hline
        \textbf{Parameter} & \textbf{Value} \\
        \hline
        $w_{\mathrm{SiO2}}$  & 1.653 mm \\
        $w_{\mathrm{MgF2}}$  & 1.291 mm \\
        \hline
        $\delta_{\mathrm{der}}$  & $-0.02^\circ$ \\
        $\delta_{\mathrm{HWP}}$  & $-0.04^\circ$ \\
        $\delta_{\mathrm{cal}}$  & \hphantom{$-$}0.03$^\circ$ \\
        \hline
    \end{tabular}
\end{table}

\begin{figure}
    \centering
    \begin{subfigure}[b]{0.48\textwidth}
        \centering
        \includegraphics[width=\textwidth]{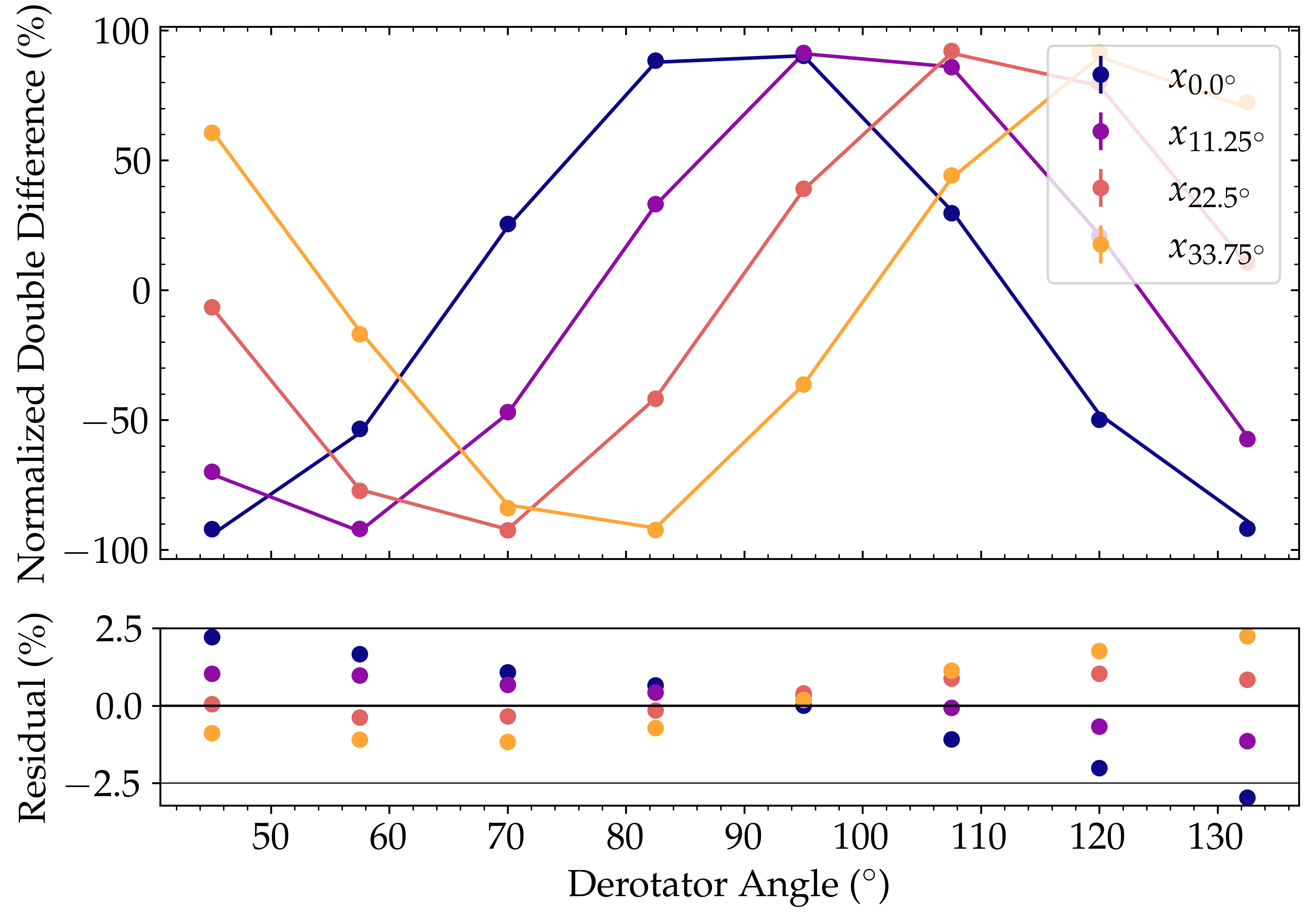}
        \caption{$\lambda=1329$ nm}
    \end{subfigure}
    \hfill 
    \begin{subfigure}[b]{0.48\textwidth}
        \centering
        \includegraphics[width=\textwidth]{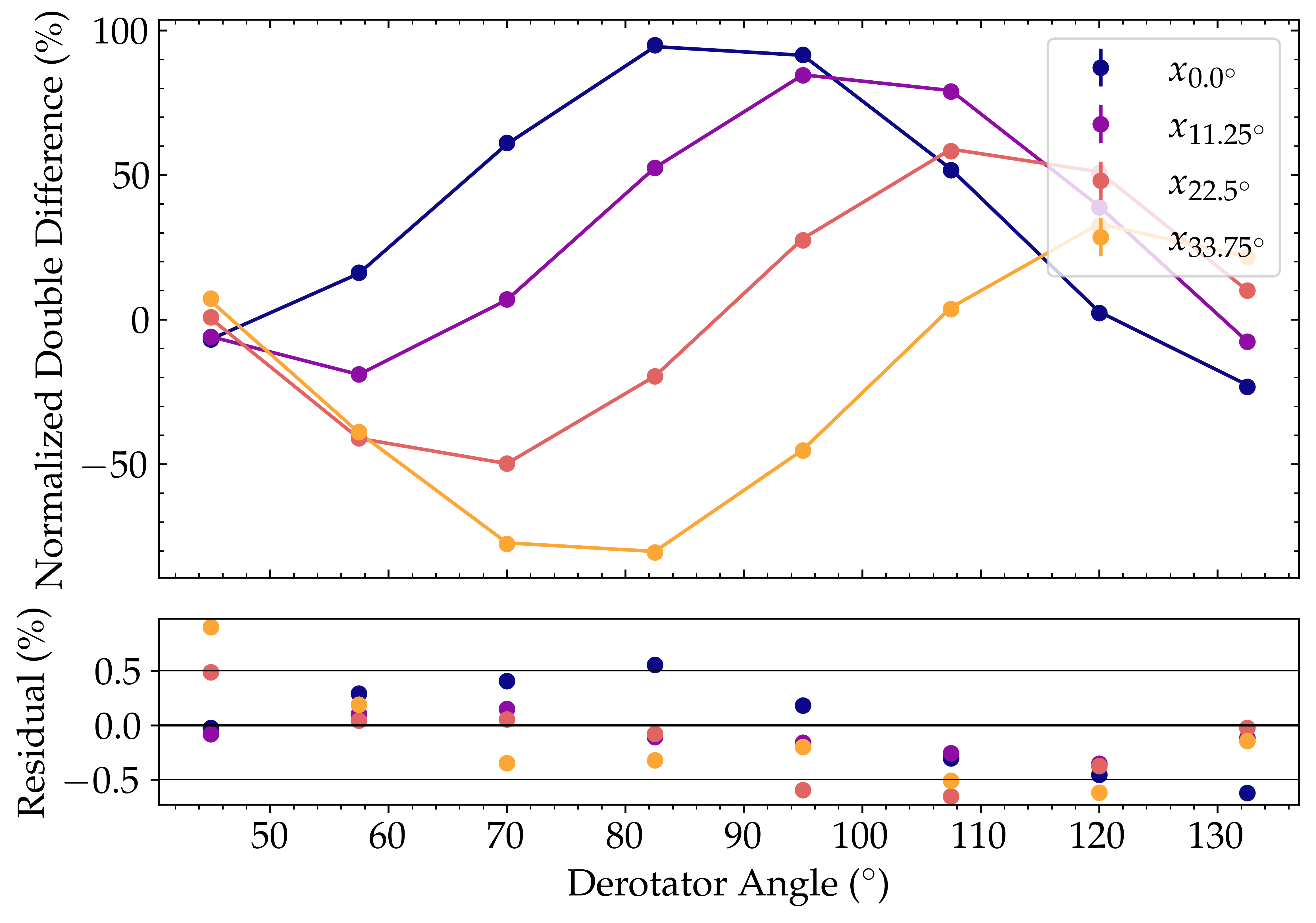}
        \caption{$\lambda$=1575 nm}
    \end{subfigure}
    
    \vspace{3pt} 
    
    \begin{subfigure}[b]{0.48\textwidth}
        \centering
        \includegraphics[width=\textwidth]{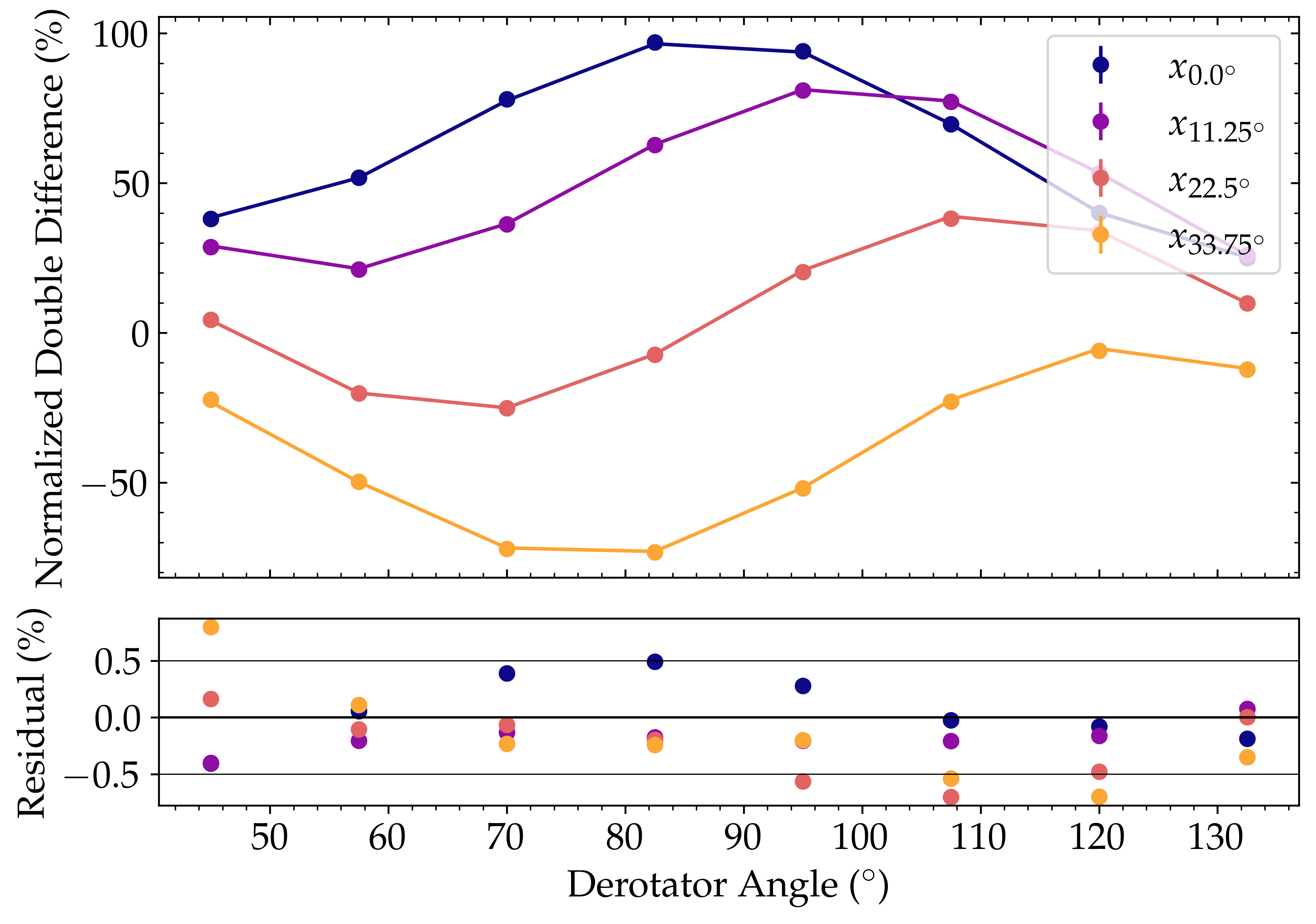}
        \caption{$\lambda=1744$ nm}
    \end{subfigure}
    \hfill
    \begin{subfigure}[b]{0.48\textwidth}
        \centering
        \includegraphics[width=\textwidth]{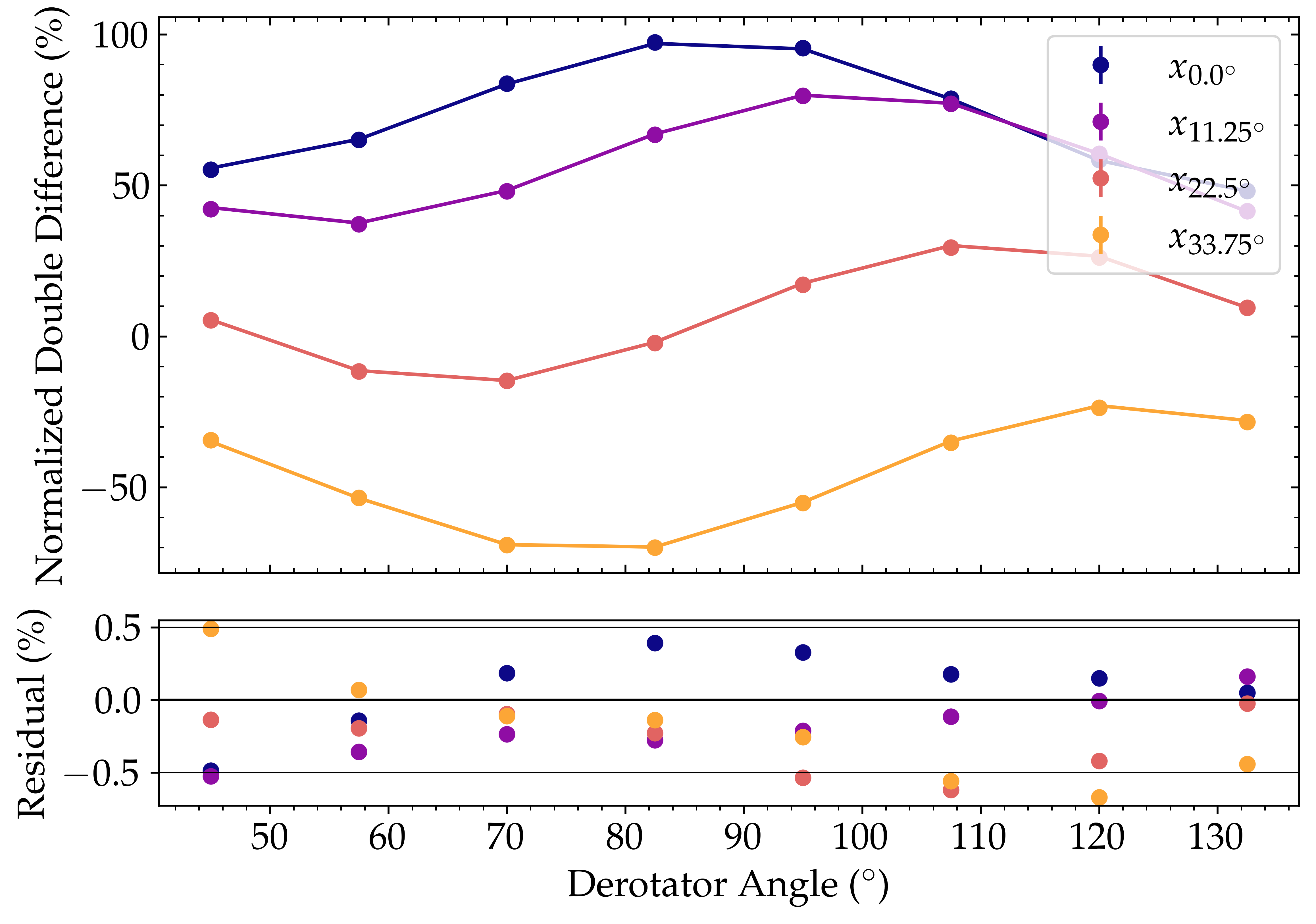} 
        \caption{$\lambda$=1932 nm}
    \end{subfigure}
    \vspace{3pt}
    \caption{Modeled and observed normalized double differences (Equation \ref{normdoublediff}) plotted as a function of the derotator angle for 4 wavelength bins. This figure uses Dataset 3 (Table \ref{tab:internal_calibration_datasets}). The normalized double differences are specified by the HWP orientation of the positive single difference. The error bars on the plots are not visible since the photon noise is negligible.}
    \label{fig:ddbyimr}
\end{figure}

\begin{figure}

    \centering
    \begin{subfigure}[b]{0.48\textwidth}
        \centering
        \includegraphics[width=\textwidth]{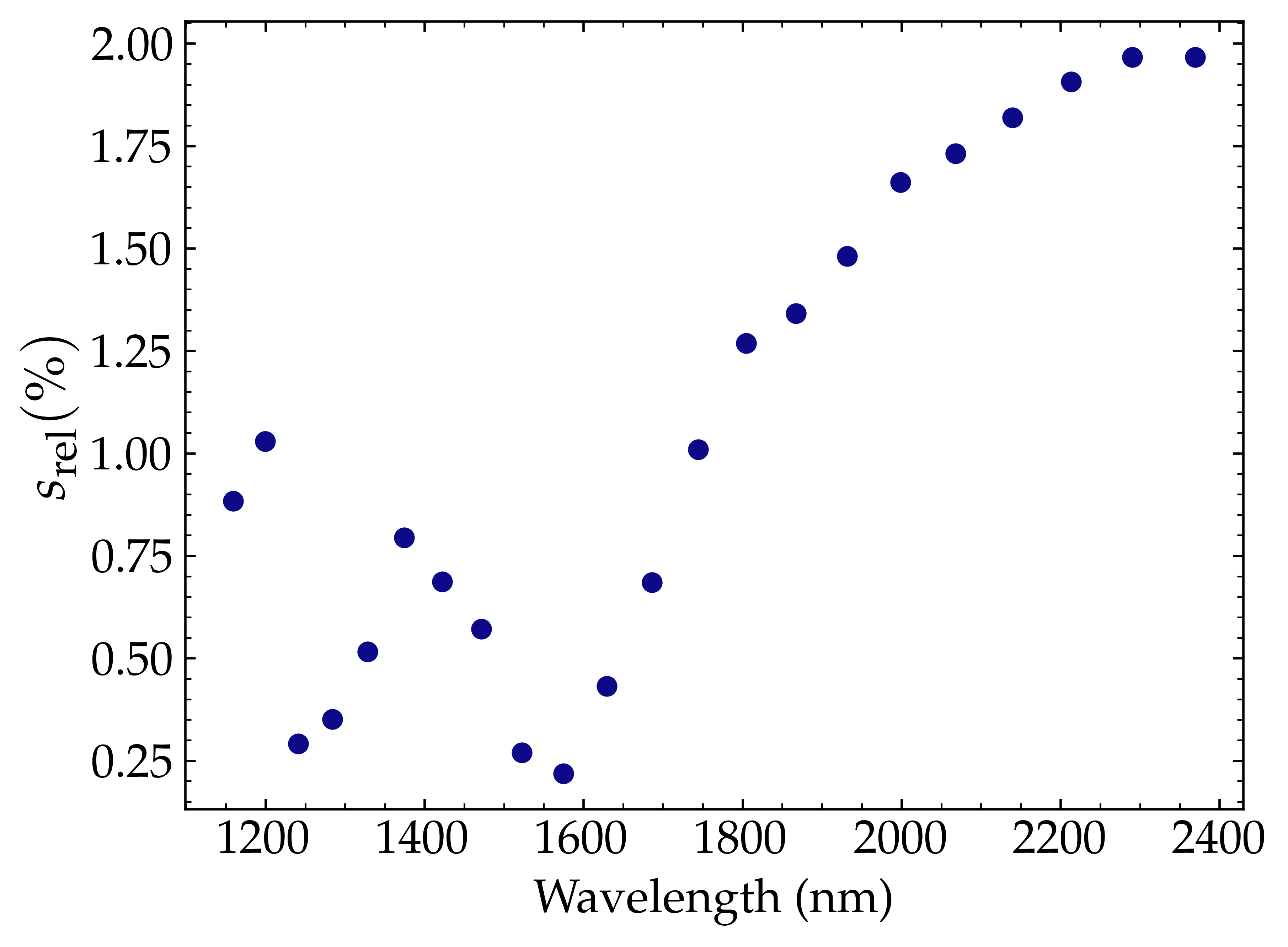}
        \caption{Linear retarder derotator model.}
        \label{fig:s_rel_hart_model}
    \end{subfigure}
    \vspace{5pt}
    \begin{subfigure}[b]{0.48\textwidth}
        \centering
        \includegraphics[width=\textwidth]{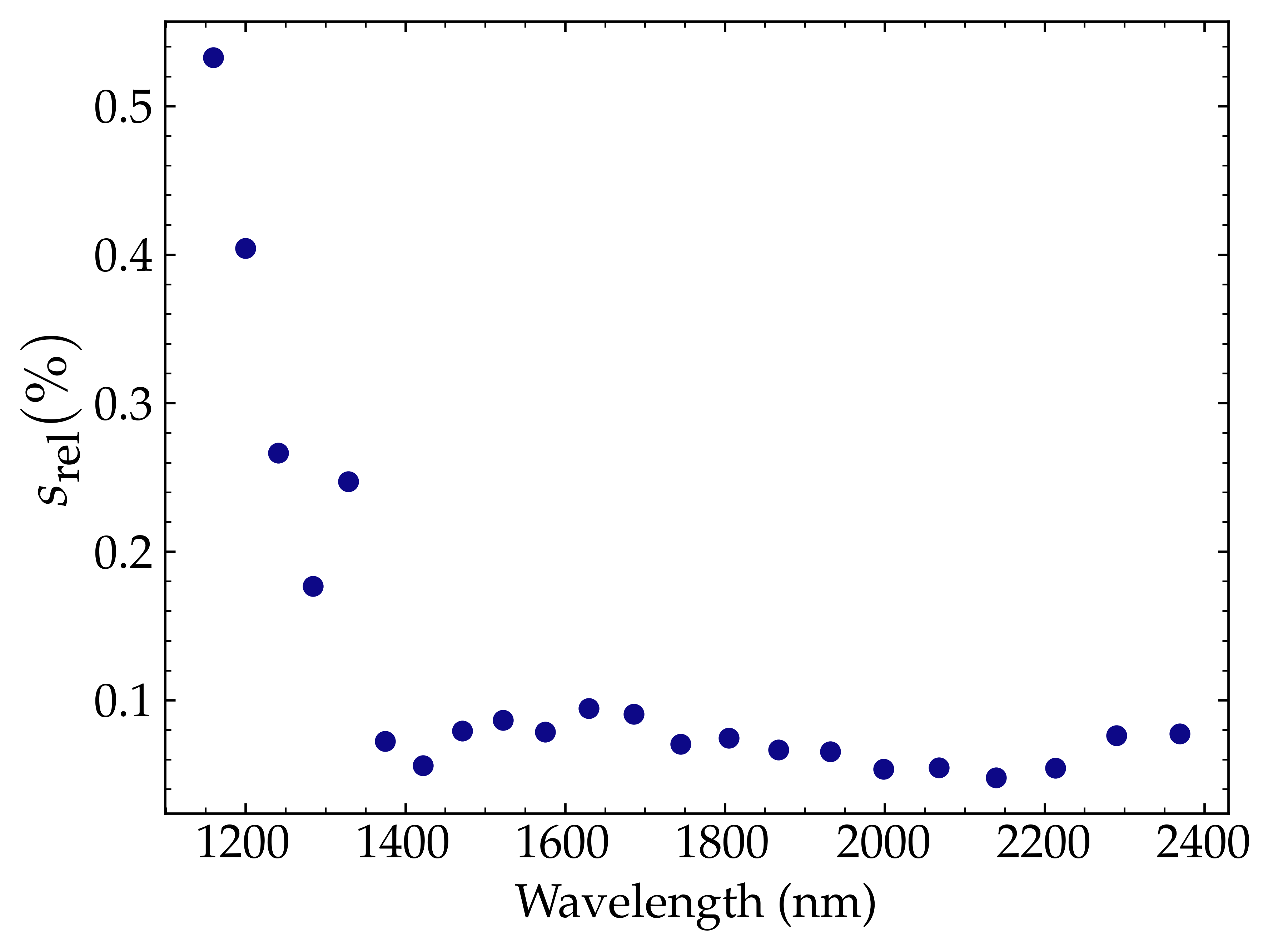}
        \caption{Elliptical retarder derotator model.}
        \label{fig:nbs_in_s_rel}
    \end{subfigure}
    \vspace{5pt}
    \caption{Relative polarimetric accuracy (see Equation \ref{eq:sem}) as a function of wavelength calculated with Dataset 3 in Table \ref{tab:internal_calibration_datasets}. We made plot (a) by globally fitting the model from Ref. \citenum{hart2021characterizationinstrumentalpolarizationeffects}, where the derotator is modeled as a horizontal linear retarder as a function of its film width. We made plot (b) by switching this derotator model for an elliptical retarder (Equation \ref{ellipticalretarder}) fit once per wavelength bin and then refitting all other model parameters.}
    \label{fig:srel_comparison_ellip}
    
\end{figure}

 Figure \ref{fig:ddbyimr} shows the observed and fitted normalized double differences. The model residuals are comparable to Ref. \citenum{hart2021characterizationinstrumentalpolarizationeffects}. The relative polarimetric accuracy $s_{\mathrm{rel}}$ (Equation \ref{eq:sem}), shown in Figure \ref{fig:nbs_in_s_rel}, is a useful metric for determining model accuracy. It scales with the sum of the residuals, allowing us to assess the residuals as a function of wavelength. We set the correction $k=9$ to account for the 5 globally fit and 4 per-wavelength fit parameters. The residuals are steady in the \textit{H}- and \textit{K}-bands, but are larger in the \textit{J}-band.

\paragraph{Elliptical Retardance from the Derotator}

A main difference between this proceeding and Refs. \citenum{hart2021characterizationinstrumentalpolarizationeffects,vanholstein2020calibrationinstrumentalpolarizationeffects,van_Holstein_2020} is the use of an elliptical retarder model (Equation \ref{ellipticalretarder}) for the derotator ($M_{\mathrm{der}}$) instead of a linear retarder (Equation \ref{linret}). The internal calibration model in Ref. \citenum{hart2021characterizationinstrumentalpolarizationeffects} reduced the derotator to a one-parameter, globally fit linear retarder model. Since the elliptical retarder has 3 parameters per wavelength bin as opposed to one physically-motivated, globally-fit parameter, our model is more at risk for overfitting. However, this increase in model complexity is necessary: Using datasets 1, 2, and 3 (no NBS/YJH 50 out, no NBS/YJH50 in, NBS/YJH50 in -- all with 100\% polarized light input) (Table \ref{tab:internal_calibration_datasets}), we found that re-fitting the model from Ref. \citenum{hart2021characterizationinstrumentalpolarizationeffects} produced poor residuals. As Dataset 1 contains neither the NBS nor the YJH50 dichroic, we know that neither of these new optics caused the poor fit; instead, we believe this is due to neglected elliptical retardance. Figure \ref{fig:srel_comparison_ellip} demonstrates this poor fit by comparing the relative polarimetric accuracy with the linear retarder derotator model from Ref. \citenum{hart2021characterizationinstrumentalpolarizationeffects} and our elliptical retarder derotator model using Dataset 3 (Table \ref{tab:internal_calibration_datasets}). 

\begin{figure}
    \centering
    \begin{subfigure}[b]{0.48\textwidth}
        \centering
        \includegraphics[width=\textwidth]{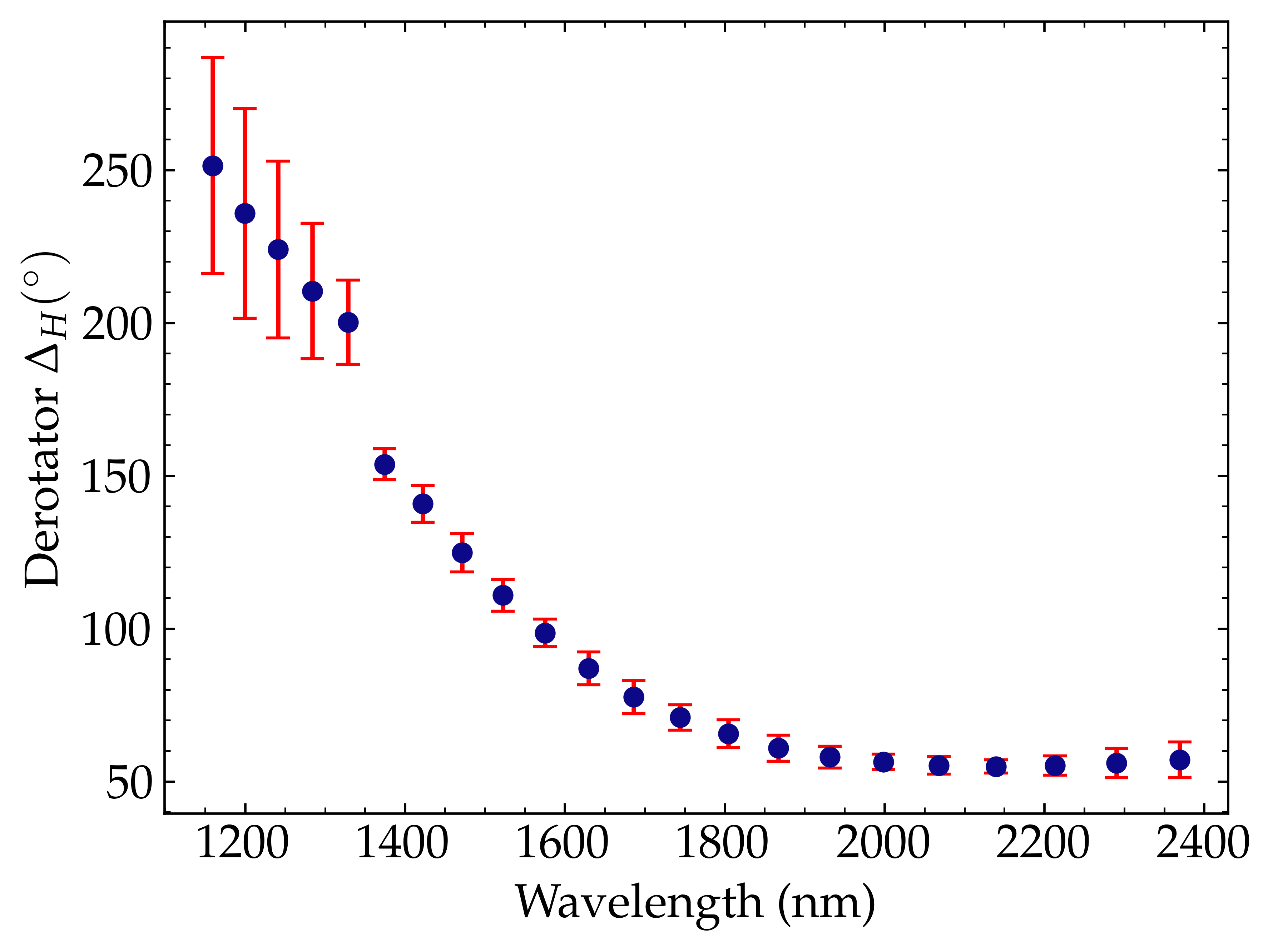}
    \end{subfigure}
    \begin{subfigure}[b]{0.48\textwidth}
        \centering
        \includegraphics[width=\textwidth]{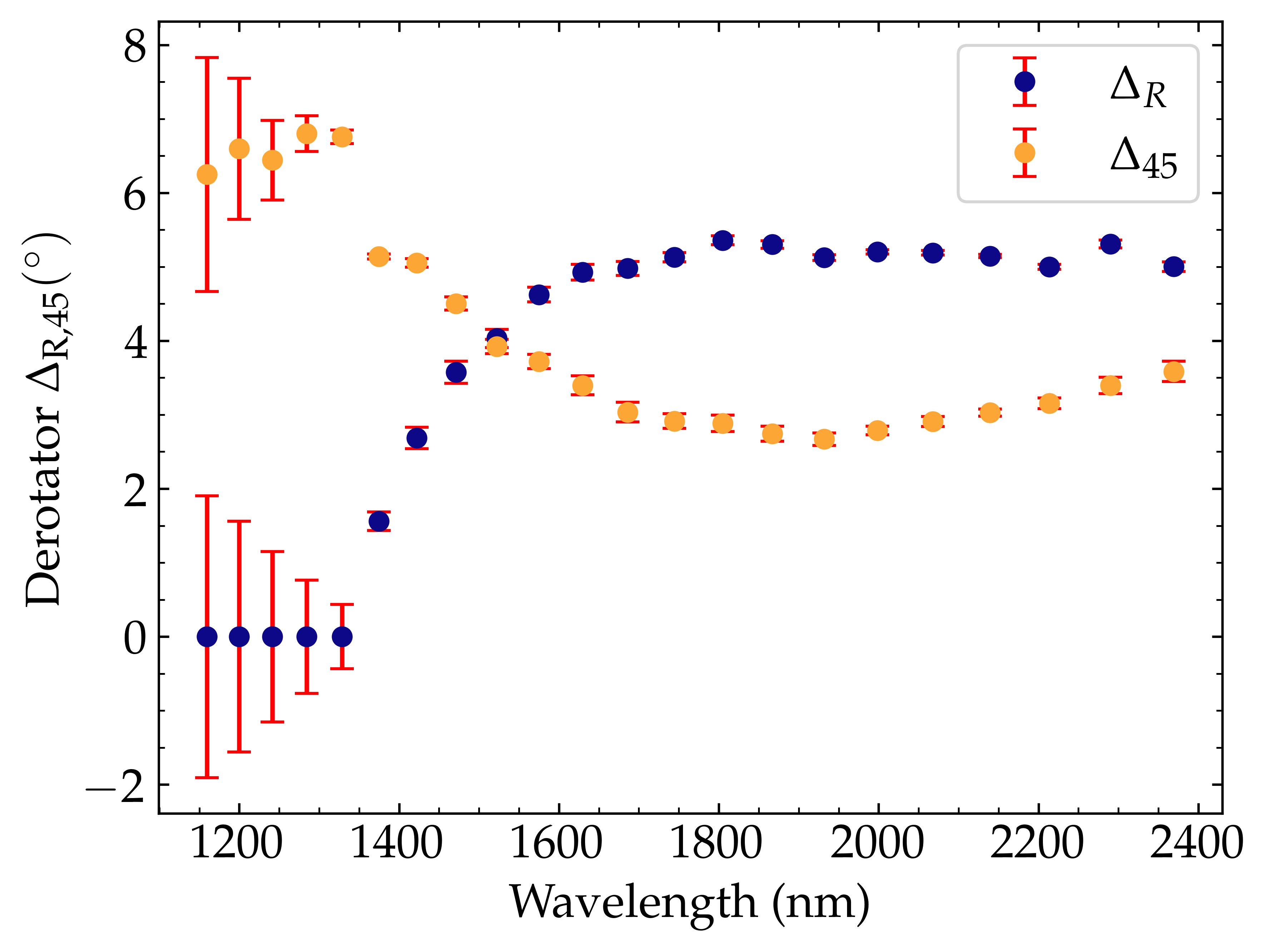}
    \end{subfigure}
    \caption{Fitted elliptical retardance components of the derotator (see Equation \ref{ellipticalretarder}). $\Delta_H$ is horizontal retardance, $\Delta_{45}$ is $45^\circ$ retardance, and $\Delta_R$ is right-handed circular retardance. We used Dataset 3 (Table \ref{tab:internal_calibration_datasets}) to fit these parameters. The component that differentiates this model from a linear retarder (as in previous CHARIS calibrations Refs. \citenum{hart2021characterizationinstrumentalpolarizationeffects} and \citenum{vanholstein2020calibrationinstrumentalpolarizationeffects}) is $\Delta_R$.}
    \label{fig:derotator_retardance}
\end{figure}

Figure \ref{fig:derotator_retardance} shows the fitted elliptical retardance components. This figure shows the fit to Dataset 3 (Table \ref{tab:internal_calibration_datasets}), and we get similar plots with the other polarized datasets (see Appendix \ref{appendix:ellipticalretfit}). The consistency of the fit across all three datasets leads us to believe that this effect is not a result of fitting artifacts. However, we are unsure why this effect was not present in previous CHARIS polarimetric calibrations. We observe elliptical retardance in Dataset 1 (Table \ref{tab:internal_calibration_datasets}), despite no known changes to the CHARIS optical path since the calibration in Ref. \citenum{hart2021characterizationinstrumentalpolarizationeffects}. One hypothesis is that the circular retardance arises from optical path components that we neglect in our model, which we discuss further in Appendix \ref{jarenappendix}. The only issue with this hypothesis is that it would imply that the effect should have been present in previous calibrations. Another possibility is changes over time in the optical path, e.g. mirror coatings oxidized between the time of this calibration and Ref. \citenum{hart2021characterizationinstrumentalpolarizationeffects} could result in form birefringence. Additionally, coatings can absorb moisture, which will change the refractive index as a function of time and degrade reflectivity. Both of these phenomena could affect the performance of the linear retarder model. Finally, since we could not reproduce the findings of Ref. \citenum{hart2021characterizationinstrumentalpolarizationeffects}, it is possible that differences in data reduction or fitting cause the discrepancy. Given this change in the derotator's retardance fit since the last CHARIS calibration, we argue that the optical path's polarization effects can change over time. This motivates regular re-calibration. The necessity of regular re-calibration is a primary motivation for our development of \texttt{pyPolCal}. 

\begin{figure}
    \centering
    \includegraphics[width=0.5\linewidth]{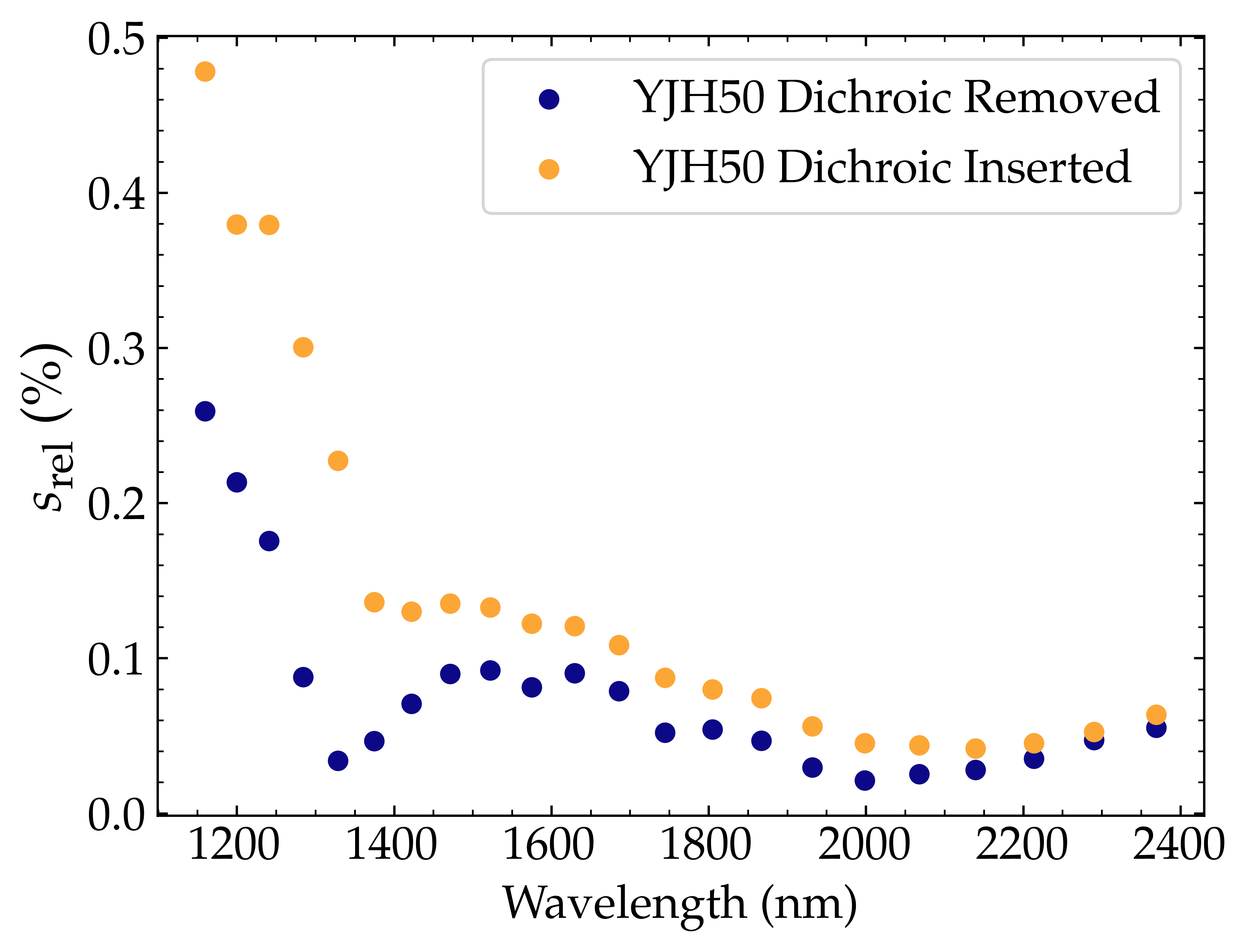}
    \caption{Relative polarimetric accuracy (see Equation \ref{eq:sem}) as a function of wavelength compared for datasets 1 and 2 (Table \ref{tab:internal_calibration_datasets}). Dataset 2 was taken with the YJH50 dichroic inserted. The model residuals are greater in the $J$-band when the dichroic is inserted.}
    \label{fig:dichroic_effect}
\end{figure}

\paragraph{Effects from the NIRWFS YJH50 Dichroic}
We use Dataset 2 (Table \ref{tab:internal_calibration_datasets}) to isolate the effects of the dichroic since it was taken before the NBS installation. Dichroics are known to have unpredictable polarization effects as a function of wavelength \cite{heath2020}. For this reason, we tried to fit a few different models to the dichroic: a linear retarder, an elliptical retarder, and a depolarizer. All of these attempts were unsuccessful; they did not reduce residuals, and the fitted parameters had large errors. Consequently, we exclude the dichroic from the model and assign the identity matrix to $M_{\mathrm{dichroic}}$. We compare the polarimetric accuracy of fits to datasets 1 and 2 (Table \ref{tab:internal_calibration_datasets}) to assess the effects of the dichroic that we could not model effectively. Figure \ref{fig:dichroic_effect} shows the relative polarimetric accuracy (Equation \ref{eq:sem}) calculated with and without the dichroic inserted. The relative polarimetric accuracy with the dichroic inserted is on average $0.18\%$ worse for wavelengths $<1400$ nm, so we conclude that the dichroic produces faint crosstalk in the \textit{J}-band that we cannot characterize with a component Mueller matrix listed in Subsection \ref{subsec:comp_mueller_matrices}. 

\paragraph{Effects of the Nasmyth beam-switcher}
By analyzing the NBS design from Ref. \citenum{zheng2022optical}, we can deduce from ray tracing that, for ideal optics, the NBS acts as a diagonal matrix $\mathrm{diag}(1,-1,1,-1)$. Normalized double differences are not sensitive to sign flips in $V$, but we include it in the model for mathematical completeness. Adding an elliptical retarder and/or a depolarizer Mueller matrix after $\mathrm{diag}(1,-1,1,-1)$ would account for effects from non-ideal optics. However, these models did not produce reliable fits. The elliptical retarder model is degenerate with the derotator, and it did not significantly reduce residuals. The depolarizer also did not reduce residuals, and its effects are likely absorbed into the Wollaston modulation efficiency ($|\eta|$) fit. The final model we implement for the NBS ($M_{\mathrm{NBS}}$) is then $\mathrm{diag}(1,-1,1,-1)$. We can compare the relative polarimetric accuracy of datasets 2 (the orange in Figure \ref{fig:dichroic_effect}) and 3 (Figure \ref{fig:nbs_in_s_rel}) to assess whether the NBS affects the polarimetric accuracy after implementing this model. The polarimetric accuracies are very similar, and we attribute the slight differences to minor changes over time in the optical path. We calculated these polarimetric accuracies using separate model parameter fits, so it is possible that effects from the NBS were absorbed into other model components. However, even if this is true, it will not impact the model performance since the NBS is permanently installed. We conclude that elliptical retardance and depolarization resulting from non-ideal optics in the NBS are either negligible or absorbed into the fits of other components.

\begin{figure}
    \centering
    \includegraphics[width=0.5\linewidth]{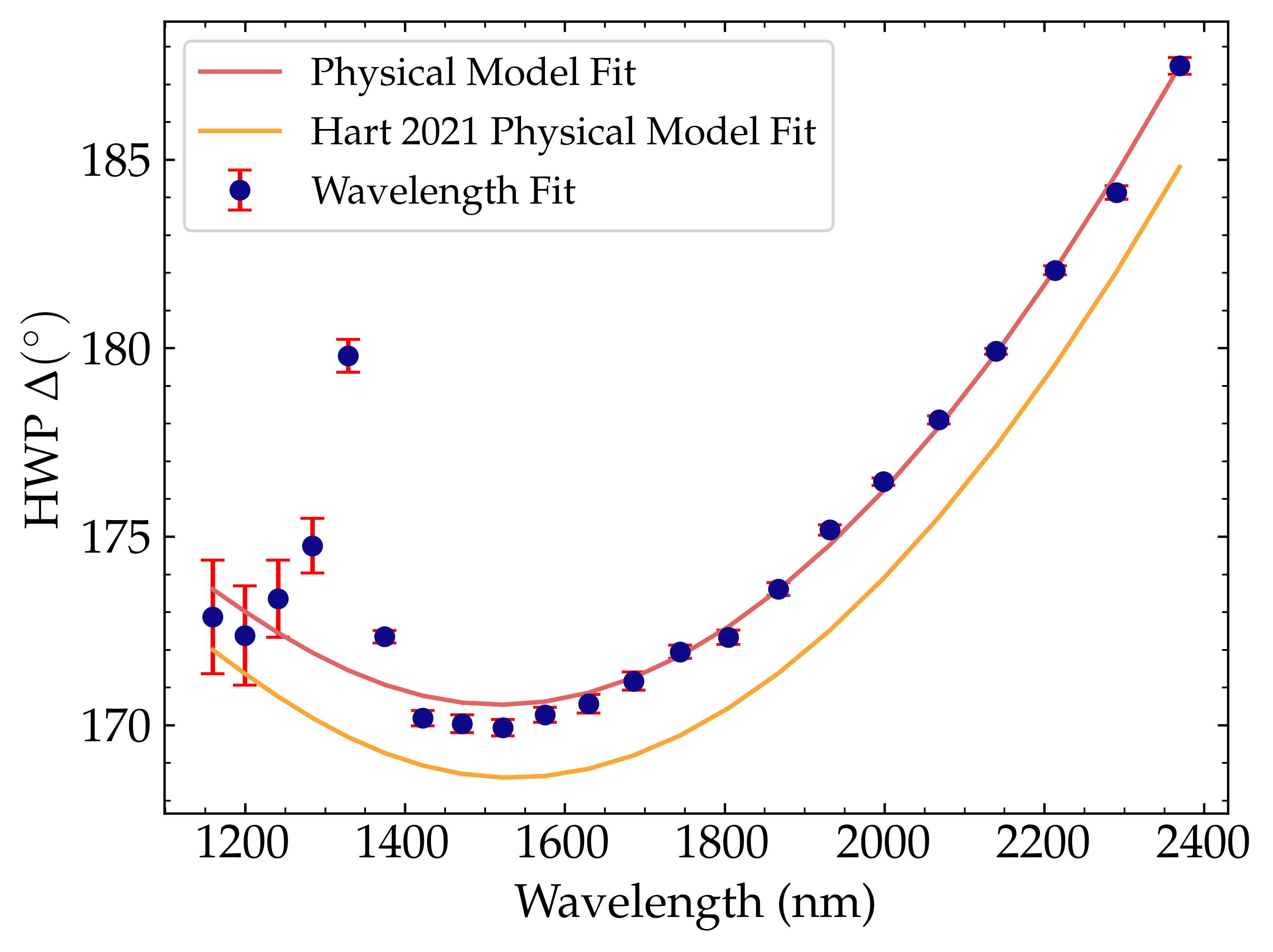}
    \caption{HWP retardance fitted per-wavelength and globally with Equation \ref{eq:hwp_physical_model}. The purple dots indicate the per-wavelength fit and the solid lines are physical model fits. The physical model fit from Ref. \citenum{hart2021characterizationinstrumentalpolarizationeffects} is in orange. Our wavelength and physical model fits in this plot use Dataset 3 (Table \ref{tab:internal_calibration_datasets}). A perfect HWP would be a horizontal line at $180^\circ$. We calculated the errors using the method from Appendix E of Ref. \citenum{van_Holstein_2020}.}
    \label{fig:hwp_retardance}
\end{figure}

\paragraph{Other Model Parameter Fits}
The globally fit parameters in Table \ref{tab:fitted_values} are sensible. The misalignment angles are near zero, which is expected since SCExAO optics are aligned precisely. The HWP fit is similar to the fit in Ref. \citenum{hart2021characterizationinstrumentalpolarizationeffects}. This is reasonable since we do not expect the HWP to change over time. Figure \ref{fig:hwp_retardance} shows the HWP wavelength fit and final global fit, overlaid with the fit from Ref. \citenum{hart2021characterizationinstrumentalpolarizationeffects}. As observed in Ref. \citenum{hart2021characterizationinstrumentalpolarizationeffects}, when $\Delta_H$ of the derotator approaches $180^\circ$ (see Figure \ref{fig:derotator_retardance}), a degeneracy arises in the per-wavelength fits. This causes the discrepancy between the wavelength and global fit around $1300$ nm.  

\begin{figure}
    \centering
    \includegraphics[width=0.5\linewidth]{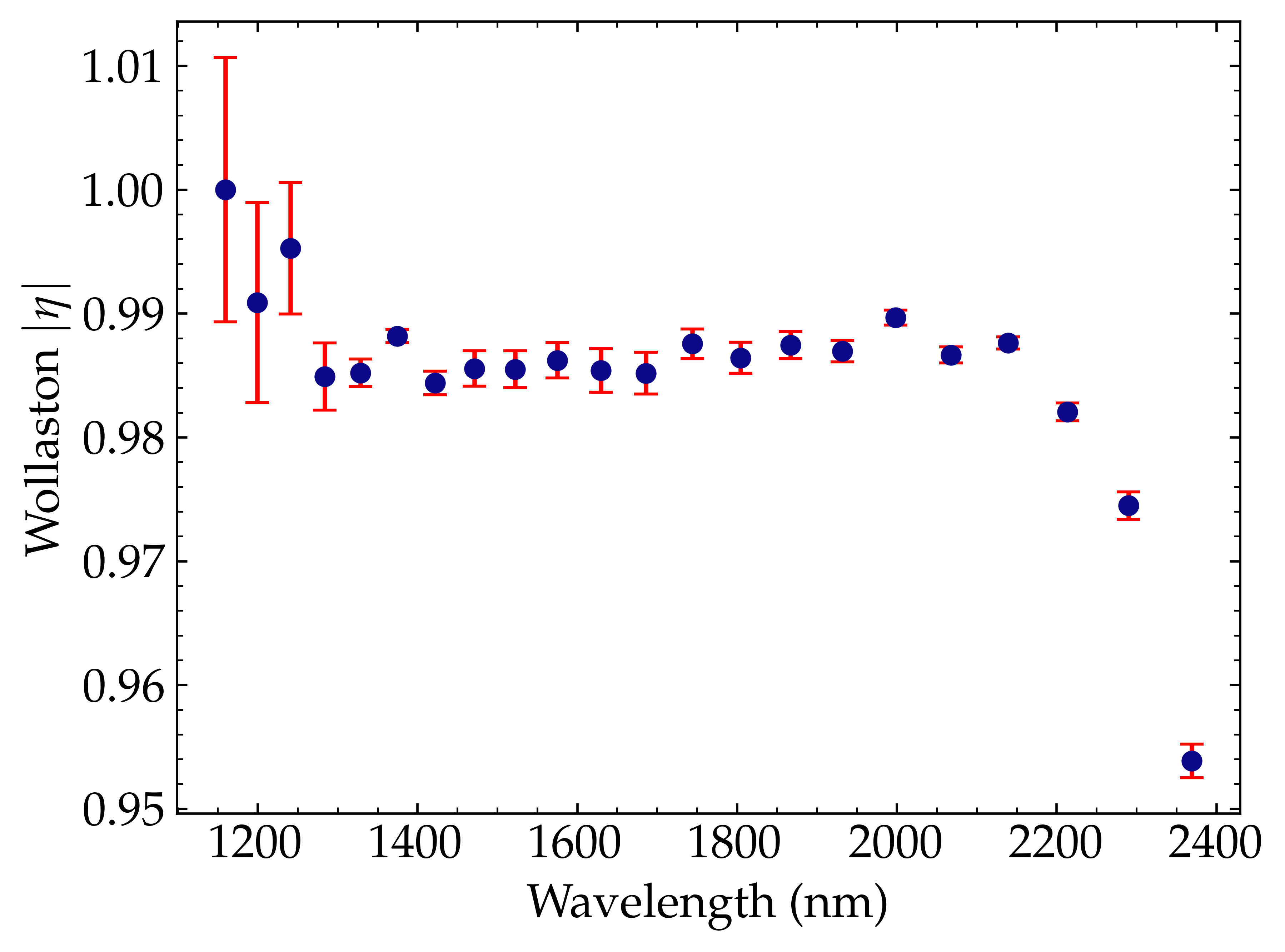}
    \caption{Fitted Wollaston prism modulation efficiency ($|\eta|$). A perfect Wollaston prism has $|\eta|=1$. This fit will also absorb depolarization effects from optics neglected in the model. We calculated the errors using the method from Appendix E of Ref. \citenum{van_Holstein_2020}.}
    \label{fig:wol_eta}
\end{figure}

Previous works fit the diattenuation of the calibration polarizer to account for a wavelength-dependent polarized intensity reduction in the data \cite{hart2021characterizationinstrumentalpolarizationeffects,vanholstein2020calibrationinstrumentalpolarizationeffects}. As shown in Figure \ref{fig:wol_eta}, we instead fit a Wollaston modulation efficiency parameter ($|\eta|$), which has the same effect. We argue that it is unlikely for a calibration polarizer to show the wavelength dependence found in Ref. \citenum{hart2021characterizationinstrumentalpolarizationeffects}, where they found that the calibration polarizer's diattenuation dropped to $\epsilon=0.95$ for some wavelengths. We obtained the datasheet for the CHARIS calibration polarizer (MOXTEK BIR05A), which details the minimum and maximum transmission coefficients for all wavelengths used in this calibration \cite{moxtek_bir_series}. We can calculate the diattenuation using Equation \ref{eq:diattenuation}. This yields $\epsilon\geq0.996$ for all CHARIS wavelength bins. It is more likely that the calibration polarizer diattenuation fit in Ref. \citenum{hart2021characterizationinstrumentalpolarizationeffects} is caused by a combination of degenerate effects: the true wavelength dependence of the calibration polarizer, depolarizing optics neglected in the Mueller matrix model, and the Wollaston modulation efficiency. The calibration polarizer is removed from the on-sky model, so any polarization effects that we attribute to the calibration polarizer would not be part of the final model. Fitting the Wollaston modulation efficiency allows this reduction in polarized intensity to be implemented into the on-sky Mueller matrix model. The fit will encapsulate both the true Wollaston modulation efficiency and the depolarizing effects from optics neglected in the Mueller matrix model. 

\subsubsection{Unpolarized Internal Calibrations}
\label{subsubsec:unpol-internal}
\begin{figure}[htpb]
    \centering
    \includegraphics[width=0.5\linewidth]{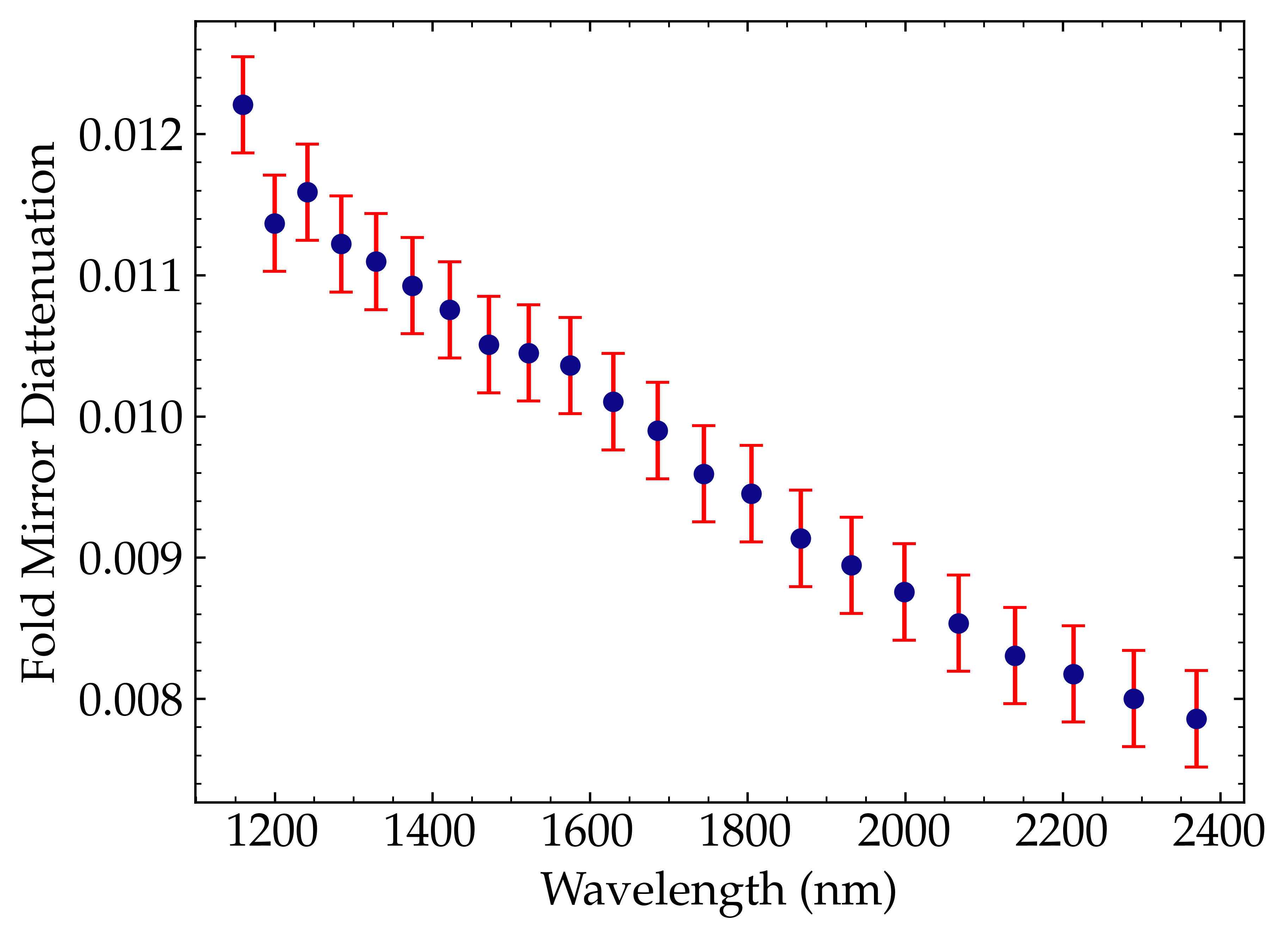}
    \caption{Fitted diattenuation of the fold mirror that directs the calibration source into the optical path. We carried out this fit with Dataset 4 (Table \ref{tab:internal_calibration_datasets}). We calculated the errors using the method from Appendix E of Ref. \citenum{van_Holstein_2020}.}
    \label{fig:foldmirror}
\end{figure}

\begin{figure}[htpb]
    \centering
    \includegraphics[width=0.5\linewidth]{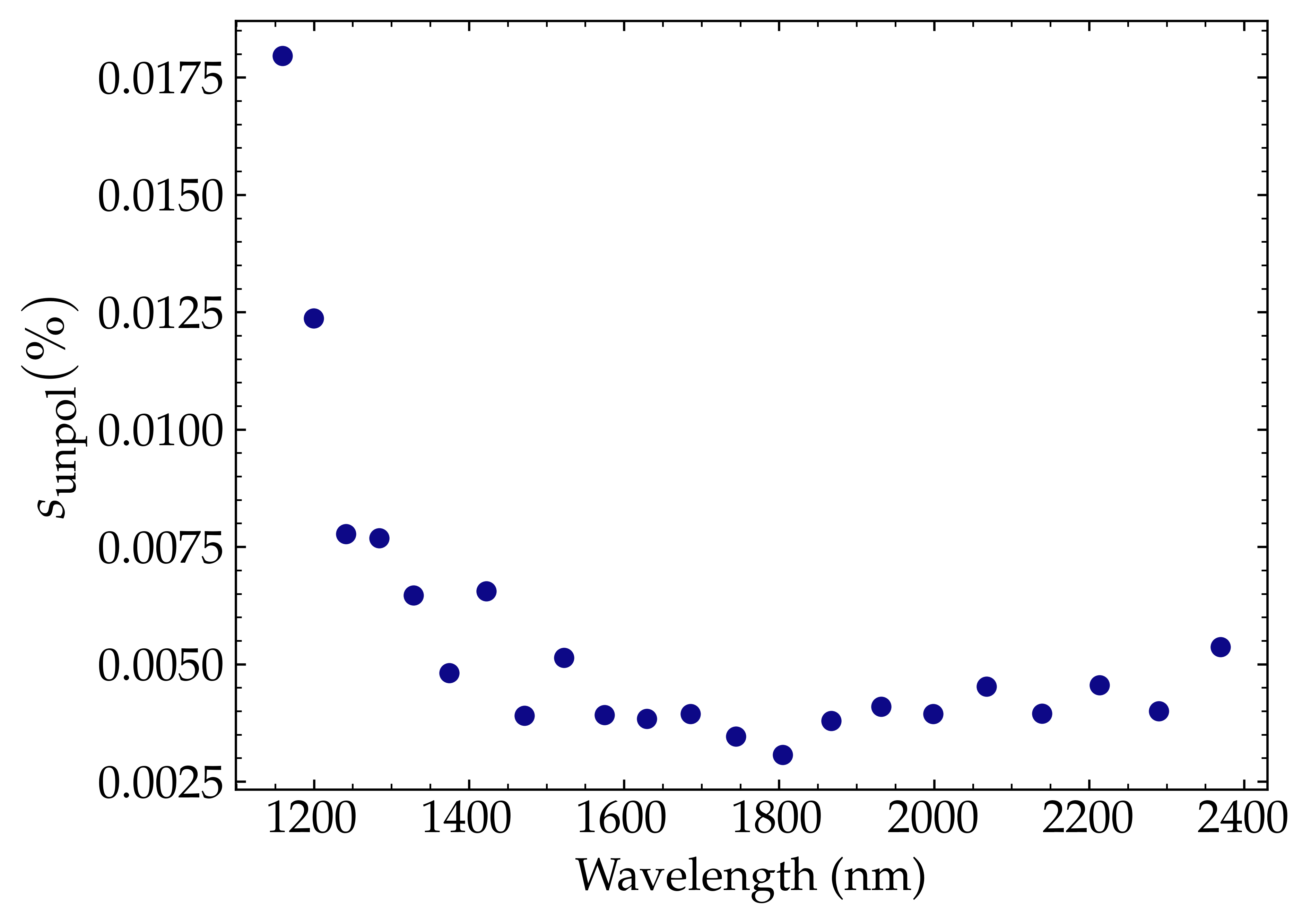}
    \caption{The polarimetric accuracy (Equation \ref{eq:sem}) of the unpolarized internal calibration as a function of wavelength, which contributes to the absolute polarimetric accuracy (Equation \ref{eq:sabs}). This plot is made with Dataset 4 (Table \ref{tab:internal_calibration_datasets}).}
    \label{fig:sres_unpol}
\end{figure}

We could not reliably fit diattenuator models to the HWP, derotator, NBS, or YJH50 dichroic. The only parameter we fit is the fold mirror diattenuation, shown in Figure \ref{fig:foldmirror}. Figure \ref{fig:sres_unpol} shows the polarimetric accuracy of the unpolarized calibration dataset ($s_{\mathrm{unpol}}$), which contributes to the calculation of the absolute polarimetric accuracy (Equation \ref{eq:sabs}). Overall, this single diattenuator model fits the data very well; $s_{\mathrm{unpol}}$ is negligible compared to $s_{\mathrm{rel}}$.

This result aligns with theoretical expectations. Equation \ref{eq:doubledif_pe} predicts that we should not observe diattenuation from the NBS or YJH50 dichroic since they are non-rotating components downstream of the HWP. We also expect that the derotator should produce negligible diattenuation because it is downstream of the HWP and static during HWP cycles. This reasoning extends to on-sky data because the derotator moves slowly enough during HWP cycles to be treated as effectively static. We find that the HWP produces negligible diattenuation, which is consistent with the VLT/SPHERE/IRDIS unpolarized internal calibrations from Ref. \citenum{van_Holstein_2020}. To corroborate our model, we obtained an independent fit of the fold mirror's diattenuation in visible wavelengths from SCExAO/VAMPIRES (see Appendix \ref{appendix:foldmirrordiat}). This fit is consistent with what we observe in the NIR (since the diattenuation of a gold mirror should decrease as a function of wavelength). We conclude that double differencing cancels out most diattenuation in the optical path.

\section{CALIBRATING THE DIATTENUATION OF M3 WITH AN UNPOLARIZED STANDARD STAR}
\label{sec:onsky-cal}
\subsection{Unpolarized Standard Star Observations and Fitting Procedure}
\label{subsec:onsky-procedure}
In November 2025, following the NBS installation, we observed an unpolarized standard star with the YJH50 dichroic inserted (Table \ref{tab:unpol_obs}) to calibrate the diattenuation of M3. We found the target using the Python package ZetaPersei, which compiles polarization standard star catalogs into searchable formats \cite{zetapersei}. We additionally observed a polarized standard star in February 2025, which we discuss in Appendix \ref{appendix:polstd}. 
\begin{table}[H]
    \centering
    \caption{Summary of the unpolarized standard star observation used to calibrate the diattenuation of M3. We obtained the DoLP measurement from Ref. \citenum{zetapersei}.}
    \label{tab:unpol_obs}
    \begin{tabular}{lcccccc}
        \toprule
        \textbf{Object} & \textbf{Date} & \textbf{Int. Time} & \textbf{Total Exp.} & \textbf{HWP} & \textbf{Altitude Range} & \textbf{DoLP} \\
         & & \textbf{(s)} & \textbf{(min)} & \textbf{Cycles} & \textbf{($^\circ$)} & \textbf{(\%)} \\
        \midrule
        HD293396 & 2025-11-3 & 10.3 & 17 & 23 & 64.0 -- 67.0 & $0.0 \pm 0.2$ \\
        \bottomrule
    \end{tabular}
\end{table}

This observation was taken with standard HWP cycles at the critical angles (0$^\circ$, 45$^\circ$, 22.5$^\circ$, 67.5$^\circ$), which enables recovery of Stokes $Q$ and $U$. With CHARIS, HWP angles are automatically offset to account for the sky rotating relative to the telescope, using the following tracking law:
\begin{equation}
\label{eq:hwptracking}
    \theta_{\mathrm{HWP}}'=\theta_{\mathrm{HWP}}+0.5p+a
\end{equation}
where $p$ is the parallactic angle, $a$ is the altitude angle, and $\theta_{\mathrm{HWP}}$ is the uncorrected HWP angle. 

We can extract normalized double differences $x_{0^\circ}$ and $x_{22.5^\circ}$ from one HWP cycle, where the subscripts denote the uncorrected HWP angle. These normalized double differences at 0$^\circ$ and 22.5$^\circ$ are equivalent to Stokes $Q$ and $U$, but we refer to them as $x_{0^\circ}$ and $x_{22.5^\circ}$ for consistency with Section \ref{sec:internal-cal}. We extracted data cubes from our raw data and flat-fielded them following standard procedures using the CHARIS DEP \cite{brandt2017datareductionpipelinecharis}. We then sky-subtracted the frames using the CHARIS data processing pipeline (DPP) \cite{currie2020onskyperformancerecentresults}. 

\begin{figure}
    \centering
    \includegraphics[width=0.5\linewidth]{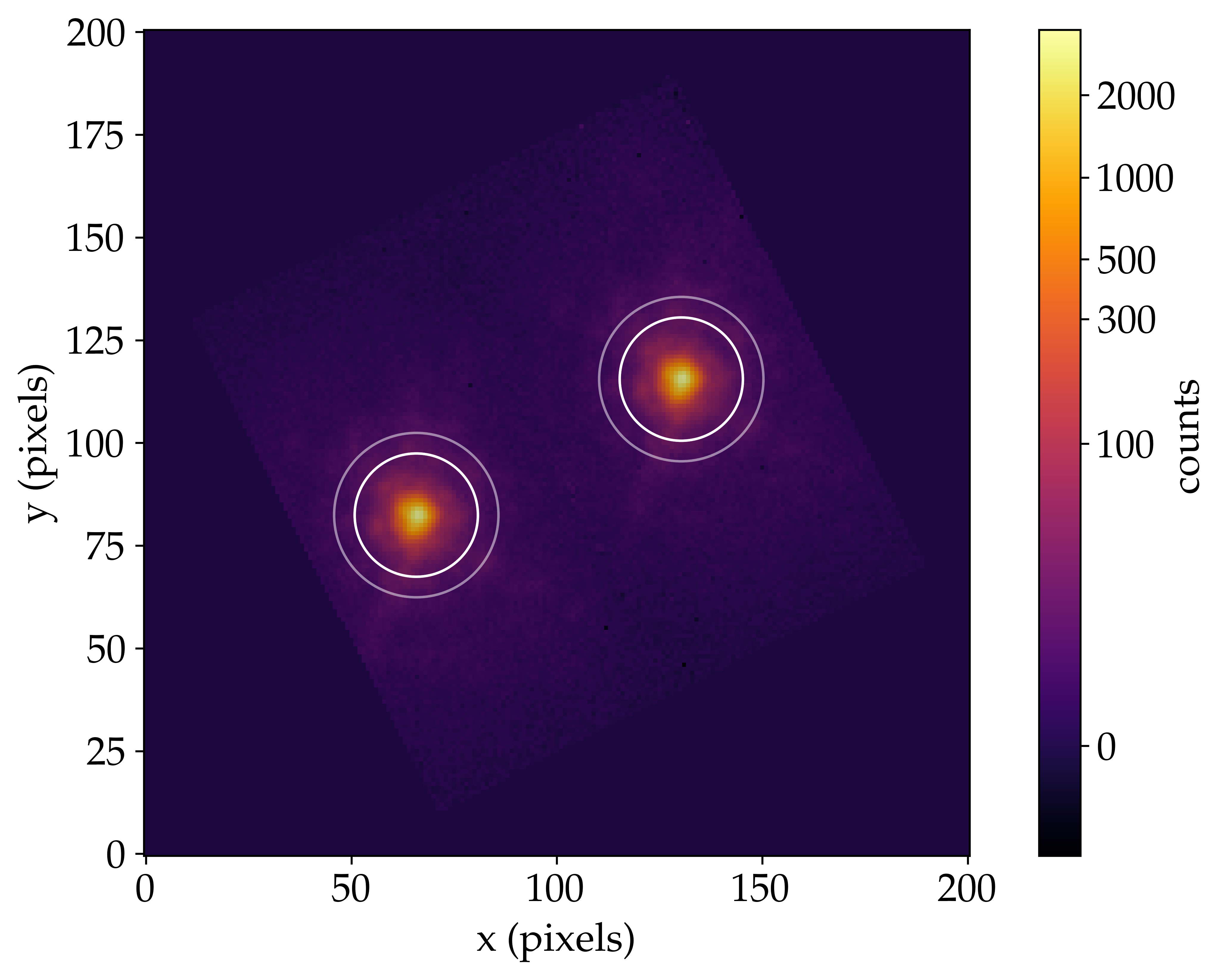}
    \caption{Data cube slice of an unpolarized standard star. The circular apertures used for aperture sums and anulli used for background subtraction are overlaid. The radii of the circular apertures are 3X the fitted FWHM of the PSF, and the annuli's outer radii are 5 pixels larger than the aperture radii.}
    \label{fig:unpolstd_frame}
\end{figure}

To calculate single differences and sums (Equations \ref{singledif} and \ref{singlesum}), we perform aperture photometry on both Wollaston beams using \texttt{Photutils} \cite{Bradley2025-oh}. An example unpolarized standard data cube slice with apertures overlaid is shown in Figure \ref{fig:unpolstd_frame}. We fit a 2D Gaussian to the point spread function (PSF) of each beam to obtain the centroid and full-width half-maximum (FWHM). We use 3 times the fitted FWHM as the radii for circular apertures, setting a maximum radius of 45 pixels to ensure that the apertures don't hit the detector edge. We additionally compute the median pixel value of an annulus with an outer radius that is 5 pixels larger than the aperture radius for additional background subtraction. We compute $I_{\mathrm{det, R}}$ and $I_{\mathrm{det, L}}$ in Equations \ref{singledif} and \ref{singlesum} by: 
\begin{equation}
    I_{\mathrm{det, L/R}}=S_{\mathrm{ap}}-\tilde{I}_{\mathrm{annulus}}A_{\mathrm{ap}}
\end{equation}
where $S_{\mathrm{ap}}$ is the aperture sum, $\tilde{I}_{\mathrm{annulus}}$ is the median pixel value of the annulus, and $A_{\mathrm{ap}}$ is the area of the aperture. We again estimate errors on the fluxes as $\sqrt{I_{\mathrm{det, L/R}}}$. We calculate the normalized double differences using the procedure in Section \ref{sec:charis-optical-path}, and we calculate their errors via Gaussian error propagation of the flux error. We extract the HWP angle using the \texttt{RET-POS1} FITS header, since it correctly implements the HWP tracking laws shown in Equation \ref{eq:hwptracking}. The header \texttt{RET-ANG1} is the uncorrected HWP angle $\theta_{\mathrm{HWP}}$ in Equation \ref{eq:hwptracking}, which we also extract to identify the critical angles for double differencing.

We use Equation \ref{onskymodel} as the Mueller matrix model, modeling the standard stars as completely unpolarized Stokes vectors. We use the fit from Section \ref{sec:internal-cal} for components downstream of Subaru's tertiary mirror (M3), and we model M3 as a diattenuator-retarder using Equation \ref{diattenuatorretarder}. We model the diattenuation as a function of wavelength and dielectric function approximation parameters ($m_1$, $m_2$, $b_1$, and $b_2$) using Equation \ref{eq:m3_physical_model}. $m_1$, $m_2$, $b_1$, and $b_2$ are free parameters. While we cannot fit the retardance directly to the unpolarized standard stars, these dielectric function approximation parameters allow us to calculate the retardance. 

We fit the physical model for M3 across all wavelength bins simultaneously using \texttt{scipy.optimize.minimize} \cite{virtanen2020scipy}. We used starting guesses of $1.94$, $2.08$, $13.70$, and $13.89$ for $m_1$, $m_2$, $b_1$, and $b_2$, respectively, derived from older calibration datasets. For bounds, we used $0.5$ times to $2$ times these older parameter fits.

\subsection{Results and Discussion}
\label{subsec:onsky-results}
\begin{figure}
    \centering
    \begin{subfigure}[b]{0.48\textwidth}
        \centering
        \includegraphics[width=\textwidth]{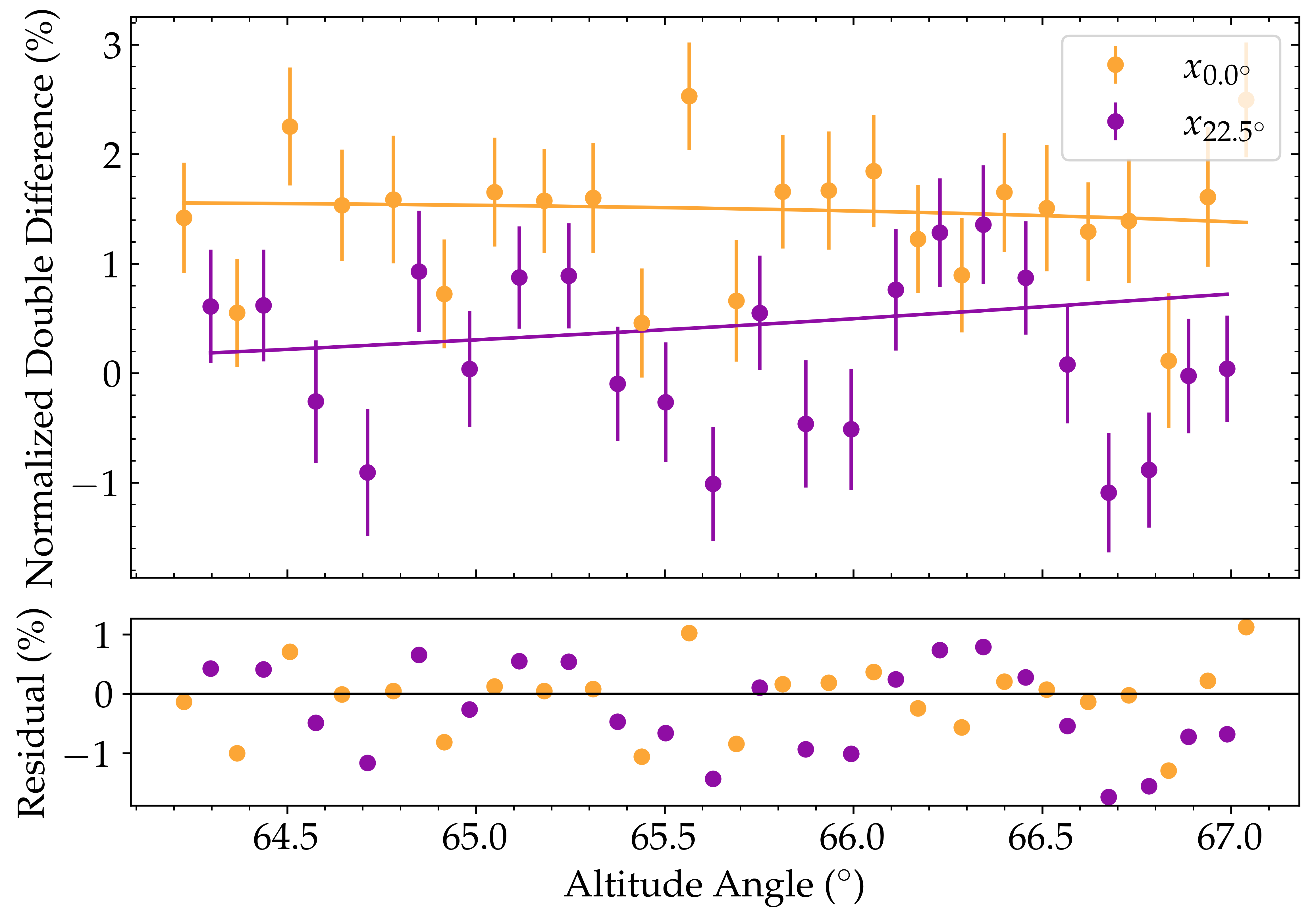}
        \caption{$\lambda=1375$ nm}
    \end{subfigure}
    \hfill 
    \begin{subfigure}[b]{0.48\textwidth}
        \centering
        \includegraphics[width=\textwidth]{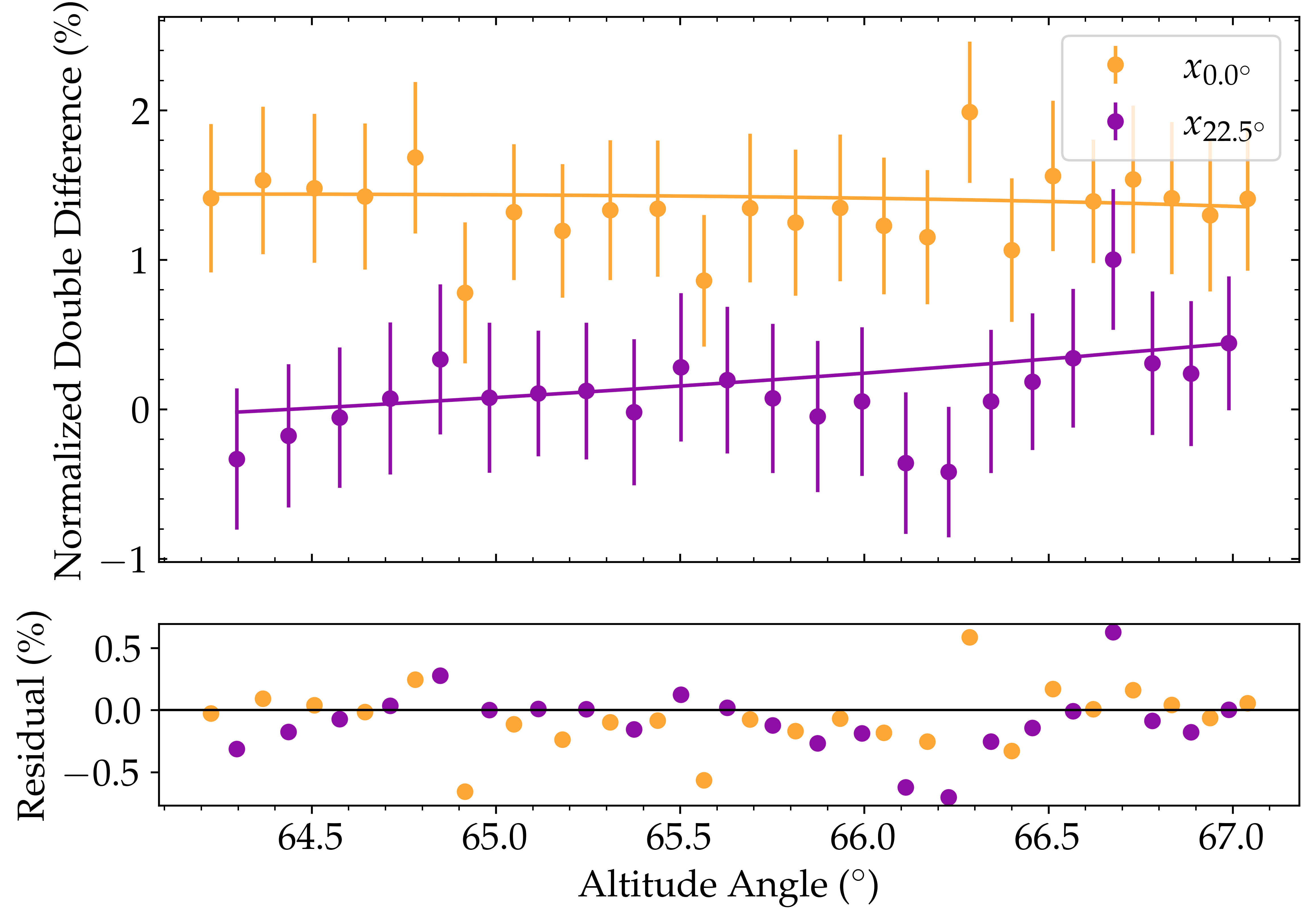}
        \caption{$\lambda$=1575 nm}
    \end{subfigure}
    
    \vspace{3pt} 
    
    \begin{subfigure}[b]{0.48\textwidth}
        \centering
        \includegraphics[width=\textwidth]{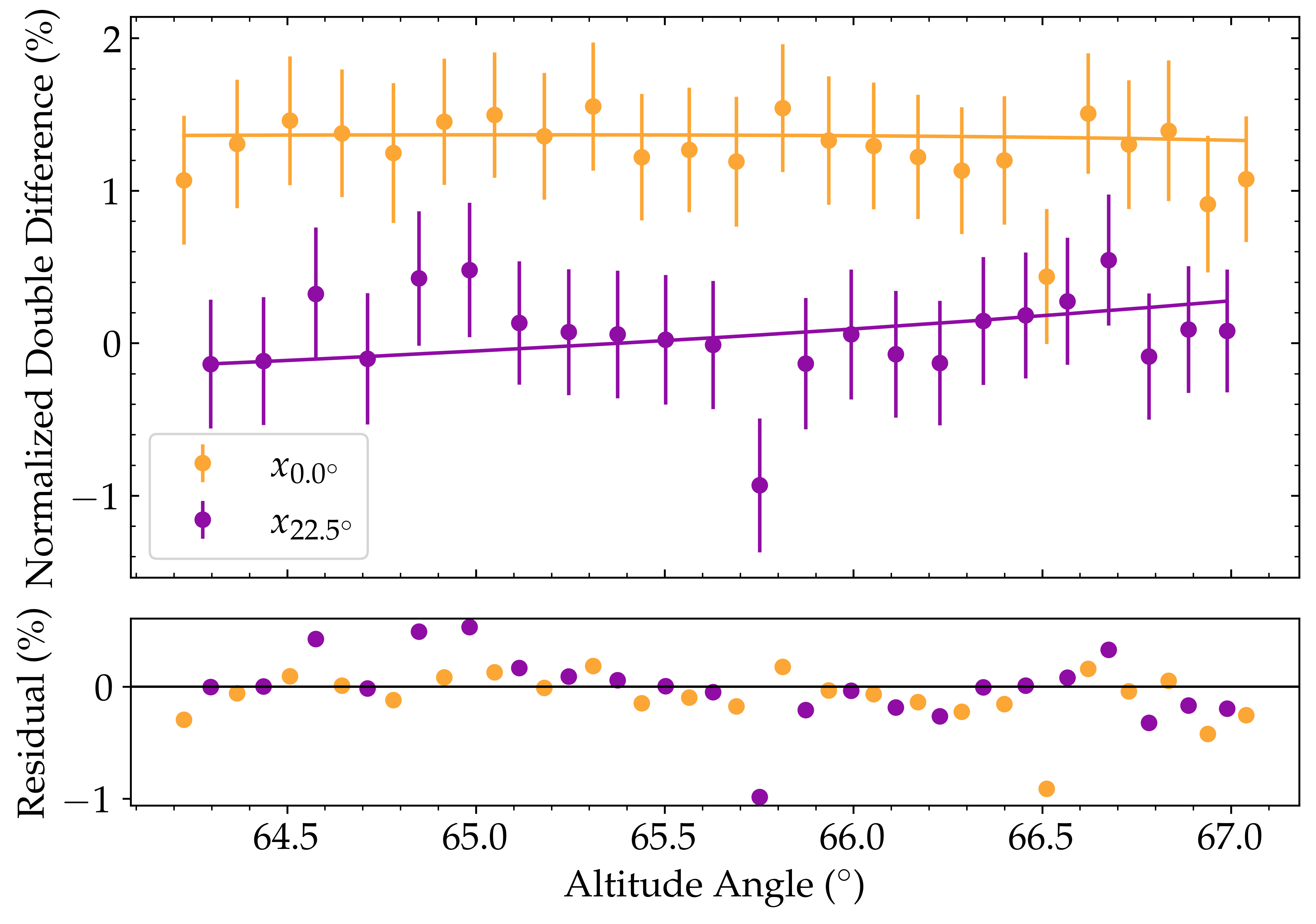}
        \caption{$\lambda=1744$ nm}
    \end{subfigure}
    \hfill
    \begin{subfigure}[b]{0.48\textwidth}
        \centering
        \includegraphics[width=\textwidth]{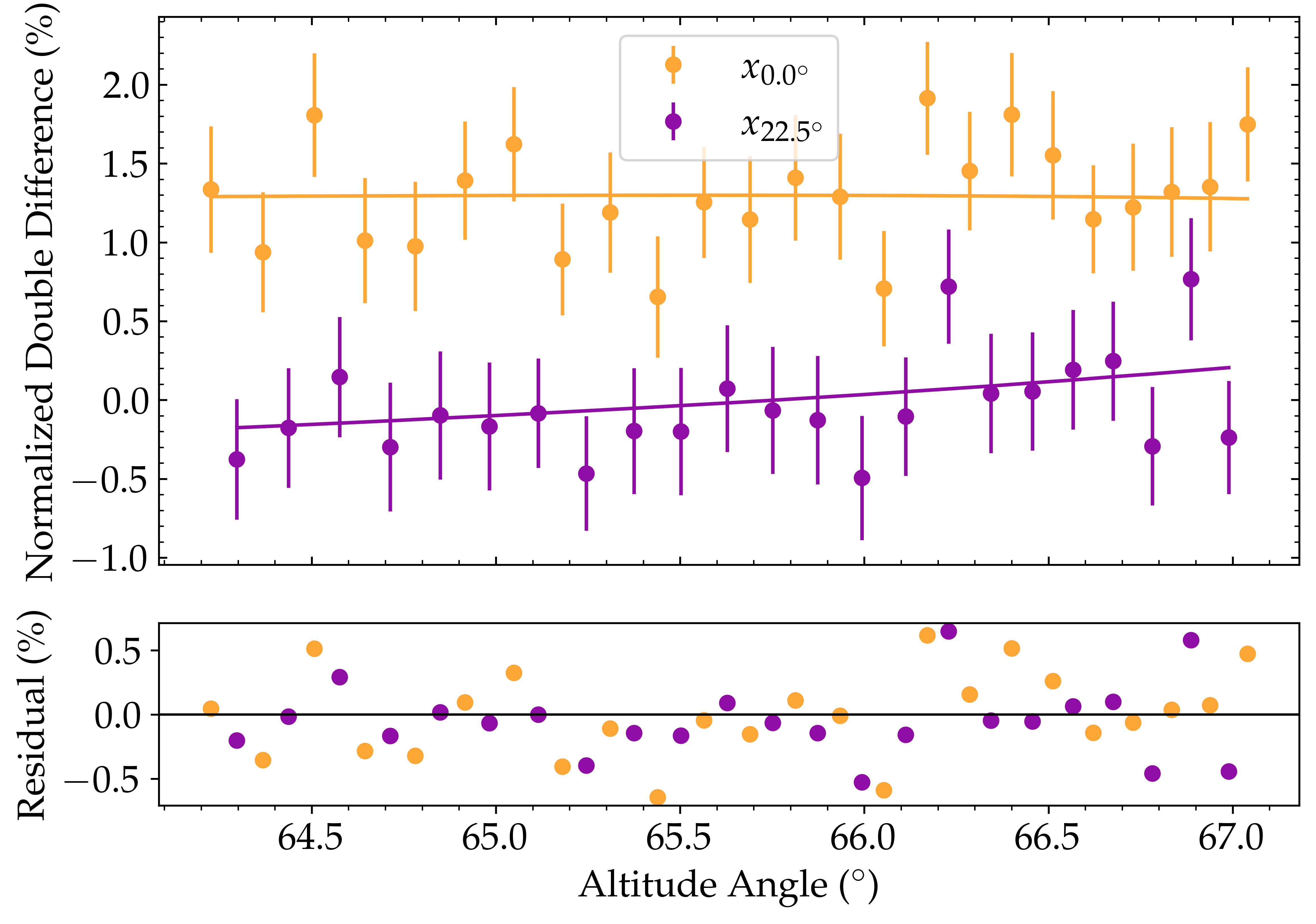} 
        \caption{$\lambda$=1932 nm}
    \end{subfigure}
    \vspace{3pt}
    \caption{Modeled and observed normalized double differences (Equation \ref{normdoublediff}) of unpolarized standard star HD293396 plotted as a function of the altitude angle for 4 wavelengths. HD293396 is modeled as completely unpolarized. The errors are estimated using photon noise estimates from the circular apertures.}
    \label{fig:ddbyalt}
\end{figure}

\begin{figure}
    \centering
    \includegraphics[width=0.5\linewidth]{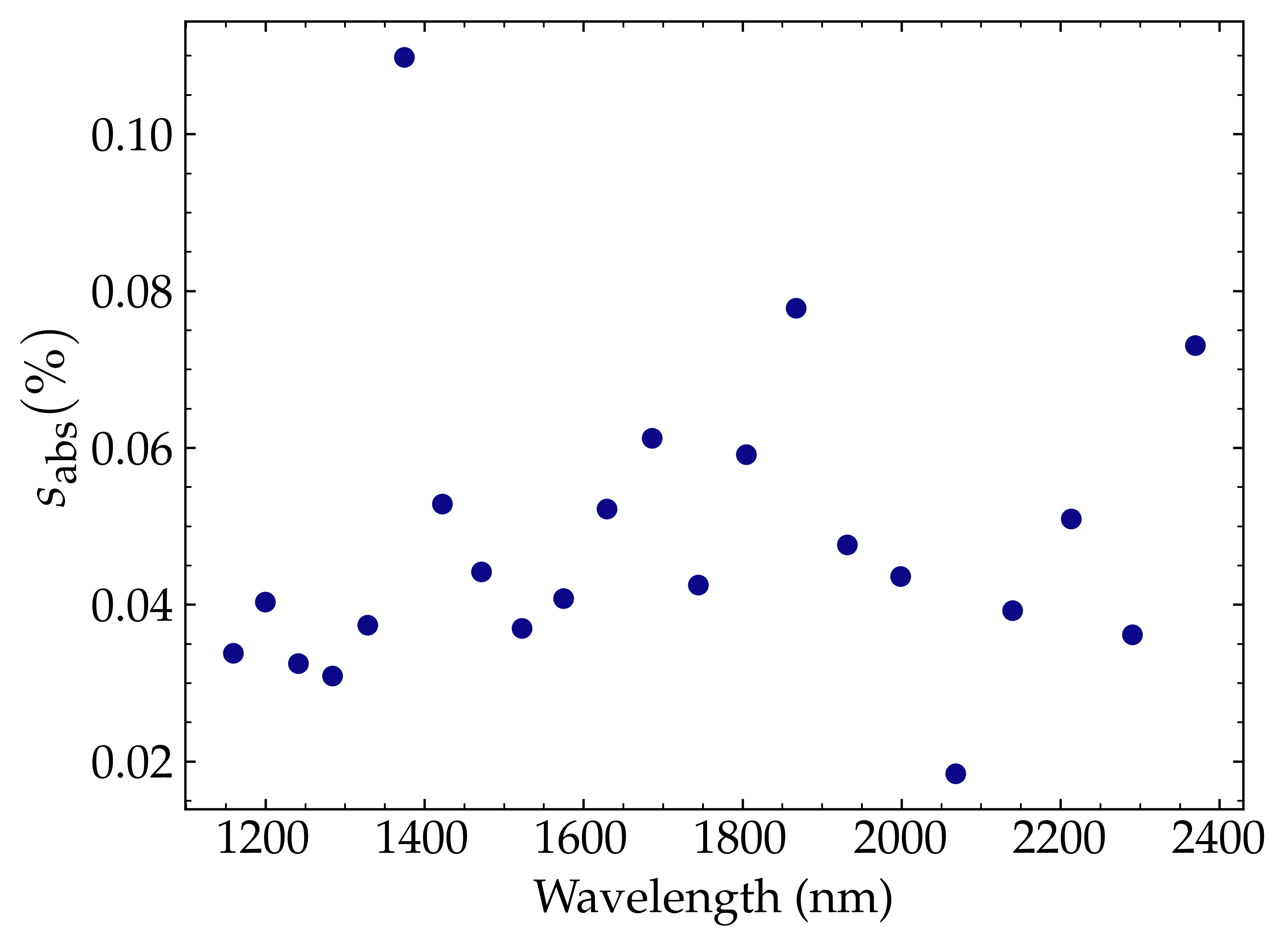}
    \caption{The absolute polarimetric accuracy (Equation \ref{eq:sabs}) of the Mueller matrix model. This demonstrates a strong wavelength dependence in the model residuals for the on-sky calibration.}
    \label{fig:s_abs}
\end{figure}

\begin{figure}
    \centering
    \begin{subfigure}[b]{0.48\textwidth}
        \centering
        \includegraphics[width=\textwidth]{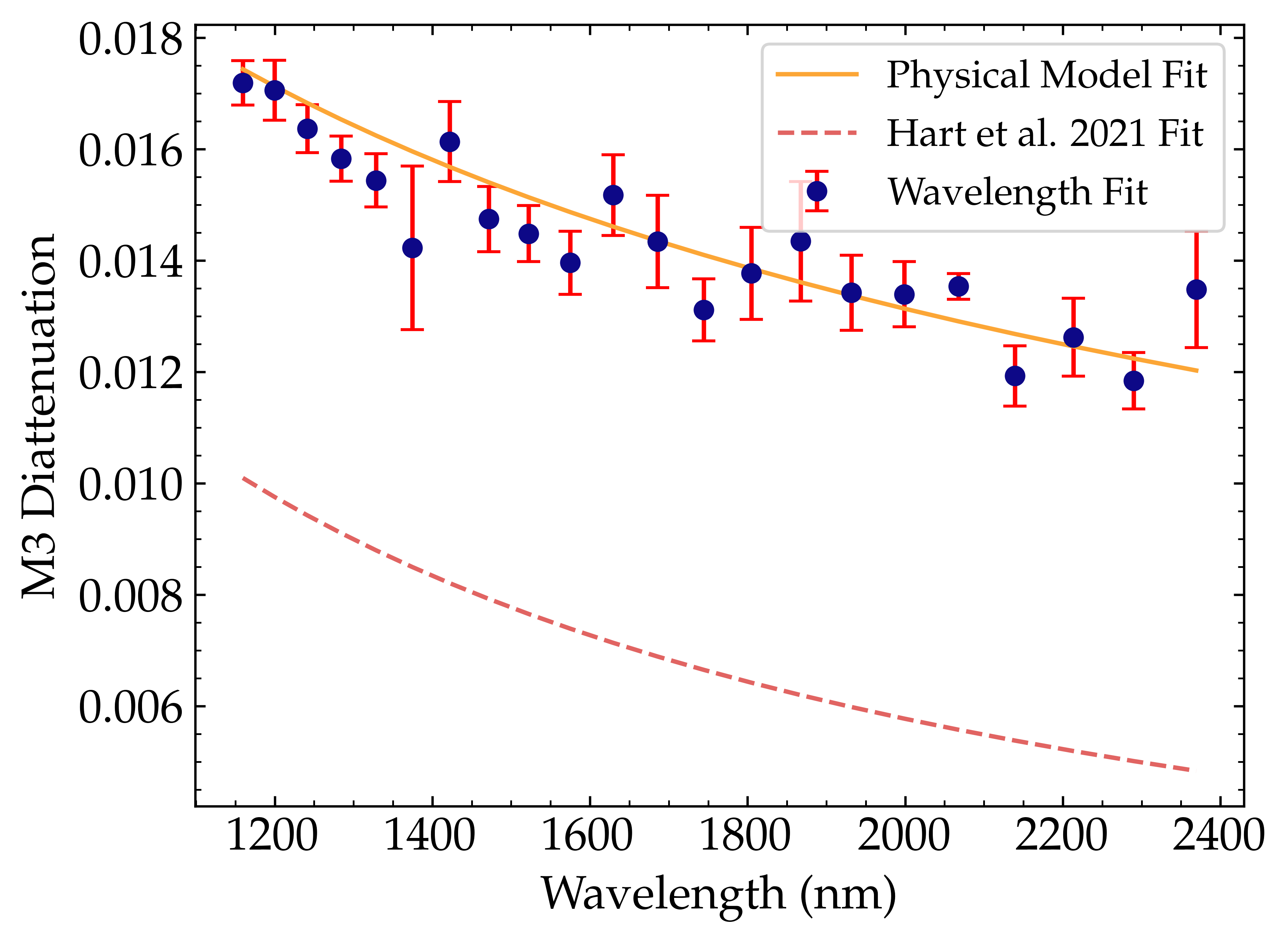}
    \end{subfigure}
    \begin{subfigure}[b]{0.48\textwidth}
        \centering
        \includegraphics[width=\textwidth]{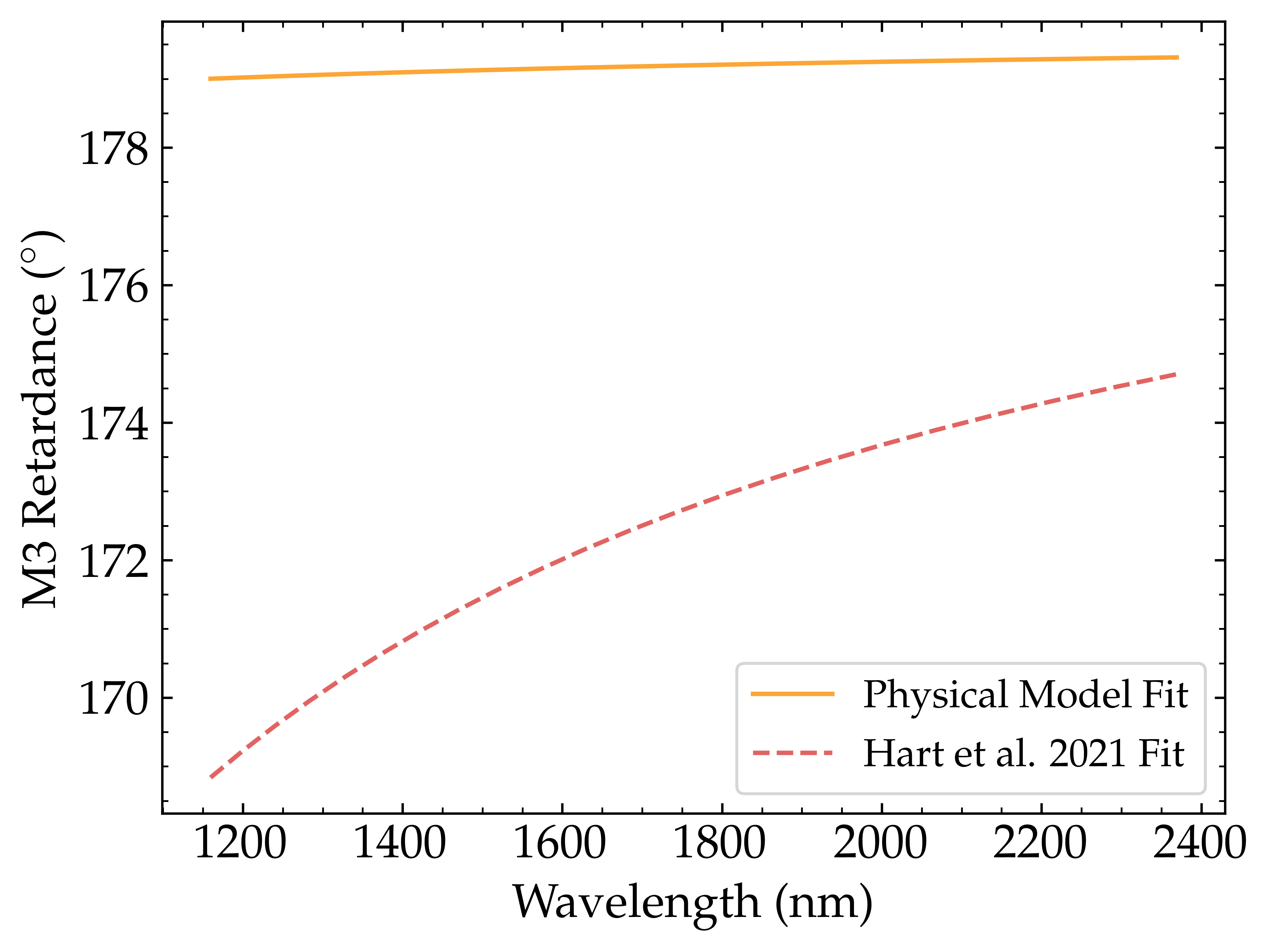}
    \end{subfigure}
    \caption{The diattenuation and retardance of M3 as a function of wavelength. The solid lines are the physical model and the dots are a separate wavelength fit. The error bars on the wavelength fits are calculated using the procedure from Appendix E of Ref. \citenum{van_Holstein_2020}. We calculate the retardance and diattenuation from the global model parameters shown in Table \ref{tab:m3_fit} using the procedure in Subsection \ref{subsec:physmodel}. The physical model fit from Ref. \citenum{hart2021characterizationinstrumentalpolarizationeffects} is also shown, which significantly deviates from our fit.}
    \label{fig:m3diat}
\end{figure}

Table \ref{tab:m3_fit} shows the fitted physical model parameters for the diattenuation of M3. As in Section \ref{sec:internal-cal}, we again could not estimate errors using the calculation from Ref. \citenum{van_Holstein_2020} due to degeneracy in the model. Multiple combinations of parameters produce the same diattenuation, so the model is poorly constrained. To help break the degeneracy, we recommend that future works include a wider range of altitude angles in their datasets, either by observing multiple unpolarized standards or observing the same standard at multiple altitude angles \cite{hart2021characterizationinstrumentalpolarizationeffects}.
\begin{table}[H]
    \centering
    \caption{M3 physical model (Equation \ref{eq:m3_physical_model}) parameters fit using observations of unpolarized standard star HD293396.}
    \label{tab:m3_fit}
    \begin{tabular}{|l|c|}
        \hline
        \textbf{Parameter} & \textbf{Value} \\
        \hline
        $m_1$ & 3.186 \\
        $m_2$ & 1.040 \\
        \hline
        $b_1$ & 15.56 \\
        $b_2$ & 9.69 \\
        \hline
    \end{tabular}
\end{table}
Figure \ref{fig:ddbyalt} shows the modeled and observed double differences as a function of altitude angle. The residuals in our fit are greater than Ref. \citenum{hart2021characterizationinstrumentalpolarizationeffects} due to noise. These residuals are the main limiting factor in our model's polarimetric accuracy. We analyze how the model residuals change as a function of wavelength by computing $s_{\mathrm{abs}}$ (Equation \ref{eq:sabs}), as shown in Figure \ref{fig:s_abs}. The contribution from $s_{\mathrm{unpol}}$ is negligible. We set the correction $k=4$ to account for the 4 parameters we fit to this dataset. The wavelength dependence of the residuals is also greater than what was observed in Ref. \citenum{hart2021characterizationinstrumentalpolarizationeffects}, which is again caused by noise in the calibration data.
Figure \ref{fig:m3diat} shows the diattenuation and retardance of M3 as a function of wavelength, with the fit from Ref. \citenum{hart2021characterizationinstrumentalpolarizationeffects} overlaid. We observe significantly more diattenuation than Ref. \citenum{hart2021characterizationinstrumentalpolarizationeffects}. Since the physical model parameters are fit using M3's diattenuation, this in turn causes an increase in the modeled retardance (see Equation \ref{eq:m3_physical_model}). This change in time of M3's polarization effects motivates regular recalibration. We also note that, since M3 was re-coated after we took this calibration data, our fit is only an estimate of the current model parameter values. The re-coating could change M3's diattenuation, and new calibration data are necessary to ensure model accuracy. \texttt{pyPolCal} will allow future investigators to easily re-fit the model for M3's diattenuation to new on-sky calibration data once it becomes available. 

\clearpage
\section{POLARIMETRIC EFFICIENCY AND ACCURACY}
\label{sec:pol-eff}
    \subsection{Polarimetric Efficiency}
\label{subsec:poleff}
\begin{figure}
    \centering
    \begin{subfigure}[b]{0.48\textwidth}
        \centering
        \includegraphics[width=\textwidth]{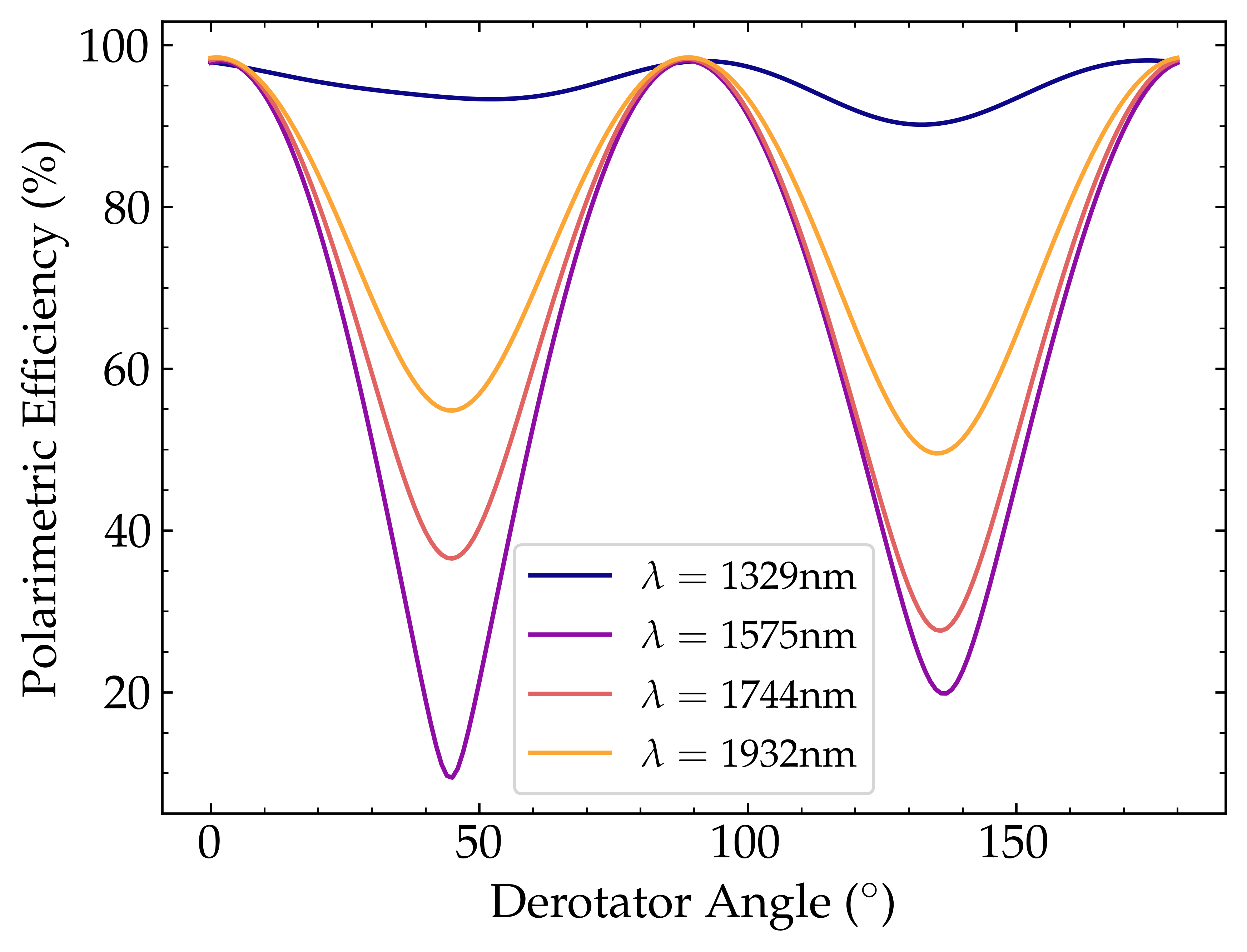}
        \caption{$p=32^\circ$ $a=48^\circ$}
    \end{subfigure}
    \hfill 
    \begin{subfigure}[b]{0.48\textwidth}
        \centering
        \includegraphics[width=\textwidth]{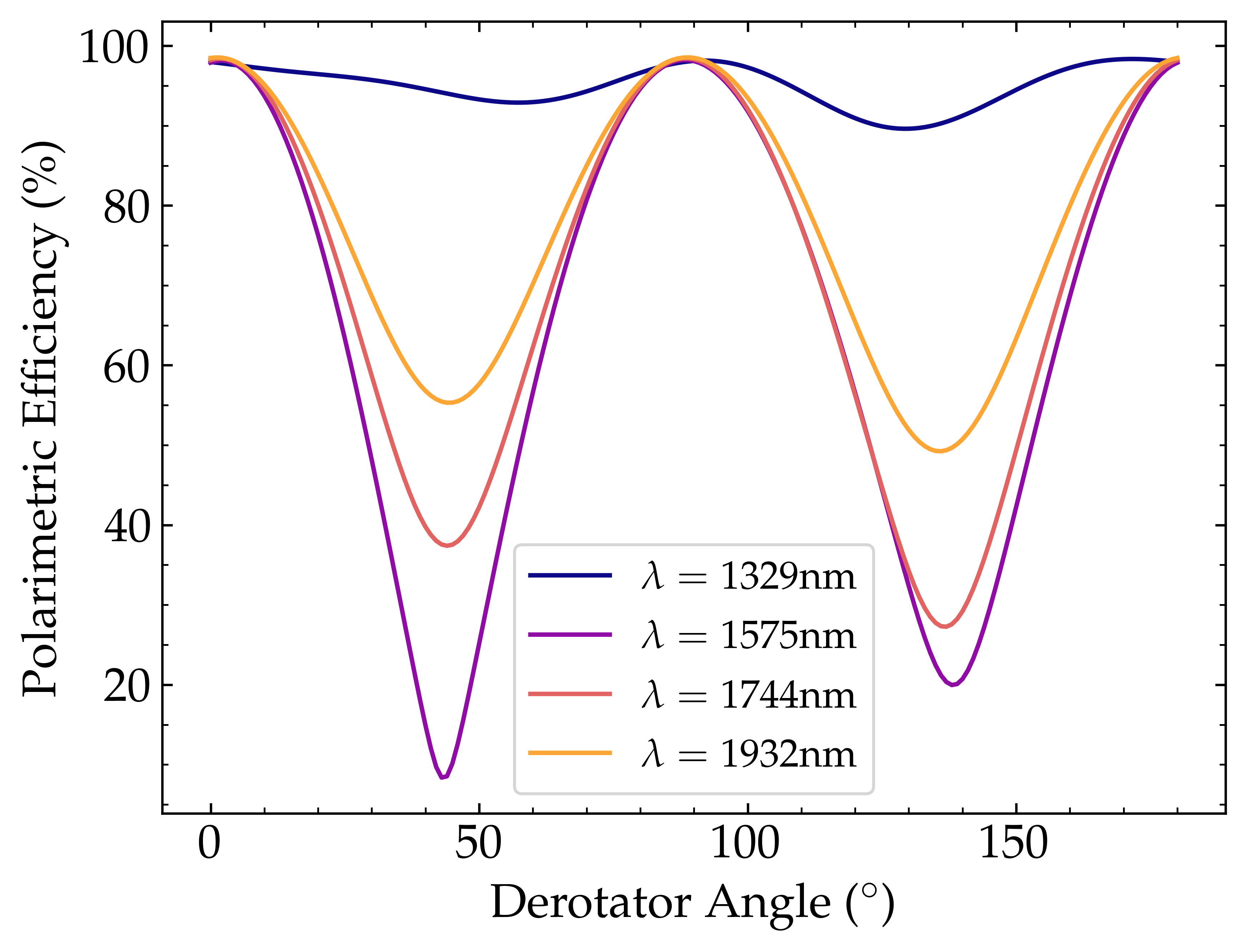}
        \caption{$p=94^\circ$ $a=60^\circ$}
    \end{subfigure}
    
    \vspace{3pt} 
    
    \begin{subfigure}[b]{0.48\textwidth}
        \centering
        \includegraphics[width=\textwidth]{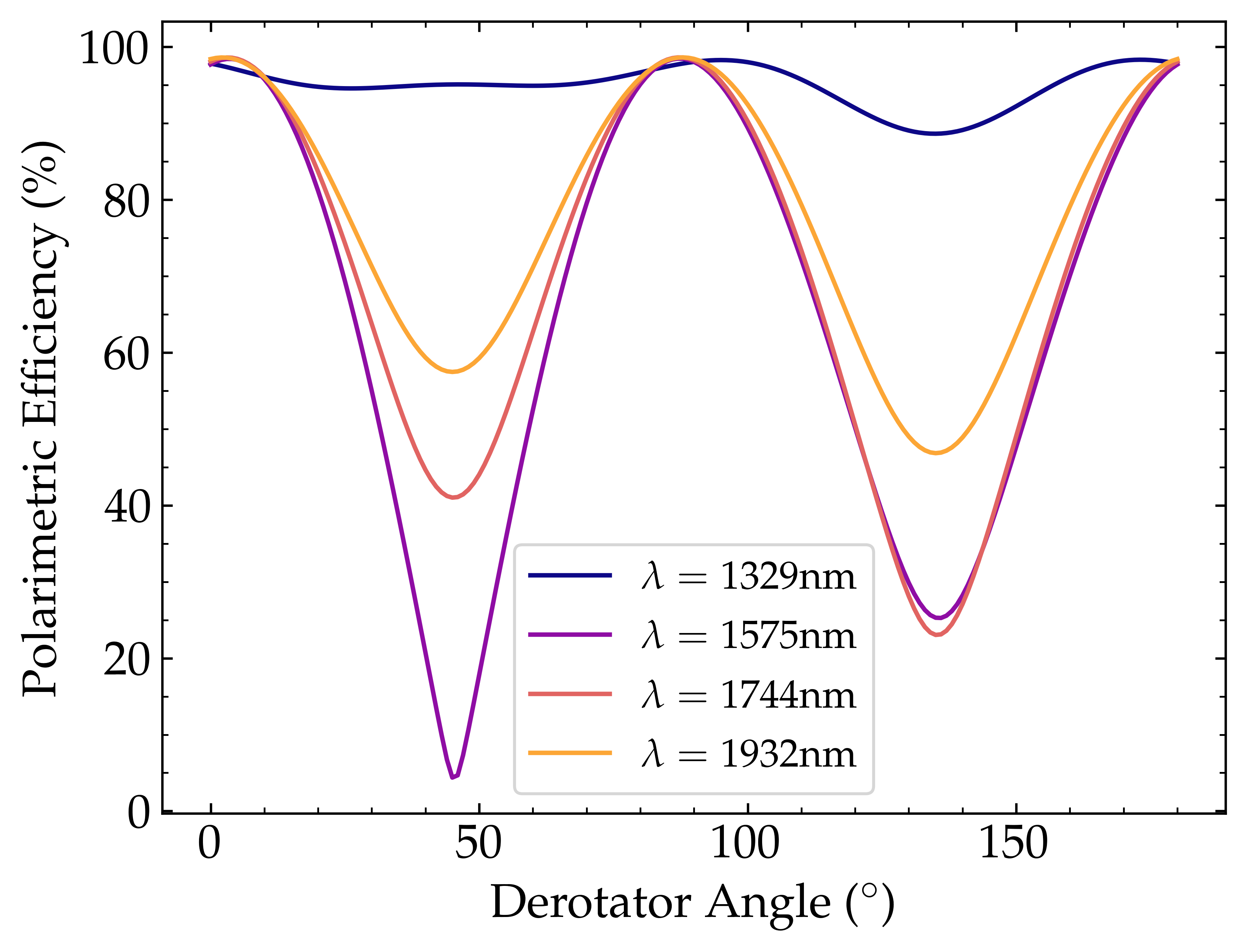}
        \caption{$p=126^\circ$ $a=60^\circ$}
    \end{subfigure}
    \hfill
    \begin{subfigure}[b]{0.48\textwidth}
        \centering
        \includegraphics[width=\textwidth]{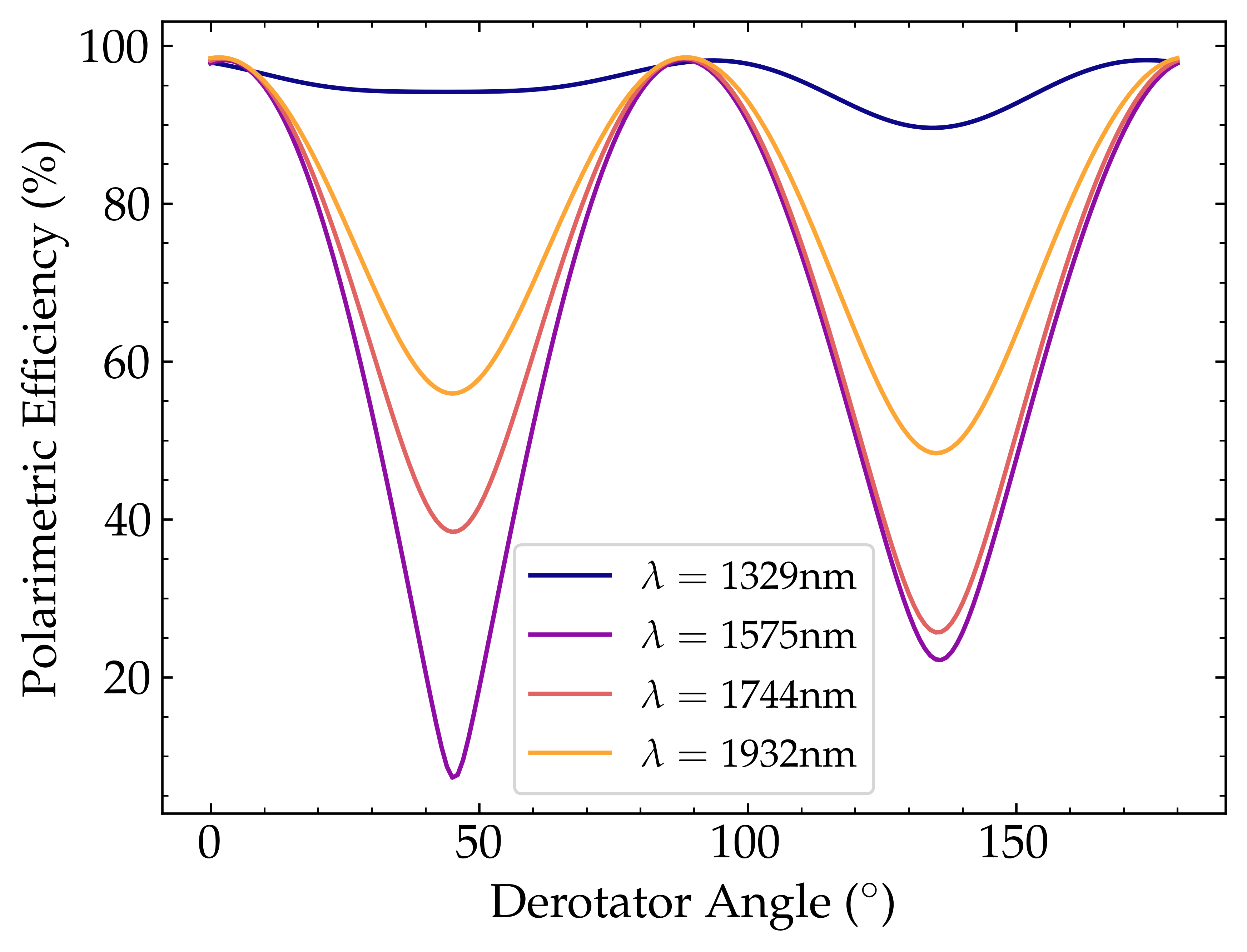} 
        \caption{$p=24^\circ$ $a=67^\circ$}
    \end{subfigure}
    \vspace{3pt}
    \caption{The polarimetric efficiency of the CHARIS Mueller matrix model. We used combinations of altitude ($a$) and parallactic ($p$) angles from real observations, which the polarimetric efficiency depends weakly on. It depends heavily on the derotator angle and wavelength.}
    \label{fig:poleffbyalt}
\end{figure}

Since CHARIS cannot measure circularly polarized light, any linearly polarized light converted into circular polarization is lost information. We model the DoLP for a $100\%$ $-Q$-polarized source after passing through the CHARIS optical path to quantify how much linearly polarized signal is lost. We interpret this modeled DoLP as the polarimetric efficiency. We account for the effects of M3's retardance, which depends on telescope pointing, using the model from Ref. \citenum{hart2021characterizationinstrumentalpolarizationeffects} and include the HWP tracking laws to simulate real targets. Figure \ref{fig:poleffbyalt} shows the polarimetric efficiency as a function of the derotator angle, plotted for different altitudes and parallactic angles. The derotator converts almost all polarized signal to circular polarization around  $45^\circ$ and $135^\circ$, minimizing the polarimetric efficiency. Changing the altitude and parallactic angle does not affect which derotator angles minimize polarimetric efficiency. It affects the magnitude of the dips in polarimetric efficiency as a function of derotator angle by a few percent. These dips depend heavily on wavelength. Though our model corrects for the optical path's polarization effects, reductions in polarimetric efficiency will decrease the signal-to-noise ratio of polarimetric data. 

\subsection{Polarimetric Accuracy of the Model}
\label{subsec:polacc-of-model}
\begin{figure}
    \centering
    \begin{subfigure}[b]{0.48\textwidth}
        \centering
        \includegraphics[width=\textwidth]{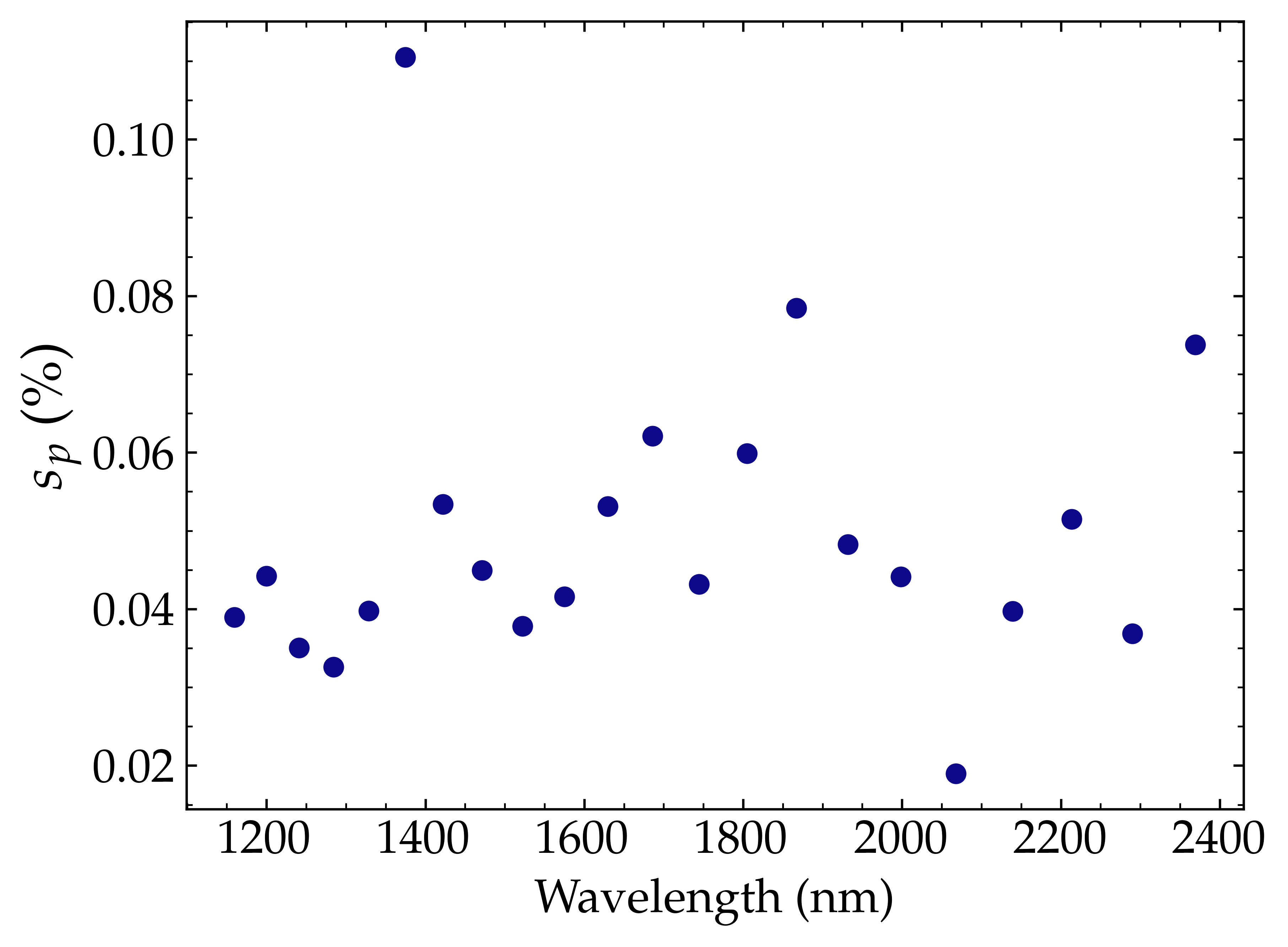}
        \caption{1\% Polarized Target}
    \end{subfigure}
    \hfill 
    \begin{subfigure}[b]{0.48\textwidth}
        \centering
        \includegraphics[width=\textwidth]{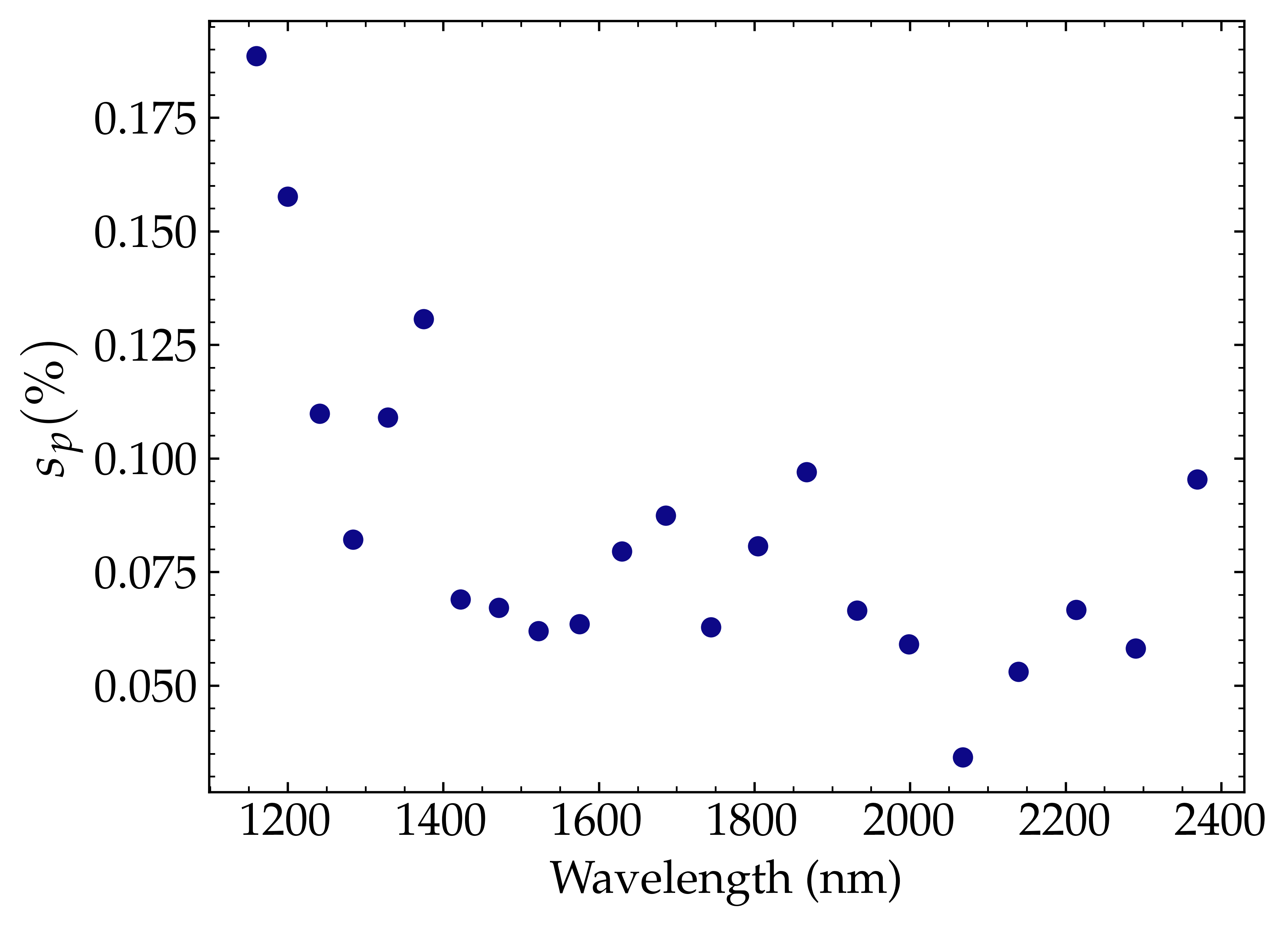}
        \caption{30\% Polarized Target}
    \end{subfigure}
    
    \vspace{3pt} 
    
    \begin{subfigure}[b]{0.48\textwidth}
        \centering
        \includegraphics[width=\textwidth]{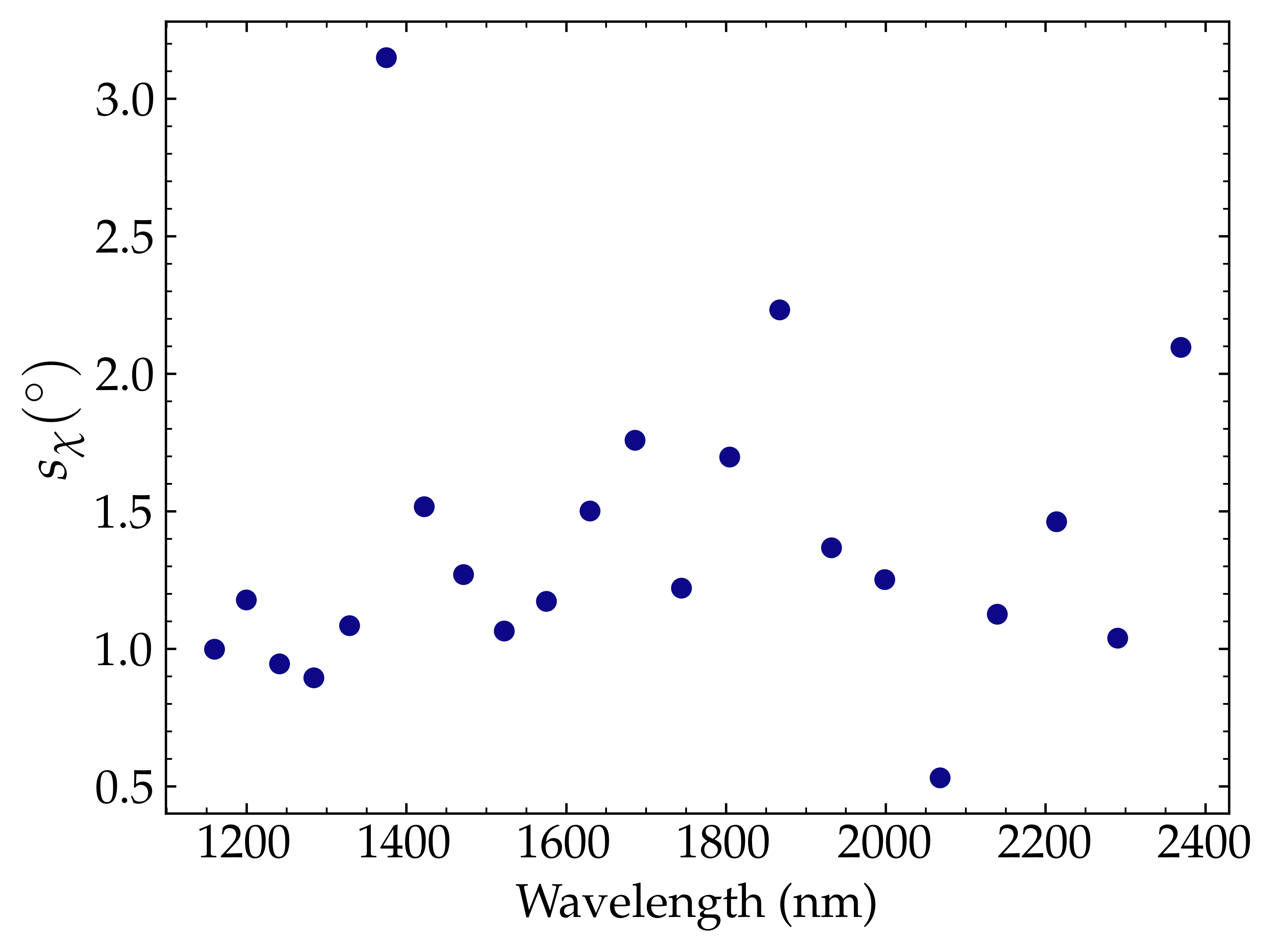}
        \caption{1\% Polarized Target}
    \end{subfigure}
    \hfill
    \begin{subfigure}[b]{0.48\textwidth}
        \centering
        \includegraphics[width=\textwidth]{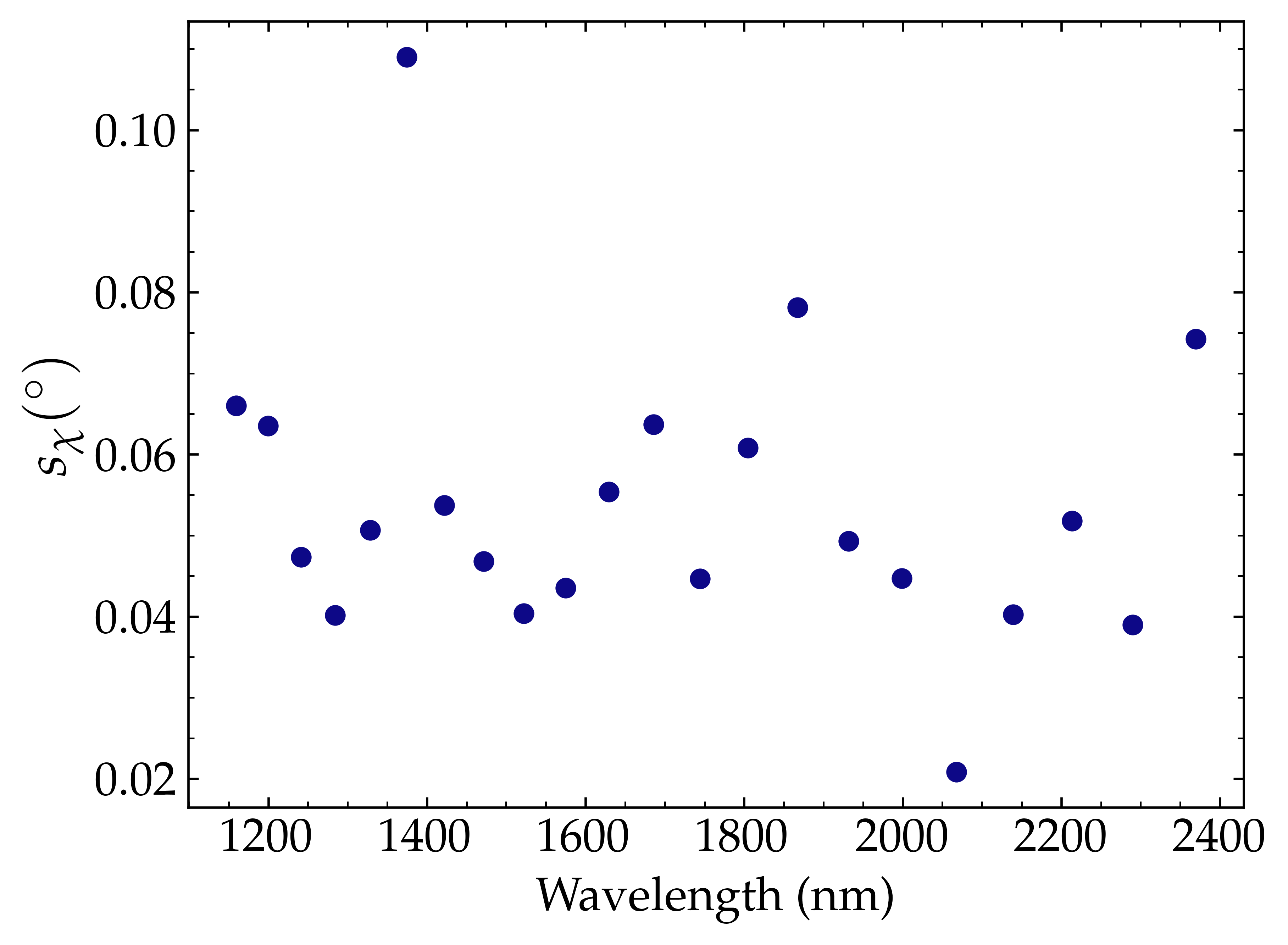} 
        \caption{30\% Polarized Target}
    \end{subfigure}
    \vspace{3pt}
    \caption{Polarimetric accuracy of the model in the DoLP ($s_p$, Equation \ref{eq:s_p}) and AoLP ($s_{\chi}$, Equation \ref{eq:s_chi}) for 1\% and 30\% polarized targets. They are strongly wavelength dependent. $s_p$ and $s_{\chi}$ have a weak dependence on the target's AoLP, which we chose to be $50^\circ$.}
    \label{fig:polacc}
\end{figure}

Using the procedure outlined in Subsection \ref{subsec:polacc}, we can calculate the polarimetric accuracy of the model in the DoLP ($s_p$) and AoLP ($s_{\chi}$) of a target. $s_{\mathrm{rel}}$ (Equation \ref{eq:sem}), the error that scales with polarized intensity, is shown in Figure \ref{fig:nbs_in_s_rel}, and $s_{\mathrm{abs}}$ (Equation \ref{eq:sabs}), the error that scales with total intensity, is shown in Figure \ref{fig:s_abs}. We calculate $s_p$ and $s_{\chi}$ for hypothetical targets with these quantities, using Equations \ref{eq:s_p} and \ref{eq:s_chi}. Figure \ref{fig:polacc} shows $s_p$ and $s_{\chi}$ for 1\% and 30\% polarized targets. 
$s_p$ is dominated by $s_{\mathrm{abs}}$ for the 1\% polarized target. $s_{\mathrm{rel}}$ becomes important for targets that are $>10\%$ polarized. Crosstalk from the YJH50 dichroic reduces the polarimetric accuracy in the \textit{J}-band for these highly polarized targets. For \textit{H-} and \textit{K}-bands, $s_\mathrm{abs}$ acts as the primary limitation on overall polarimetric accuracy for all targets, regardless of their polarization fraction. Since $s_{\mathrm{abs}}$ is dominated by residuals from the on-sky calibration, the noise in the on-sky calibration data is the main limiting factor in our model's accuracy. 

The re-coating of M3 in December 2025 and future re-coatings will impact our model accuracy. The diattenuation fit for M3 heavily contributes to the polarimetric accuracy results, and it is likely that M3's diattenuation has changed. Consequently, we interpret our polarimetric accuracy calculations as estimates for CHARIS's current state. Re-fitting the M3 model parameters with new calibration data and re-calculating $s_{\mathrm{abs}}$ would yield better estimates of $s_p$ and $s_\chi$. 

\section{CONCLUSIONS AND RECOMMENDATIONS}
\label{sec:conclusion}
In this work, we have updated the Mueller matrix model for SCExAO/CHARIS's polarimetric mode using internal source and on-sky calibration data. Due to poor fits of the previous model from Ref. \citenum{hart2021characterizationinstrumentalpolarizationeffects} to the polarized internal source calibration data, we changed the image derotator model from a linear retarder to an elliptical retarder. We characterized the polarization effects of new components in CHARIS's optical path: We find that the YJH50 dichroic produces \textit{J}-band crosstalk that we cannot characterize with our modeling capabilities, which manifests as a decrease in \textit{J}-band polarimetric accuracy for highly polarized targets. We additionally find that the NBS flips the sign of $Q$ and $V$ polarized light. Using unpolarized internal calibration data, we find that the optical path produces minimal diattenuation. From unpolarized standard star observations, we find that M3's diattenuation has increased since the previous CHARIS calibration in Ref. \citenum{hart2021characterizationinstrumentalpolarizationeffects}. We also observe increased noise in our on-sky calibration data compared to Ref. \citenum{hart2021characterizationinstrumentalpolarizationeffects} that leads to an increase in model residuals. Using our model, we report a polarimetric efficiency as a function of the derotator angle similar to that in Ref. \citenum{hart2021characterizationinstrumentalpolarizationeffects}. It is strongly wavelength-dependent and minimized around $45^\circ$ and $135^\circ$ derotator angles. From the residuals of the model, we calculate a wavelength-dependent polarimetric accuracy ranging from 0.02\% to 0.12\% in the DoLP and $0.5^\circ$ to $3.2^\circ$ in the AoLP for a 1\% polarized target. The main limiting factor for our model's accuracy is the residuals from the on-sky calibration. Finally, we provide the source code for this calibration as part of an open-source polarimetric calibration package: \texttt{pyPolCal} \cite{pypolcal}. We plan to implement this model update into the CHARIS-DPP to enable observers to easily carry out accurate quantitative polarimetry.

Our recommendations for future works are as follows:
\begin{itemize}
    \item \textbf{New unpolarized standard star observations to re-calibrate the model for M3.} 
    This is our strongest recommendation, since M3's December 2025 re-coating could change its diattenuation and retardance. We recommend observations at a wide range of altitude and parallactic angles to better fit M3's diattenuation model, which can be achieved by observing multiple unpolarized standards or the same standard at multiple altitude angles.
    \item \textbf{Regular re-calibration using the procedure outlined in this proceeding and in the \texttt{pyPolCal} tutorial notebooks.} 
    We have shown that the optical path's polarization effects can change over time by demonstrating a change in the derotator and M3 fits compared to the previous calibration. Therefore, it is necessary to regularly re-visit the model to ensure accuracy.
    \item \textbf{Characterizing other NIRWFS dichroics.} Obtaining polarized internal calibration datasets with different dichroics inserted and following the procedure from Subsection \ref{subsec:internal-procedure} would determine their effects on model accuracy.
    \item \textbf{New polarized standard star observations to verify the model.} 
    The procedure is described in Appendix \ref{appendix:polstd}. Confirming that the polarimetric accuracy is close to what is predicted by Equation \ref{eq:s_p} would help confirm this model's accuracy.

\end{itemize}

\acknowledgments 
This research is based [in part] on data collected at the Subaru Telescope, which is operated by the National Astronomical Observatory of Japan. We are honored and grateful for the opportunity of observing the Universe from Maunakea, which has the cultural, historical, and natural significance in Hawaii.

The development of SCExAO and AO3k is supported by the Japan Society for the Promotion of Science (Grant-in-Aid for Research \#23340051, \#26220704, \#23103002, \#19H00703, \#19H00695 and \#21H04998), the Subaru Telescope, the National Astronomical Observatory of Japan, the Astrobiology Center of the National Institutes of Natural Sciences, Japan, the Mt Cuba Foundation and the Heising-Simons Foundation. The development of the CACAO software is supported by the National Science Foundation under award 2410616. CHARIS was built at Princeton University under a Grant-in-Aid for Scientific Research on Innovative Areas from MEXT of the Japanese government (\#23103002). The authors wish to recognize and acknowledge the very significant cultural role and reverence that the summit of Maunakea has always had within the indigenous Hawaiian community, and are most fortunate to have the opportunity to conduct observations from this mountain. 

B.L.L. acknowledges support from the National Science Foundation Astronomy \& Astrophysics Postdoctoral Fellowship under Award No. 2401654. Any opinions, findings, and conclusions or recommendations expressed in this material are those of the author(s) and do not necessarily reflect the views of the National Science Foundation. J.N.A was supported by NASA through the NASA Hubble Fellowship grant \#HST-HF2-51547.001-A awarded by the Space Telescope Science Institute, which is operated by the Association of Universities for Research in Astronomy. This research made use of Photutils, an Astropy package for
detection and photometry of astronomical sources \cite{Bradley2025-oh}.

\bibliography{report} 
\bibliographystyle{spiebib} 

\appendix

\section{Degeneracies and Covariance in the Polarized Internal Calibration Model}
\label{appendix:mcmc}
\begin{figure}
    \centering
    \includegraphics[width=0.5\linewidth]{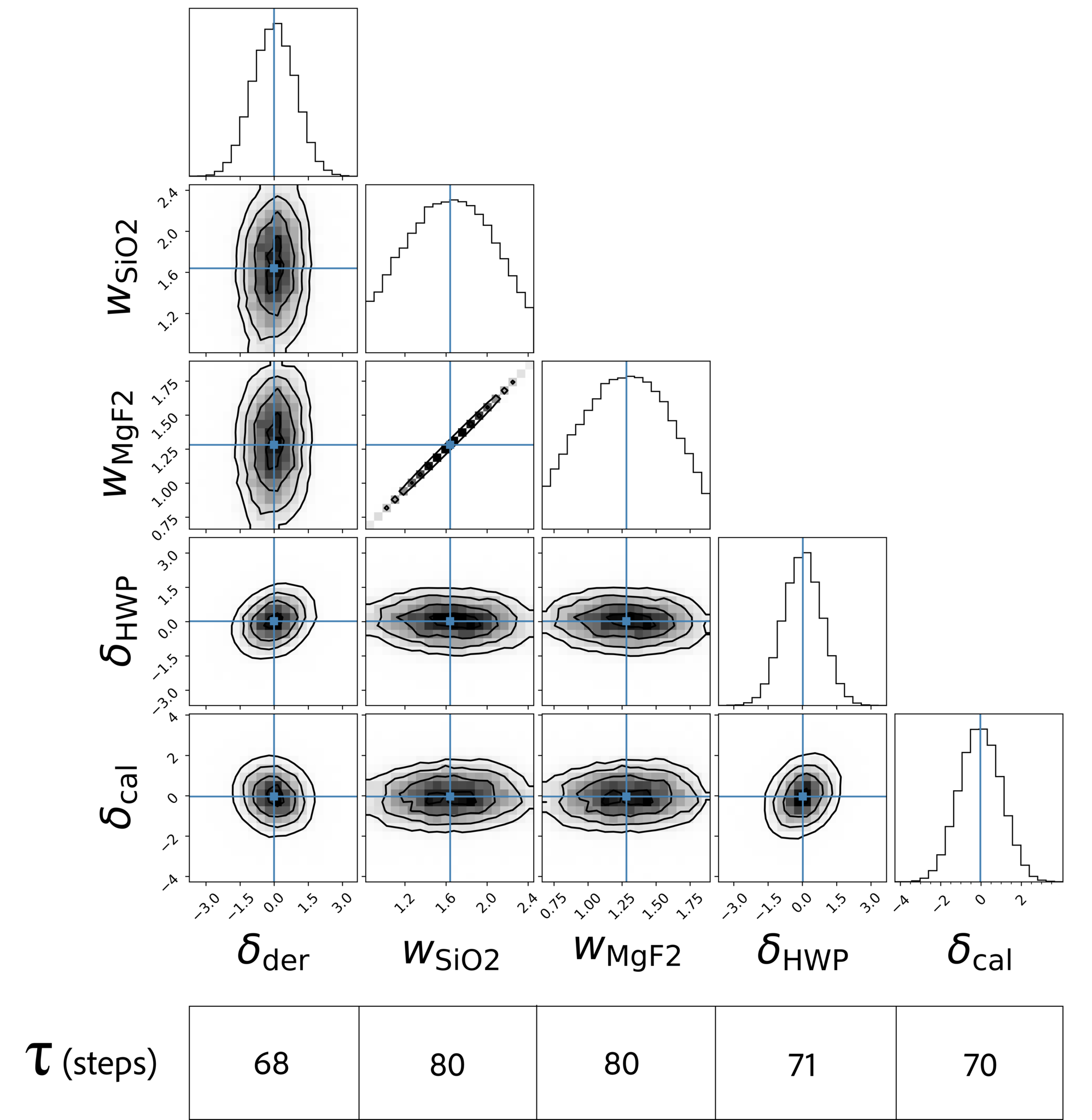}
    \vspace{5pt}
    \caption{Corner plot of the global fit for the internal calibrations. The autocorrelation time ($\tau$) of each parameter is on the bottom row. Our 10,000 step chains are 125 times the length of the longest autocorrelation time. The dataset used in this simulation is Dataset 3 (Table \ref{tab:internal_calibration_datasets}). Descriptions of all parameters can be found in Table \ref{tab:internal_parameters}. We used uniform priors for $w_{\mathrm{SiO2}}$, $w_{\mathrm{MgF2}}$ and Gaussian priors for misalignment angles.}
    \label{fig:corner}
\end{figure}

To characterize the degeneracy between the HWP layer thicknesses and fitted misalignment angles in the internal calibration model, we perform a 10,000 step Markov Chain Monte Carlo (MCMC) simulation using \texttt{emcee.EnsembleSampler}. We use Dataset 3 (Table \ref{tab:internal_calibration_datasets}) and fit across all wavelength bins simultaneously. We use the same model parameters as the \texttt{scipy.optimize.minimize} global fit described in Subsection \ref{subsec:internal-fitting}. We use uniform priors for the HWP layer thicknesses with the same bounds as the \texttt{scipy.optimize.minimize} fit and Gaussian priors with mean $\mu=0^\circ$ and standard deviation $\sigma=1^\circ$ for misalignment angles. We strictly use the MCMC simulation to qualitatively characterize degeneracy within the model and do not implement the results in our final parameter fits. 
The results of the MCMC simulation with global model parameters using Dataset 3 (Table \ref{tab:internal_calibration_datasets}) are shown in Figure \ref{fig:corner}. We performed this MCMC simulation across all wavelength bins and fit the same parameters shown in Table \ref{tab:fitted_values}. The HWP layer widths are perfectly degenerate with each other, and they are also degenerate with the misalignment angles. This is reasonable since half-wave retardance rotates the polarization state. This degeneracy is the reason why we set misalignment angles to $0^\circ$ in the wavelength fits. It is also the reason that we exclude error estimates for the globally fit parameters. Because these parameters are not independent, we cannot estimate the error on the global fits in Table \ref{tab:fitted_values} using the calculation from Ref. \citenum{hart2021characterizationinstrumentalpolarizationeffects}. We also cannot use estimates from the posterior width since the measurement errors are not accurate. The residuals (using the posterior medians as model parameters) are slightly worse than the \texttt{scipy.optimize.minimize} results in Table \ref{tab:fitted_values}, so we do not implement the results of this simulation in the final model. 
\section{Elliptical Retardance Plots from Before the NBS Installation}
\begin{figure}[H]
    \centering
    \begin{subfigure}[b]{0.48\textwidth}
        \centering
        \includegraphics[width=\textwidth]{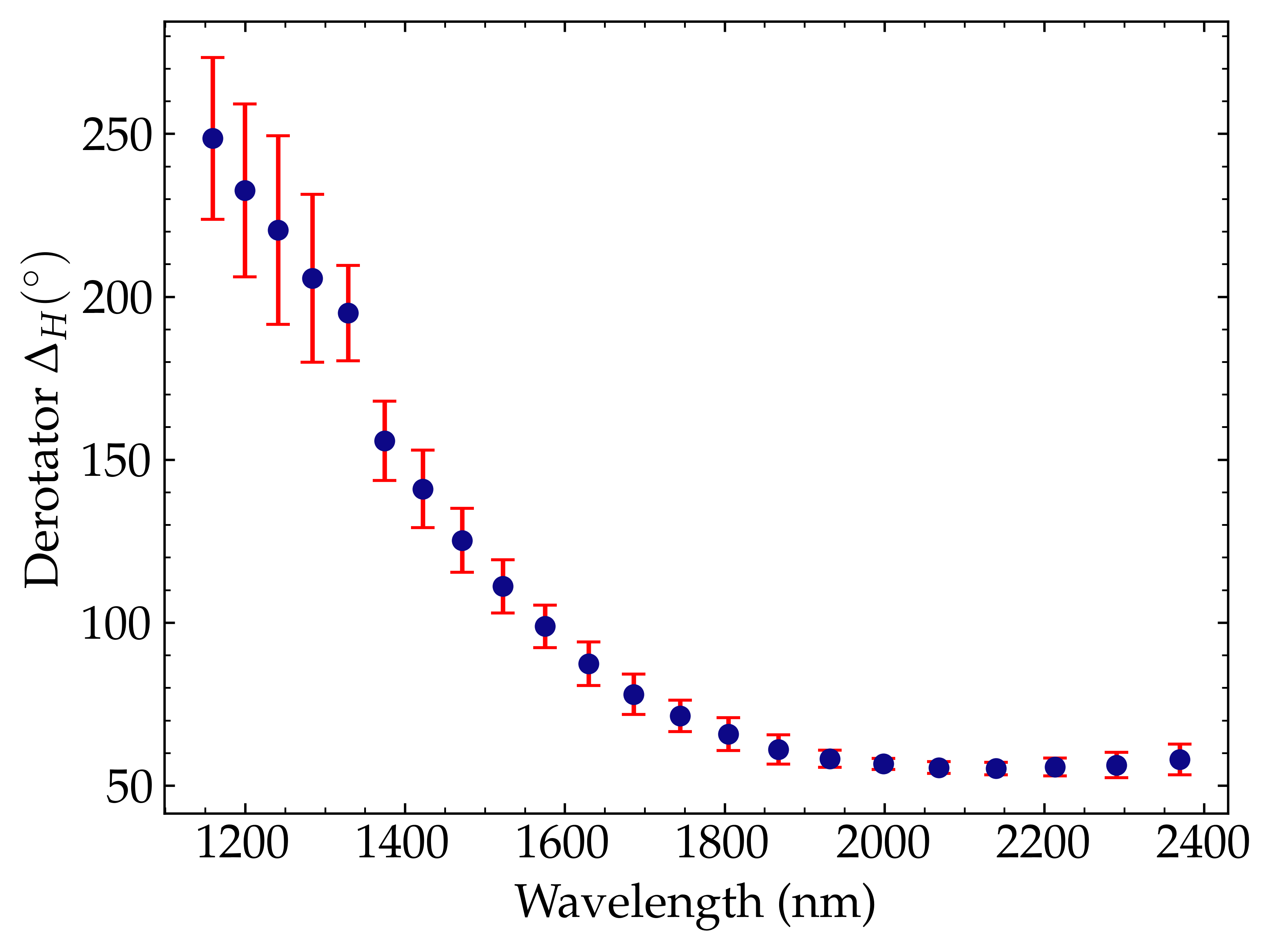}
    \end{subfigure}
    \begin{subfigure}[b]{0.48\textwidth}
        \centering
        \includegraphics[width=\textwidth]{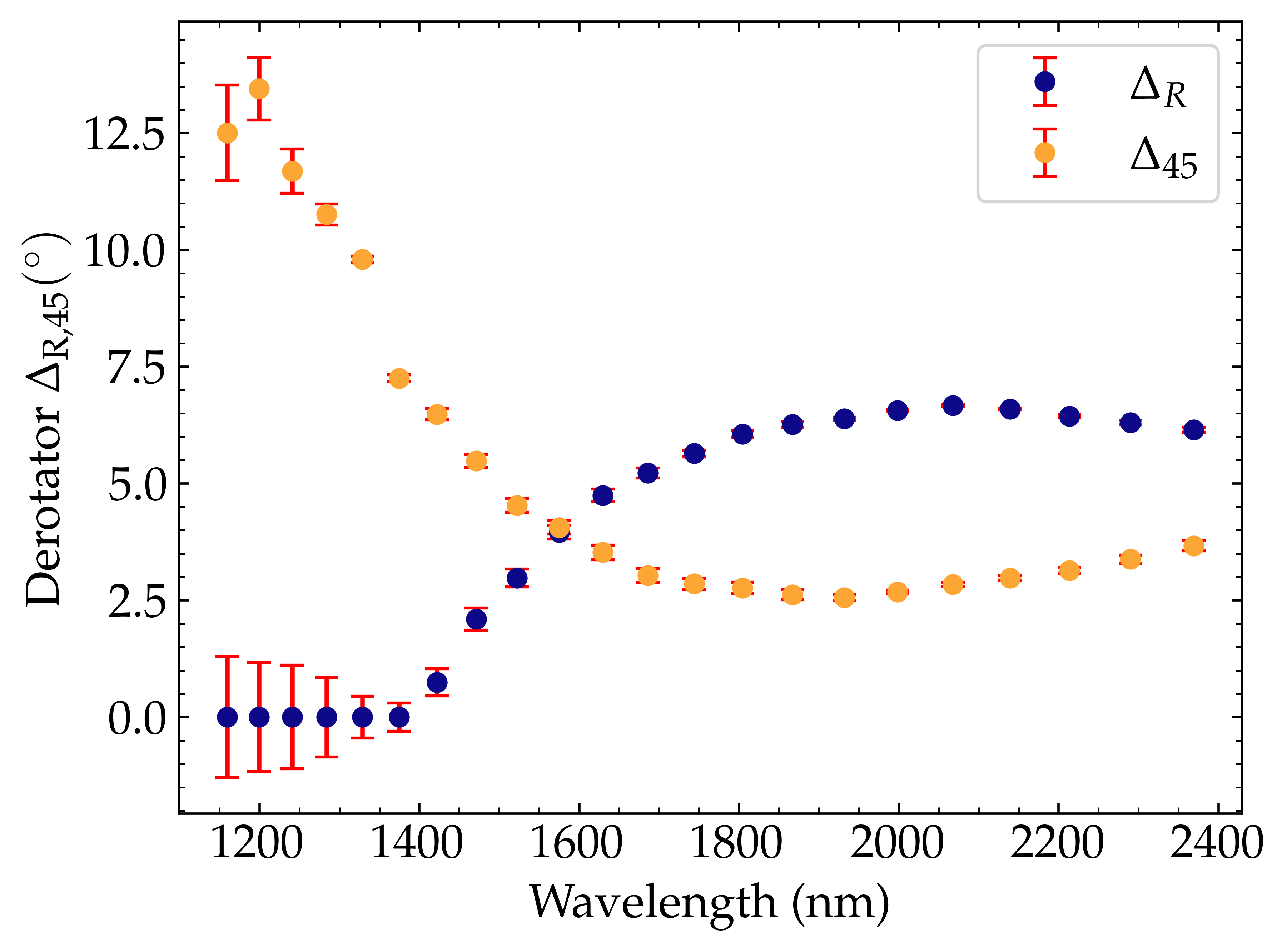}
    \end{subfigure}
    \caption{Elliptical retardance components (see Equation \ref{ellipticalretarder}) of the derotator fit using Dataset 2 (Table \ref{tab:internal_calibration_datasets}). $\Delta_H$ is horizontal retardance, $\Delta_{45}$ is $45^\circ$ retardance, and $\Delta_R$ is right-handed circular retardance. The NBS is not installed and the YJH50 dichroic is inserted in this dataset. We calculated the errors using the method from Appendix E of Ref. \citenum{van_Holstein_2020}.}
    \label{fig:derotator_retardance_nbs_out_pick_in}
\end{figure}

\begin{figure}[H]
    \centering
    \begin{subfigure}[b]{0.48\textwidth}
        \centering
        \includegraphics[width=\textwidth]{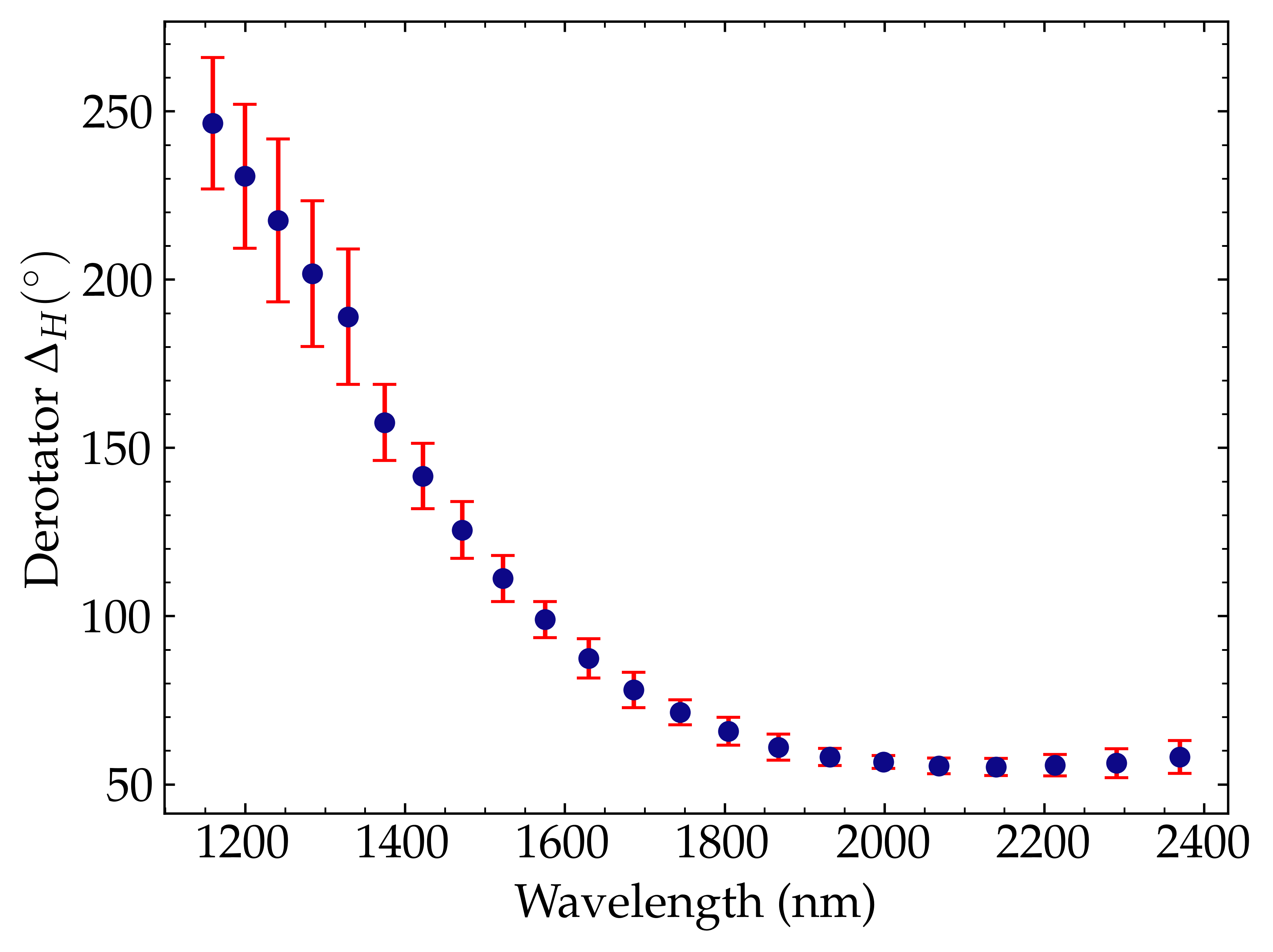}
    \end{subfigure}
    \begin{subfigure}[b]{0.48\textwidth}
        \centering
        \includegraphics[width=\textwidth]{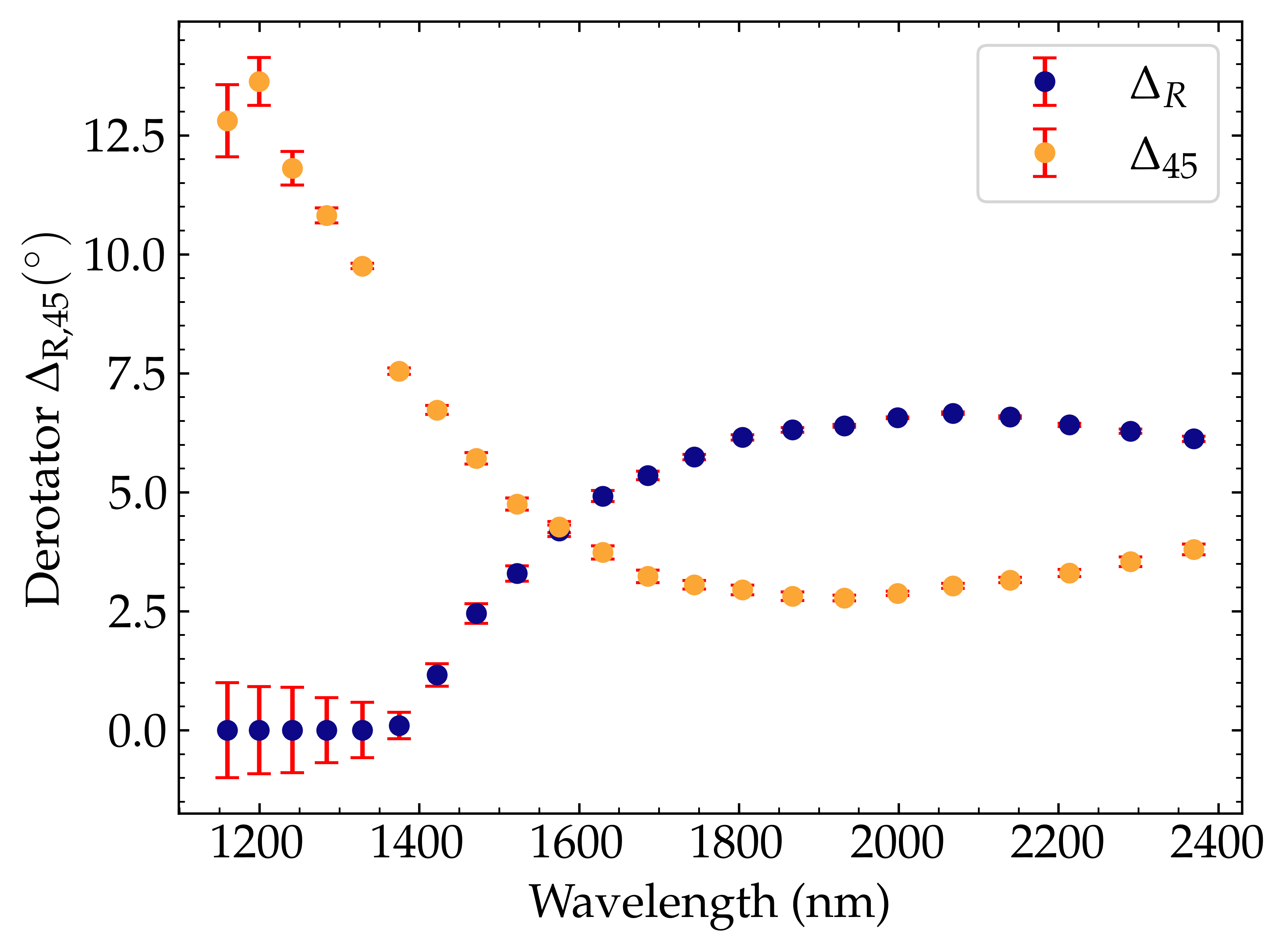}
    \end{subfigure}
    \caption{Elliptical retardance components (see Equation \ref{ellipticalretarder}) of the derotator fit using Dataset 1 (Table \ref{tab:internal_calibration_datasets}). $\Delta_H$ is horizontal retardance, $\Delta_{45}$ is $45^\circ$ retardance, and $\Delta_R$ is right-handed circular retardance. The NBS is not installed and the YJH50 dichroic is not inserted in this dataset. We calculated the errors using the method from Appendix E of Ref. \citenum{van_Holstein_2020}. This shows that the elliptical retardance is not due to the new optics, since neither are present in this dataset.}
    \label{fig:derotator_retardance_nbs_out_pick_out}
\end{figure}

Figures \ref{fig:derotator_retardance_nbs_out_pick_in} and \ref{fig:derotator_retardance_nbs_out_pick_out} show the fit of an elliptical retarder model (Equation \ref{ellipticalretarder}) to the derotator ($M_{\mathrm{der}}$ in Equation \ref{internalmodel}) using internal calibration datasets 2 and 1 (Table \ref{tab:internal_calibration_datasets}), respectively. Both datasets are from before the NBS installation. Dataset 2 contains the YJH50 dichroic while Dataset 1 does not. The plots are identical, showing that retardance from the dichroic is not absorbed into the derotator fit.
\label{appendix:ellipticalretfit}
\section{ON ELLIPTICAL POLARIZATION}
\label{jarenappendix}

One area where this study differs from prior investigations is on the inclusion of the K-mirror as an elliptical retarder. While there is no physical reason for the K-mirror to be an elliptical retarder (in the absence of substantive birefringence on the coating), we know that it arises in optical systems. We motivate this with Pauli spin matrices in Equation \ref{eq:pauli_spin},

\begin{equation}
    \sigma_1 = 
    \begin{pmatrix}
        1 & 0 \\
        0 & -1 \\
    \end{pmatrix}
    \phantom{00000}
    \sigma_2 = 
    \begin{pmatrix}
        0 & 1 \\
        1 & 0 \\
    \end{pmatrix}
    \phantom{00000}
    \sigma_3 = 
    \begin{pmatrix}
        0 & -i \\
        i & 0 \\
    \end{pmatrix}.
    \label{eq:pauli_spin}
\end{equation}

These three matrices represent a half-wave linear retarder at $0^\circ$ ($\sigma_1$), $45^\circ$ ($\sigma_2$), and a half-wave circular retarder ($\sigma_3$). The commutator relationships of these matrices are given in Equation \ref{eq:commutator},

\begin{equation}
    [\sigma_i, \sigma_j] = 2i \sigma_k.
    \label{eq:commutator}
\end{equation}

Working this out for two misaligned linear retarders (i.e. $\sigma_1$ and $\sigma_2$), then yields a circular retarder, as shown in Equation \ref{eq:commutator_s3}

\begin{equation}
    \sigma_1 \sigma_2 - \sigma_2 \sigma_1 = 2i \sigma_3.
    \label{eq:commutator_s3}
\end{equation}

In other words, a cascade of misaligned linear retarders can manifest in circular retardation. If these retarders are not perfectly misaligned, the result is a combination of linear and circular retardance -- also known as elliptical retardance. Any mirror at non-zero angle of incidence in an optical system will apply both diattenuation and retardance. Adaptive optics systems are nearly entirely composed of such mirrors, and as polarized light propagates through them, it will slowly accrue circular retardance. The model used in this paper to fit to the CHARIS data is representative of the state of the art of polarimetric calibration, but is incapable of generating circular retardance due to the 6 optics considered in the model. In actuality, the cascade of optics that feed CHARIS is composed of tens of individual refractive or reflective surfaces, including off-axis parabolic relays, and fold mirrors. We hypothesize that each of these applies a small amount of mis-aligned linear diattenuation which ultimately arises as circular retardance.

\section{Fold Mirror Diattenuation in Visible Light}
\label{appendix:foldmirrordiat}
\begin{figure}[H]
    \centering
    \includegraphics[width=0.5\linewidth]{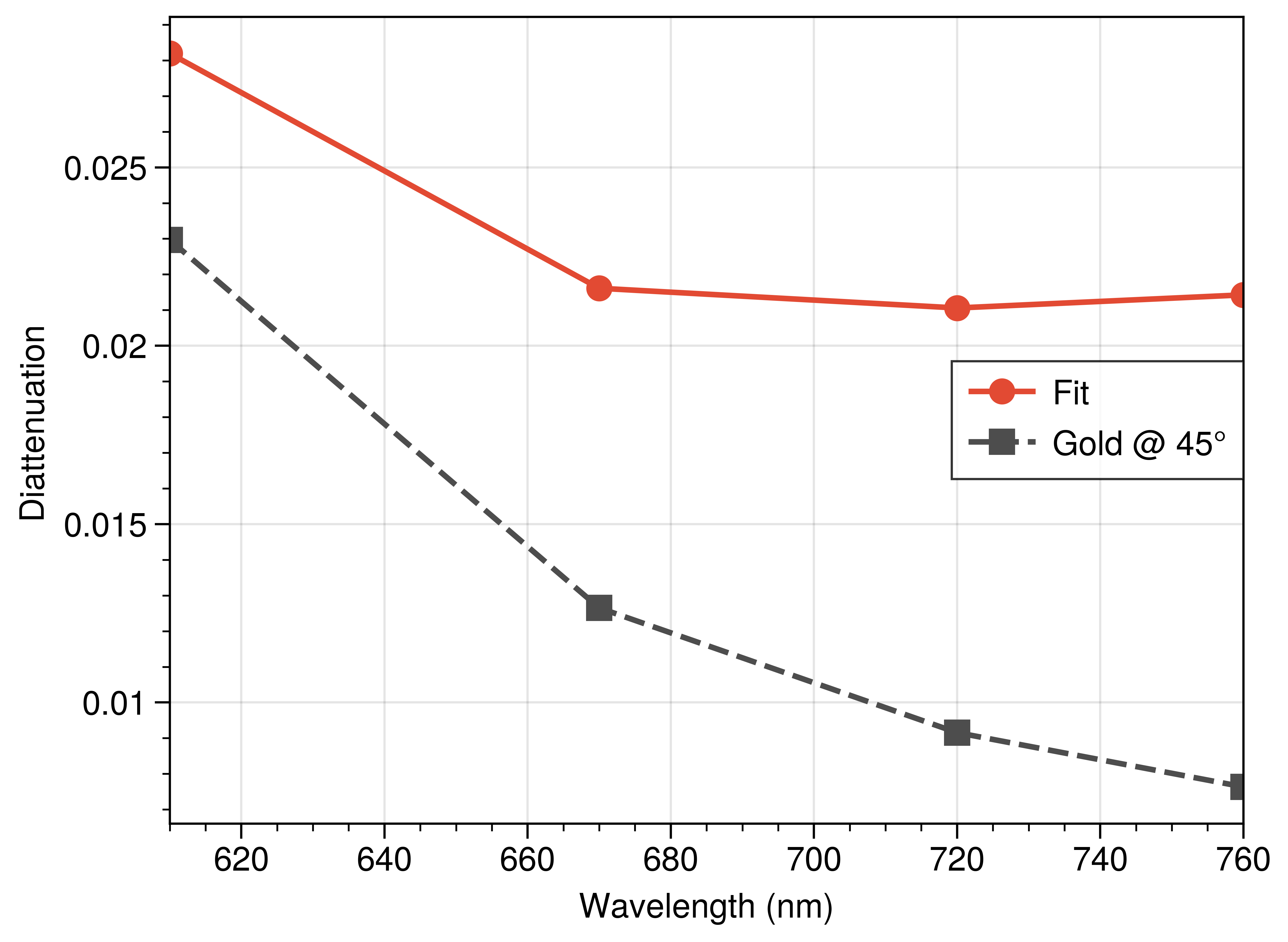}
    \caption{Diattenuation of the fold mirror as a function of wavelength in visible light, fit using unpolarized calibration data from SCExAO/VAMPIRES. The dotted line is the theoretical diattenuation of a gold mirror at a $45^\circ$ angle of incidence, calculated using the index of refraction.}
    \label{fig:vampiresmirror}
\end{figure}

Figure \ref{fig:vampiresmirror} shows the fold mirror diattenuation in visible light. We carried out this fit with SCExAO/VAMPIRES unpolarized calibration data. The theoretical diattenuation of a gold mirror at a $45 ^\circ$ angle of incidence is also plotted, which is lower than the fold mirror fit for all wavelengths. A number of factors could cause this discrepancy, such as a protective coating on the mirror or degradation of the gold over time.  However, it shows that the scale of the fitted diattenuation is roughly that of a gold mirror.

\section{Testing the Model with a Polarized Standard}
\label{appendix:polstd}
\begin{figure}[H]
    \centering
    \includegraphics[width=0.5\linewidth]{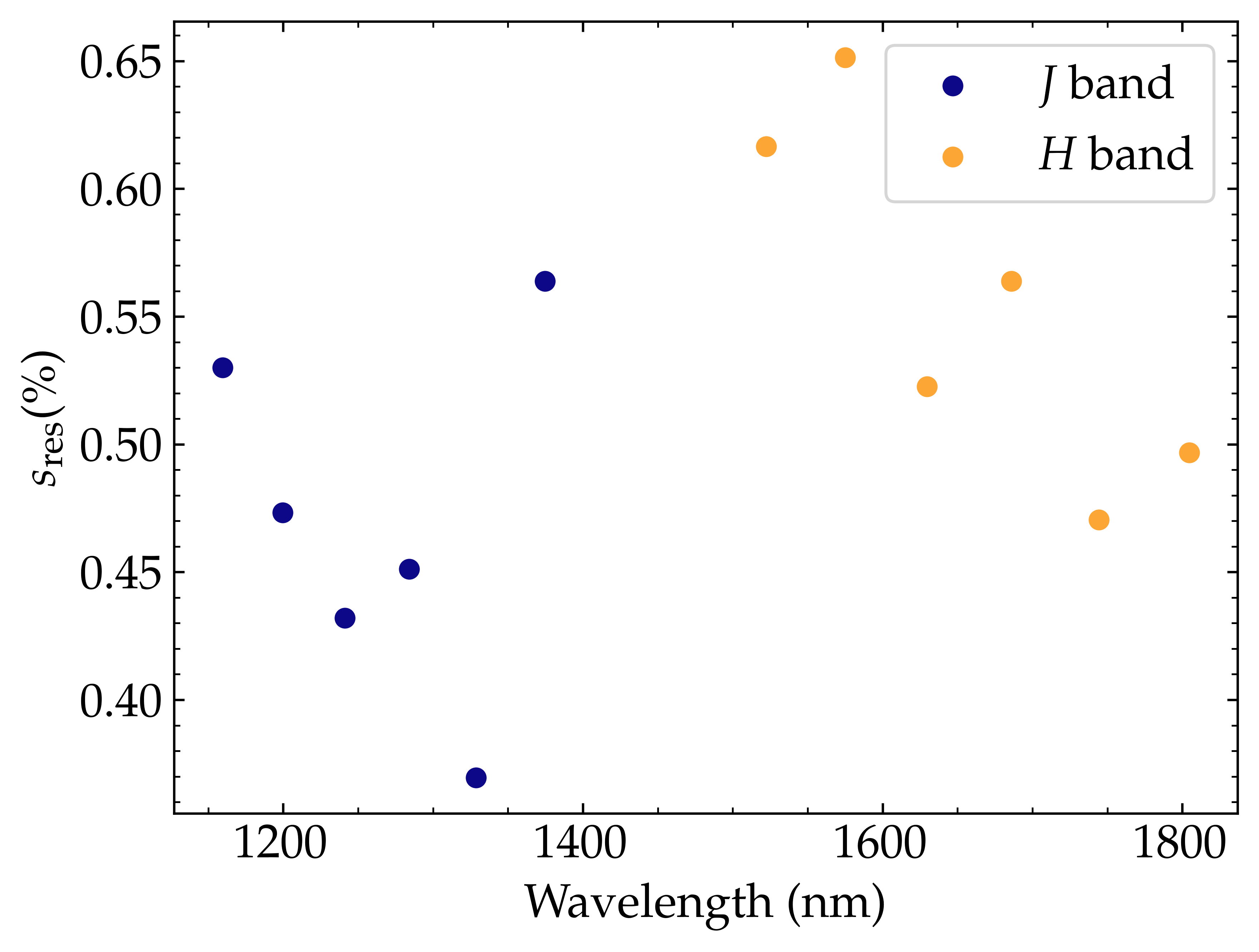}
    \caption{The standard deviation of the residuals per wavelength for the polarized standard HD30675.}
    \label{fig:polstd}
\end{figure}

In February 2025, prior to the NBS installation, we observed polarized standard star HD30675 \cite{zetapersei}. Serkowski law parameters, which are known for this star, allow us to extract the star's DoLP and AoLP as a function of wavelength.  It is $1-2\%$ polarized in the $J$- and $H$-bands. The error on the Serkowski law parameters for HD30675 in the $K$-band is large, so we neglect this band. We model the input Stokes vector and apply the Mueller matrix model. We use the model parameters fit in this proceeding except M3's diattenuation, since the calibration data is old and M3's diattenuation appears to change over time. For the M3 model parameters, we use a fit from February 2025 with an unpolarized standard. Comparing this model with the observed double differences is a good test for the Mueller matrix model. Since this data and the M3 fit are from February 2025, this is not a meaningful test of the current model. However, it does provide evidence that the M3 retardance model from Ref. \citenum{hart2021characterizationinstrumentalpolarizationeffects} works generally. We currently do not have any other concrete tests of M3's retardance. This test also serves as a sanity check for the internal calibration model. Figure \ref{fig:polstd} shows the standard deviation of the residuals, which we interpret as the DoLP measurement uncertainty as a function of wavelength. Variable seeing ($\sim1$ arcsec) and uncertainty in the Serkowski law parameters contribute to the error. Given these factors, the uncertainty is well within reason, confirming that the Mueller matrix model effectively corrects for instrumental polarization effects in on-sky data. 
\end{document}